\documentclass{aa}
\usepackage{braket}
\usepackage{graphicx}\usepackage{txfonts}
\usepackage{lipsum}
\usepackage{subcaption}
\usepackage{lscape}
\usepackage{placeins}
\newcommand{\litiosette}{\ensuremath{^{7}\mathrm{Li}}}
\newcommand{\berilliosette}{\ensuremath{^{7}\mathrm{Be}}}
\newcommand{\borootto}{\ensuremath{^{8}\mathrm{B}}}
\newcommand{\ancap}{\ensuremath{\bar{\nu}_{e}}}
\begin{document}
\title{Revisiting $^7$Be weak and radiative transition rates in Big Bang nucleosynthesis: Implications for the primordial lithium problem}
\author{Simone Taioli\inst{1,2,3*} \and Francesca Triggiani\inst{4}
\and Stefano Simonucci\inst{5,6*}
}
\institute{
European Centre for Theoretical Studies in Nuclear Physics and Related Areas (ECT*), Bruno Kessler Foundation, Strada delle Tabarelle, 286, 38123 Trento, Italy
\and Trento Institute for Fundamental Physics and Applications (TIFPA-INFN), Via Sommarive, 14, 38123 Trento, Italy
\and INAF, Observatory of Abruzzo, Via Mentore Maggini snc, 64100 Teramo, Italy \and National Institute for Nuclear Physics (INFN), Sezione di Milano, Via Celoria 16, 20133 Milano, Italy.
\and School of Science and Technology -- Physics Division, University of Camerino, Via Madonna delle Carceri, 9B, 62032 Camerino, Italy
\and National Institute for Nuclear Physics (INFN), Sezione di Perugia, Via Alessandro Pascoli, 23c, 06123 Perugia, Italy\\
\email{stefano.simonucci@unicam.it}\\
\email{taioli@ectstar.eu}
}
  \abstract
{The primordial \litiosette~abundance predicted by standard Big Bang nucleosynthesis (BBN) exceeds the value inferred from old metal-poor stars by a factor of approximately three to four. In this model, most primordial $^{7}$Li is produced as $^{7}$Be in the early Universe and is subsequently converted by electron capture (EC). However, any additional production or destruction channels for $^{7}$Be, such as proton capture (PC) or antineutrino capture (AC) during BBN, may affect the final $^{7}$Li yield.}
{We quantify the effect on $^{7}$Be depletion of (i) in situ EC, including the corresponding AC channel, during the Big Bang epoch; (ii) a positron decay open channel from the Be nuclear excited state; and (iii) PC via both the standard radiative channel $^{7}$Be$(p,\gamma)^{8}$B, including the possibility of stimulated emission (SE) due to the dense $\gamma$ photon background present during the nuclear statistical equilibrium (NSE) epoch, as well as by analysing the possible contribution of a three-body `Auger-like' variant in which the capture energy is transferred to a continuum electron.}
{We calculated decay rates using first-order perturbation theory, modelling weak decay with a Fermi contact interaction and factorising hadronic and leptonic currents in the transition operator matrix elements. The field operators describing hadrons and leptons were determined using mean-field methods. 
Thermally averaged nuclear rates were obtained by folding the relevant cross sections with Maxwell–Boltzmann distributions and accounting for the number density of the particle triggering the nuclear reaction in the temperature range 1 keV < $k_{\rm B}T$ < 275 keV, which also lies within the range characterising the NSE epoch.
For $^{7}$Be$(p,\gamma)^{8}$B, we used the astrophysical $S_{17}$ factor and included Debye screening. The SE channel was addressed using the quantum field theory formalism. The Auger-like cross section was derived using a plane-wave electron approximation and evaluated numerically at representative BBN temperatures and densities.}
{The calculated $^{7}$Be EC rate decreases rapidly as the Universe expands and cools over time. Including the corresponding AC channel enhances the total weak destruction rate at early times, although its contribution rapidly decreases as the temperature falls.
The thermally averaged $^{7}$Be$(p,\gamma)^{8}$B rate is only marginally enhanced when the nuclear SE mechanism and plasma screening are considered (1–3\% at $k_{\rm B}T \simeq 87$ keV).
The Auger-like three-body capture cross section is approximately $4\times10^{-3}$ of the radiative channel at $k_{\rm B}T=100$ keV, and it decreases to about $10^{-10}$ of the radiative channel by $k_{\rm B}T=10$ keV. Our first-principles weak rates differ substantially from previous estimates based on $\log(ft)$ systematics, leading to a revision of the predicted $^{7}$Be half-life by nearly one order of magnitude under BBN conditions.}
{Within the approximations adopted here, \(^7\)Be EC, PC, and $\beta^+$ decay provide percent-level corrections to the fastest and dominant depletion channel, $^{7}\mathrm{Be}(n,p)^{7}\mathrm{Li}$, which is followed by the rapid destruction $^{7}\mathrm{Li}(p,\alpha)\,^4\mathrm{He}$. On their own, these decays are unlikely to resolve the cosmological lithium discrepancy with observations, as first-principles simulations lead to a correction of about one order of magnitude in the half-life (approximately 2 days) compared with phenomenological approaches based on $\log(ft)$, which does not help close the gap with observations of old halo stars. Our first-principles study paves the way for a complete reassessment of the network of all nuclear transitions in order to accurately quantify the abundance of \(^7\)Li within established nuclear and atomic physics models, but beyond semi-empirical approximations based on $\log(ft)$ systematics, and before considering interpretations that assume physics beyond the Standard Model.}
\keywords{Big Bang nucleosynthesis, early Universe, electroweak decays, strong interaction, isotopic abundances of light nuclei, the cosmological lithium problem}
\titlerunning{The primordial lithium problem}
\maketitle
\section{Introduction}
Standard Big Bang nucleosynthesis (BBN) is one of the cornerstones of modern cosmology. It describes the synthesis of the lightest nuclei during the first few minutes after the Big Bang. Together with observations of the cosmic microwave background (CMB), which provide a precise determination of the primordial baryon density, BBN constitutes one of the most stringent tests of the standard cosmological model and of the physics of the early Universe \citep{Alpher1948,Wagoner1967,Yang1984,KolbTurner1990,Peebles1993,Bertulani,Miranda}. Over recent decades, increasingly precise cosmological observations, improved nuclear reaction measurements, and refined theoretical calculations have transformed BBN into a precision discipline \citep{Cyburt2016,Pitrou2018,Fields2020}.
Within the standard cosmological framework, the primordial abundances depend essentially on a single free parameter, namely the baryon-to-photon ratio ($\eta$), whose value is accurately determined from CMB observations. Modern BBN calculations are routinely performed using state-of-the-art numerical codes such as \texttt{PArthENoPE} \citep{Pisanti2008,Gariazzo2022} and \texttt{PRIMAT} \citep{Pitrou2018}, which incorporate updated nuclear reaction networks and the latest cosmological and particle physics inputs. The remarkable agreement between theoretical predictions and the observed primordial abundances of D, $^{2}$H, $^{3}$He, and $^{4}$He is one of the major successes of modern cosmology.
Despite this overall success, one long-standing discrepancy remains unresolved. Standard BBN calculations predict a primordial $^{7}$Li abundance systematically higher than that inferred from spectroscopic observations of old metal-poor sub-solar-mass Population II halo stars. These long-lived stars formed shortly after the Big Bang from gas that had not yet been significantly enriched by successive generations of stars, making them the best available tracers of the primordial chemical composition. The nearly constant $^{7}\mathrm{Li}$ abundance observed over a wide range of effective temperatures and metallicities, known as the Spite plateau \citep{Spite1982}, together with reported detections of the more fragile $^{6}\mathrm{Li}$ isotope of cosmic-ray origin \citep{2022MNRAS.509.1521W}, has generally been interpreted as evidence that these stars have largely preserved their primordial lithium abundance, although modest stellar depletion cannot be ruled out completely \citep{articolobonifacio}.
As stated, the primordial $^{7}$Li abundance inferred from observations is lower than the standard BBN prediction, by approximately a factor of three to four:
\[
\left(^{7}\mathrm{Li}/\mathrm{H}\right)_{\mathrm{BBN}}
\simeq (4{-}5)\times10^{-10},
\qquad
\left(^{7}\mathrm{Li}/\mathrm{H}\right)_{\mathrm{obs}}
\simeq (1.3{-}1.6)\times10^{-10},
\]
as inferred by combining CMB constraints from the Wilkinson Microwave Anisotropy Probe and Planck missions with BBN calculations \citep{primordial_nucleosynthesis_Coc_Vangioni,articoloindiano2.27}. In contrast, the $^{7}$Li abundance in the interstellar medium is higher than the primordial value expected from BBN, reflecting additional Galactic production mechanisms \citep{casuso2003origin}. The resulting cosmological lithium problem, for which $^{7}$Li is the only significant BBN outlier at a level exceeding $5\sigma$ \citep{Cyburt2016,Pitrou2018}, has motivated extensive investigations in nuclear physics, stellar evolution, and possible extensions of the Standard Model \citep{Cyburt2016,Fields2020}.
Several classes of explanations have been proposed. Astrophysical mechanisms include surface depletion of $^{7}$Li through rotationally induced mixing, cosmic ray-related processes, thermohaline transport, magnetic effects, and microscopic diffusion during stellar evolution \citep{michaud1986lithium,charbonnel2010thermohaline,andrews1988rotational,eggenberger2010effects,michaud1991lithium}. These processes may transport lithium to layers at temperatures of a few million kelvin, where it is destroyed by proton capture. Systematic uncertainties associated with non-local thermodynamic equilibrium (NLTE) and three-dimensional effects in stellar atmospheres may also modify the inferred lithium abundance \citep{mishra2024plasma}. Other possibilities involve uncertainties in nuclear reaction rates, including $^{3}\mathrm{He}(\alpha,\gamma)^{7}\mathrm{Be}$ and weak processes involving thermally populated nuclear or electronic excited states. Finally, extensions beyond the Standard Model may alter nuclear reaction rates, fundamental constants, or the baryon distribution during BBN without necessarily spoiling the successful prediction of the primordial deuterium abundance.\\
\indent Beyond its cosmological relevance, lithium is also an important tracer of transport and mixing processes in stellar astrophysics. Observations of the Sun and solar analogues indicate substantial lithium depletion, with the photospheric abundance being approximately two orders of magnitude lower than the meteoritic value \citep{asplund2009chemical}. In stars less massive than the Sun, the convective envelope remains sufficiently extended for lithium destruction to continue during the main sequence. Pronounced lithium depletion is also observed in younger Population I F-type stars with masses of approximately $1.2$--$1.5\,M_{\odot}$ and effective temperatures between $6400$ and $6900$ K. This feature, known as the lithium dip, is attributed to non-standard mixing processes such as rotationally induced transport and microscopic diffusion \citep{boesgaard1986alithium,boesgaard1986blithium,balachandran1995lithium}. Although relevant to stellar evolution, the lithium dip is distinct from the determination of the primordial lithium abundance from Population II halo stars. In higher-mass stars, the outer convective regions contract and no longer reach layers hot enough to efficiently destroy lithium.\\
\indent
A particularly important issue is the determination of reliable electron-capture (EC) and proton-capture (PC) rates on $^{7}$Be under astrophysical temperature and density conditions, which may differ substantially from terrestrial environments and may vary by orders of magnitude between stellar envelopes and interiors. These rates directly affect the production and destruction pathways of $^{7}$Li. In stellar plasmas, their temperature and density dependence can also be influenced by electron degeneracy and screening. In the solar core, the temperature is sufficiently high and the density sufficiently low that electrons can be treated approximately as a classical gas. In denser environments produced, for example, during gravitational contraction, degeneracy may instead become important. Under such conditions, the widely used Debye--Hückel approximation for electron screening \citep{bahcall1962beta,bahcall20187be} may become questionable when the electron de Broglie wavelength is comparable to the interparticle separation.
\\
\indent
A temperature- and density-dependent EC rate for $^{7}$Be in stars ascending the red giant branch and the asymptotic giant branch was developed using a mean-field ab initio approach based on the numerical solution of the quantum-mechanical equations of motion \citep{Simonucci_2013,palmerini2016lithium}. The electron density at the nucleus, calculated within a Hartree--Fock framework, leads to a more effective EC process and to a $^{7}$Be lifetime shorter than that obtained using the Debye--Hückel treatment adopted in the Bahcall rate \citep{bahcall1962beta,bahcall20187be}, on which the solar-extrapolated rates of \citep{adelberger2011solar} are based. When applied to low-mass stars of approximately $2$--$3\,M_{\odot}$ and solar metallicity undergoing deep mixing, the revised rate yields lower equilibrium abundances of $^{7}$Be in the layers above the hydrogen-burning shell \citep{palmerini2016lithium}.
\\
\indent
A shorter $^{7}$Be EC lifetime produces a maximum difference of approximately $4\%$ in the efficiency of the $^{7}$Be channel relative to previous estimates \citep{vescovi2019effects,iben1967effect}. This modification also affects the competition between EC and the PC reaction $^{7}\mathrm{Be}(p,\gamma)^{8}\mathrm{B}$, which determines the branching between the $^{7}$Be and $^{8}$B solar-neutrino production channels and is therefore relevant to the interpretation of terrestrial measurements of the $^{8}$B neutrino flux.\\
\indent
Electroweak decay rates are also expected to depend sensitively on the ionisation state and on the local population of atomic and nuclear levels. The NLTE conditions may modify both the electron density at the nucleus and the effective plasma-screening potential \citep{mascali2022novel,mascali2023new,agodi2023nuclear}. The impact of NLTE populations has, for example, been assessed in low-density laboratory magnetoplasmas and compared with high-density collision-driven LTE plasmas, revealing potentially significant differences \citep{mishra2024plasma}. A precise description of these effects is therefore desirable for modelling lithium production and destruction in astrophysical environments and for developing a consistent picture of Galactic $^{7}$Li nucleosynthesis. Although the present treatment is restricted to LTE conditions, ionisation and thermal excitation are explicitly included.\\
\indent
Thus, despite extensive theoretical \citep{Simonucci_2013,palmerini2016lithium,vescovi2019effects}, experimental \citep{damone20187}, and observational efforts \citep{annurev:/content/journals/10.1146/annurev-nucl-102010-130445,Richard_2005,Fields2022}, no universally accepted solution has been found for the discrepancy between the primordial lithium abundance predicted by standard BBN and that inferred from metal-poor halo stars. The discordance persists despite updated experimental nuclear reaction rates and increasingly precise cosmological parameters.
\\
\indent
Nearly $95\%$ of the primordial $^{7}$Li is produced indirectly as $^{7}$Be during BBN. The surviving $^{7}$Be subsequently undergoes EC after the nucleosynthesis epoch, once atomic electrons become available, thereby producing $^{7}$Li. Consequently, an accurate description of the weak and strong interaction channels involving $^{7}$Be is essential for assessing whether conventional nuclear and atomic physics can alleviate the cosmological lithium discrepancy. In particular, it is important to determine whether weak destruction channels operating during BBN and processes neglected or treated semi-empirically in previous calculations can modify the surviving $^{7}$Be abundance.
The aim of this work is to reassess selected weak and strong reaction rates involving $^{7}$Be and investigate previously unexplored plasma-induced processes that may have affected the primordial \litiosette~abundance. Although the dominant BBN reaction network is well established, some weak-interaction channels and finite-temperature effects have received comparatively less attention or have been treated using semi-empirical extrapolations whose accuracy under early-Universe conditions remains uncertain. A first-principles treatment of relativistic atomic and nuclear dynamics can therefore provide reaction rates more closely matched to the highly ionised finite-temperature plasma present during BBN.
Within our ab initio framework \citep{morresi2018relativistic,taioli2021relativistic,taioli2022theoretical}, we investigate weak processes that convert \berilliosette~into \litiosette, including EC, antineutrino capture (AC), and positron emission from thermally populated nuclear excited states. We also calculate the PC rate of $^{7}$Be during BBN and examine two additional mechanisms: (i) a three-body Auger-like process in which the capture energy is transferred to a continuum electron rather than emitted as a real photon and (ii) nuclear stimulated emission (SE) induced by the dense thermal photon background.
Previous studies \citep{fuller1982,fuller_smith} estimated several of these weak channels using $\log(ft)$ values derived from laboratory systematics \citep{nudat} and extrapolated to early-Universe conditions. These estimates yielded half-lives of several days for some of the relevant transitions \citep{fuller_smith}, which is substantially longer than the characteristic duration of primordial nucleosynthesis. However, BBN occurred in a hot, dense, and highly ionised relativistic plasma, in which atomic charge states, continuum electrons, nuclear excitations, and lepton distributions can significantly modify weak decay rates.
Experimental evidence that atomic ionisation can strongly affect $\beta$ decay rates has been obtained in storage-ring measurements of highly charged ions \citep{Litvinov_2011}. A striking example is fully ionised $^{187}\mathrm{Re}^{75+}$, whose decay rate is enhanced by approximately nine orders of magnitude relative to neutral $^{187}\mathrm{Re}$, which has a half-life of about $42$ Gyr. This enhancement arises from bound-state $\beta$ decay, a channel that is energetically inaccessible in the neutral atom and in which the emitted electron is created directly in a vacant bound orbital. Thermal population of nuclear excited states may likewise modify decay rates under astrophysical conditions.
Plasma-induced modifications of $\beta$ decay, such as those described by the Takahashi--Yokoi model \citep{TAKAHASHI1983578,TAKAHASHI1987375}, are also among the motivations for the PANDORA facility at INFN--LNS in Catania \citep{mascali2022novel}. PANDORA employs an electron-cyclotron-resonance magnetoplasma to enable direct measurements of nuclear decay rates in a controlled plasma environment.
In the present work, we recalculate the relevant $^{7}$Be weak-interaction rates using first-principles simulations that incorporate finite-temperature and density effects, atomic ionisation, and both atomic and nuclear excitations. These calculations provide inputs more closely matched to cosmological conditions and enable a new assessment of the influence of EC, AC, positron decay, radiative PC, SE, and Auger-like capture on the primordial \litiosette~abundance.
\section{Reassessment of the cosmological \litiosette~problem}
The early thermal history of our Universe, relevant to baryon evolution and primordial nucleosynthesis, can be described as a sequence of freeze-outs, because the nuclear reaction rates $\Gamma$ decrease as the Universe expands and cools. In particular, detailed balance between reactants and products of nuclear reactions leading to nucleosynthesis is maintained only while the approximate freeze-out condition \citep{KolbTurner1990}
\begin{equation}\label{equilibrium}
\Gamma \gtrapprox \frac{1}{t}
\end{equation}
is satisfied, where $t$ is the cosmic age, defined as the time elapsed since the singularity at the beginning of the Universe. It is worth noting that this freeze-out is not an instantaneous event; rather, it represents a smooth evolutionary transition during which the reaction rate falls below the Hubble expansion rate, leading to the gradual decoupling of light elements from the thermal plasma.
According to BBN, for example, in a reaction such as $\gamma + \gamma \leftrightarrow e^{-} + e^{+}$, where lepton-antilepton pairs $e^{\pm}$ are produced by and in equilibrium with $\gamma$ photons, the interaction rate is given by $\Gamma = \langle n_{\gamma} v_{\rm{rel}} \sigma \rangle$, where $n_{\gamma}$ is the $\gamma$ photon density, $v_{\rm{rel}}$ is the relative velocity, and $\sigma$ is the interaction cross section. When condition (\ref{equilibrium}) is no longer satisfied, the expansion of the Universe outpaces the reaction rate, pair creation can no longer balance annihilation, so the species decouple and their abundance relative to photons becomes fixed (freeze-out).\\
\indent 
The cosmic age in Eq.~(\ref{equilibrium}) is related to the photon temperature approximately by
$T \equiv T_\gamma = C\,t^{-1/2}$
\citep{perkins}. This temperature can therefore also serve as a timescale to indicate the various epochs of the early Universe. For the indicative timescales reported throughout this work, including in the figures and tables, we adopt
$C = 1.8\times10^{9}\ {\rm K\,s^{1/2}},$
obtained by normalising this relation to the present CMB temperature and age of the Universe. The resulting time–temperature correspondence should be regarded solely as an indicative cosmological timescale. All physical calculations reported in this work are performed directly as functions of temperature and are therefore independent of this approximate conversion between temperature and cosmic time.\\
\indent At $k_{\rm{B}}T \gtrsim 200\,\mathrm{MeV}$, the plasma is a quark–gluon fluid. Between $k_{\rm{B}}T \simeq 200$ and $40\,\mathrm{MeV}$, baryons (antibaryons) $b$ ($\bar b$) remain in equilibrium with photons via $\gamma + \gamma \leftrightarrow b + \bar b$ \citep{2025EPJST.234.1125R}, while below $k_{\rm{B}}T \lesssim 100\,\mathrm{MeV}$, only light nucleons (antinucleons), such as protons $p$ (antiprotons $\bar p$) and neutrons $n$ (antineutrons $\bar n$), survive. Their pair production freezes out at $k_{\rm{B}}T \approx 20\,\mathrm{MeV}$, while transient deuterium formation is promptly erased by photodisintegration. In this context, we note that the WMAP baryon asymmetry, $(n_{b} - n_{\bar{b}})/n_{\gamma} = 6.11 \times 10^{-10}$, reflects the primordial matter–antimatter imbalance.\\
\indent
Electrons, positrons, and neutrinos remain coupled to radiation through $\gamma + \gamma \leftrightarrow e^- + e^+ \leftrightarrow \nu + \bar{\nu}$ until neutrino decoupling at $k_{\rm{B}}T \approx 3\,\mathrm{MeV}$ and $e^\pm$ decoupling at $k_{\rm{B}}T \approx 1\,\mathrm{MeV}$. Weak interactions maintain $n/p$ equilibrium down to $k_{\rm{B}}T \approx 700\,\mathrm{keV}$ (weak freeze-out), after which only $\beta^-$ decays can change the $n/p$ ratio until $k_{\rm{B}}T \approx 100\,\mathrm{keV}$, when neutrons become bound in $\alpha$ particles and the $n/p$ ratio freezes (the so-called Nuclear Statistical Equilibrium, NSE). The breakdown of $n + p \leftrightarrow {}^{2}\mathrm{H} + \gamma$ at $k_{\rm{B}}T \approx 50$–$60\,\mathrm{keV}$ allows deuterium to survive and enables the synthesis of light nuclei, such as \litiosette~and \berilliosette. \\
\indent
In this work, we calculate the reaction rates of the channels contributing to the formation and depletion of ${}^{7}\mathrm{Li}$ during the Big Bang in the temperature range 275 to 1 keV both above and below NSE when conditions allowed these nuclei to form. All of these reactions proceed through ${}^{7}\mathrm{Be}$ as the intermediate nucleus \citep{fuller_smith,pdg_bbn}. In the BBN model, ${}^{7}$Be is produced primarily by
${}^{3}{\rm He}(\alpha,\gamma){}^{7}{\rm Be}$\footnote{
Note that the reaction ${}^{6}{\rm Li}(p,\gamma){}^{7}{\rm Be}$ makes a much
smaller contribution due to the very low primordial ${}^{6}$Li abundance.} at temperatures of 50–60 keV, while
the dominant production channel of primordial lithium is the
${}^{7}{\rm Be}(n,p){}^{7}{\rm Li}$ reaction ($Q$-value = 1.64 MeV), which is characterised by a large cross section due to the still high free neutron density, followed by fast burning via
${}^{7}{\rm Li}(p,\alpha){}^{4}{\rm He}$. However, by the time BBN reaches the stage where \berilliosette~is abundant ($t \approx 300$ s, corresponding to temperatures well below NSE), free neutrons have dropped significantly, having been trapped in bound states, mostly as $\alpha$ particles. Consequently, the slower pace of the ${}^{7}{\rm Be}(n,p){}^{7}{\rm Li}$ reaction is no longer sufficient to destroy all Be.
Moreover, photodisintegration of ${}^{7}$Be is negligible because photons with $E_{\gamma} \geq 1.586$ MeV (for ${}^{7}{\rm Be} +\gamma \rightarrow {}^{3}{\rm He} + {}^{4}{\rm He}$) or $E_{\gamma} \geq 5.60$ MeV (for ${}^{7}{\rm Be} +\gamma \rightarrow {}^{6}{\rm Li} + p$) are exponentially rare at BBN temperatures $\lesssim 0.1\ \mathrm{MeV}$. As a result, most \berilliosette~survives and decays to \litiosette~ through various electroweak processes producing the cosmological excess in BBN compared to that observed in low-metallicity halo stars.\\
\indent
In this context, our analysis examines three electroweak processes responsible for the production of ${}^{7}\mathrm{Li}$ under conditions above and below NSE.
Specifically, we performed the ab initio calculation of the EC process,
\begin{equation}\label{eqn1}
{}^{7}{\rm Be} + e^- \rightarrow {}^{7}{\rm Li} + \nu_e,
\end{equation}
and as it is driven by the same interaction Hamiltonian, we also calculated the AC rate,
\begin{equation}\label{eqn2}
{}^{7}{\rm Be} + \bar{\nu}_{e} \rightarrow {}^{7}{\rm Li} + e^{+},
\end{equation}
as well as
\begin{equation}\label{eqn2_2}
{}^{7}{\rm Be^*}  \rightarrow {}^{7}{\rm Li} + e^{+} + \nu_{e},
\end{equation}
where ${}^{7}{\rm Be^*}$ denotes the first excited nuclear state of Be.
We emphasise that although neutrino interactions are typically elusive, we assessed whether the high densities of electron antineutrinos in the early Universe could make this channel relevant for ${}^{7}$Be destruction (or \litiosette\ production). As the Universe expands and cools, the antineutrino background becomes diluted, causing the contribution from this channel to become rapidly negligible.
\\
\indent We note that the energetics of the weak channels introduced in Eqs.~(\ref{eqn1})–(\ref{eqn2_2}) differ substantially and largely determine their relative importance during BBN.
Electron capture on the $^{7}\mathrm{Be}$ ground state is exothermic, with $Q = 861.815\ \text{keV}$ for the transition to the $^{7}\mathrm{Li}$ ground state and $384.215\ \text{keV}$ for the transition to the first excited state (see Fig. \ref{weakrates_scheme}); thus, no threshold exists and the available phase space is relatively large.
In contrast, the inverse AC reaction has a negative $Q$-value. For the ground-state transition, $Q = -160.10\ \text{keV}$, implying a threshold antineutrino energy
$E_{\bar{\nu}}^{\rm th} = |Q| = 160.10\ \text{keV}$,
neglecting the small nuclear recoil. Consequently, only the high-energy tail of primordial antineutrinos with energies significantly higher than the average thermal energy of the system contributes to this process. As the Universe cools below $k_{\rm B}T \sim 100\ \text{keV}$, the occupation probability above the threshold decreases approximately as
\begin{equation}
f_{\bar{\nu}}(E > E_{\nu}^{\rm th})
\propto
\exp\!\left(
-\frac{E_{\nu}^{\rm th}}{k_{\rm B}T_\nu}
\right),
\end{equation}
that is, the population of antineutrinos above the threshold decreases exponentially. This results in a rapid suppression and a strong temperature dependence of the AC rate despite the still large relic neutrino density, and consequently the AC rate falls much more rapidly than the corresponding EC rate.
For transitions originating from the thermally populated first excited state of $^{7}\mathrm{Be}$ ($E^* = 429.2\ \text{keV}$), the additional excitation energy increases the effective $Q$-value by $429.2\ \text{keV}$. As a result, the ground-state AC channel becomes effectively threshold-free ($E_{\bar{\nu}}^{\rm th} = 0$), while the larger positive $Q$-values also enhance the phase space for EC and make $\beta^+$ decay energetically allowed. Nevertheless, these channels remain suppressed overall because the thermal population of the excited state is only at the percent level around $k_{\rm B}T \simeq 100\,\text{keV}$. \\
\indent From a microscopic perspective, the AC process is the counterpart of ordinary EC. Both reactions are governed by the same charged-current weak interaction and involve the same nuclear Gamow--Teller matrix elements, differing only in the leptonic states and the available phase space. Unlike charged-particle capture reactions, the incoming antineutrino experiences no electromagnetic interaction with the nucleus and is therefore completely unaffected by the Coulomb barrier and the associated Gamow suppression factor. Consequently, the AC rate is determined by the weak interaction matrix element, the relic antineutrino distribution function, and the available phase space.
A simple order of magnitude estimate illustrates the magnitude of the AC contribution. The reaction rate may be approximated as
\begin{equation}
\lambda_{\rm AC}
\approx
n_{\bar{\nu}_e}
\left\langle
\sigma_{\rm weak}v
\right\rangle .
\end{equation}
Although the elementary weak cross section below 1 MeV is extremely small,
\begin{equation}
\sigma_{\rm weak}
\approx
10^{-45}\text{--}10^{-44}\ {\rm cm}^{2},
\end{equation}
the relic antineutrino density during BBN remains enormous,
\begin{equation}
n_{\bar{\nu}_e}
\approx
10^{27}\text{--}10^{28}\ {\rm cm}^{-3},
\end{equation}
around \(k_{\rm B}T\sim100~{\rm keV}\). As a consequence, the resulting capture rate naturally reaches values of order of
\begin{equation}
\lambda_{\rm AC}
\approx
10^{-7}\text{--}10^{-6}\ {\rm s}^{-1},
\end{equation}
in agreement with our detailed numerical calculations.
Accordingly, the relatively large AC rate should not be interpreted as evidence of an unusually strong weak interaction; rather, it is the natural consequence of the exceptionally high relic antineutrino density during BBN, combined with the absence of Coulomb suppression.
\\
\indent Furthermore, we contrast this with bound-state EC. While it shares the same nuclear weak matrix elements with AC, EC is additionally constrained by the atomic electron wave function and proceeds predominantly through capture from the atomic \(1s\) ($K$-shell) orbital, whose finite probability density at the nucleus maximises the overlap with the nuclear charge distribution. From the nuclear-structure perspective, the transition matrix element is described within the shell model by the overlap between the initial and final many-body wave functions. In \({}^{7}\mathrm{Be}\), the \(1s_{1/2}\) shell forms an inert closed core, whereas the weak transition is dominated by the valence nucleons occupying the \(1p_{3/2}\) shell. Consequently, the Gamow--Teller matrix element is governed primarily by the valence-shell configuration and the overlap between the initial and final nuclear states, while the degree of forbiddenness reflects the corresponding angular-momentum and parity selection rules.\\
\indent
Moreover, among the possible reactions leading to \berilliosette~destruction (and thus to a lower production of \litiosette), there is also the strong force-driven PC, for which we compute the rate here under Big Bang conditions both above and below NSE ($1$ keV $< k_{\rm{B}}T < 275$ keV) to determine its impact on primordial abundance and assess whether this can reconcile the cosmological lithium problem. Specifically, we studied the following mechanism:
\begin{equation}\label{Be_to_B}
{}^{7}{\rm Be} + p \rightarrow {}^{8}{\rm B} + \gamma,
\end{equation}
which is a non-resonant radiative PC process driven by both the strong and electromagnetic interactions and dominated by direct E1 transitions.
In this regard, we first used classical non-relativistic methods supplemented by experimentally determined cross sections. Second, more accurate quantum mechanical calculations were performed to account for the likely presence of a highly energetic photon background, which may alter the rate through a mechanism similar to SE in lasers between nuclear levels.
Moreover, we considered a three-body variant of the process (\ref{Be_to_B}) in which the PC energy gain is transferred to a free electron in the surrounding plasma via a virtual photon (an Auger-like process in the continuum):
\begin{equation}\label{radiative_b}
{}^{7}{\rm Be} + p + e^{-} \rightarrow {}^{8}{\rm B} + e^{-*}.
\end{equation}
This energy transfer mechanism, typically neglected in astrophysical scenarios, can modify PC under NSE conditions, analogous to internal conversion in atomic systems. The calculations of the PC cross sections for all these mechanisms are presented in Appendix A.\\
\indent Finally, the resulting new weak and radiative interaction rates for these processes, including temperature, density and charge state effects are essential inputs for recalculating nuclide abundances within the BBN model. Implementing them in BBN reaction networks will allow their impact on the predicted primordial abundances to be quantitatively assessed.

\section{Electron and antineutrino capture on \(^7\)Be: Theoretical and computational methods}
Our initial calculations address EC (see Eq.~(\ref{eqn1})), AC (see Eq.~(\ref{eqn2})), and excited-state positron emission (see Eq.~(\ref{eqn2_2})) in \berilliosette\ over the range $1$ keV $< k_{\rm B}T < 275$ keV, including thermal conditions under which Be begins to form ($10$ keV $< k_{\rm B}T < 100$ keV).
Typically, the EC rate for a given transition between nuclear levels of parent and daughter nuclei is determined by obtaining the nuclear transition matrix elements from the $\log(ft)$ values, where $f$ is the lepton phase-space factor describing the phase space accessible to the electron and neutrino during the decay, and $t \equiv t_{1/2}$ is the half-life \citep{PhysRev.124.495}. ${\rm Log}(ft)$s are typically estimated by analogy to laboratory decays of nearby nuclei with similar transitions, which can result in a large range of values, the so-called systematics.
Finally, the transition rate can be obtained by including the shape factor $S$, which accounts for forbiddenness, and the Fermi function $F(Z, w)$, where $Z$ is the atomic number of the decaying nucleus and $w$ is the ratio of the electron momentum to its energy \citep{fuller_smith}. The Fermi function accounts for Coulomb distortion of the incoming electron or escaping positron wave functions. In this context, the Coulomb repulsion between electrons in the plasma and the electron-nuclear attraction are treated using an average Coulomb wave correction factor, which is also included in the Fermi function. Earlier works \citep{fuller_smith}, based on systematics, assumed LTE and that weak interaction rates depend polynomially on plasma (electron) temperature, with a Fermi-Dirac occupation of the orbitals for both electron and neutrino fermionic species.
While this standard approach may work well for predicting the lineshape of allowed and forbidden unique transitions, it fails, for example, for forbidden non-unique transitions, where there is no simple relation for $S$ and the systematics provide too broad a range of values, reflecting uncertainty about the nuclear levels involved. This approach was used to estimate isotope half-life in highly ionised plasma in the seminal work by Takahashi and Yokoi \citep{TAKAHASHI1987375,PhysRevC.36.1522}, which, after 35 years, remains the most widely used method for calculating primordial and stellar nucleosynthesis, despite the semi-empirical treatment of the hadronic and leptonic currents. \\
\indent
We did not rely on semi-empirical approaches based on the determination of $\log(ft)$ values and $F(Z,w)$. Instead, we developed a theoretical and computational framework \citep{taioli2021relativistic,taioli2022theoretical} using first-principles simulations to calculate
the EC rates by determining the local electron density at the nucleus, to which the rate is proportional, modulo some weighting factors. This method fully incorporates relativistic (and quantum electrodynamical) effects.
Within this framework, we can account for several factors influencing the EC, such as electron-electron interaction, charge state distribution, and electron and nuclear excitation of the parent and daughter atoms at the same level of theory.
A significant difference in our approach concerns the treatment of the leptonic current, which is typically included using a semi-empirical Fermi function. In contrast, we use first-principles simulations based on the Dirac-Hartree-Fock (DHF) method to determine the electronic wave functions that enter the leptonic current, accounting for the population-level distribution and charge state of the parent Be ion undergoing beta decay and the Li daughter nucleus. This includes non-orthogonality between initial and final states, shake-up, shake-off, the energy continuum and the integration of the electronic degrees of freedom within the nuclear range. Temperature enters into the assessment of the chemical potential of the species and influences the local density of bound and continuum electrons at the Be nucleus. In this respect, we have demonstrated \citep{taioli2022theoretical} the importance of accurately including the leptonic part, which may decrease, for example, the half-life of beta decaying $^{134}$Cs isotopes by 20\% in the range [0, 15] keV compared to the frozen-electron-only beta decay rate. In contrast, the rate may increase by a factor of 15 between 10 and 100 keV compared to room temperature conditions, where $\log(ft)$ values are measured, due to the inclusion of fast-decaying nuclear excited states.\\
\indent
Moreover, we assess the nuclear transition matrix elements that enter the hadronic currents using a mean-field approach, similar to the independent-particle nuclear shell model, by solving the Dirac equation for the nucleons with a relativistic Woods-Saxon nuclear potential. This mean-field approach can, of course, be systematically improved by employing more correlated many-body techniques without altering the foundation of our method.
\\
\indent The EC rate on ${}^{7}$Be was calculated to the first order in time-dependent perturbation theory using Fermi's golden rule in atomic units
\citep{morresi2018relativistic}:
\begin{equation}\label{FGR}
\Gamma_{i \rightarrow f} =
2\pi \big|\langle f|\hat{V}|i\rangle\big|^{2}
\,\rho_{i}\,\rho_{f}\,\delta(E_{i}-E_{f}),
\end{equation}
where $\hat{V}$ is the weak interaction Hamiltonian, $\rho_{i}$ and
$\rho_{f}$ are the initial and final densities of states, $|i\rangle$ and $|f\rangle$ label the initial (parent nucleus) and final (daughter nucleus) states, and the delta
function enforces energy conservation between the initial and final state energies, $E_i$ and $E_f$, respectively.
The weak interaction is described within the Fermi theory of beta decay by
the effective potential
\begin{eqnarray}
\label{inter_ham}
\hat{V}&=&\frac{G_{\mathrm F}}{\sqrt{2}}\int d^{3}r \,
\Big[
\hat{\bar{\psi}}_{p}(\vec{r})\gamma^{\mu}(1 - x\gamma^{5})
\hat{\psi}_{n}(\vec{r})
\Big]\times \nonumber \\
&&\Big[
\hat{\bar{\psi}}_{e}(\vec{r})\gamma_{\mu}(1-\gamma^{5})
\hat{\psi}_{\nu}(\vec{r})
\Big] + {\rm h.c.},
\end{eqnarray}
where $G_{\mathrm F} = 1.16637 \times 10^{-5}$ GeV$^{-2}$ is the Fermi coupling constant and $x = C_A / C_V = 1.26 \pm 0.02$ is the ratio of the axial vector ($C_A$) to vector ($C_V$) contributions to the hadronic current, and $\gamma^{\mu}~ (\mu=1,...,5)$ are the Dirac matrices. In Eq. (\ref{inter_ham}), $\hat{\bar{\psi}}_{p}(\vec{r})$, $\hat{\psi}_{n}(\vec{r})$, $\hat{\bar{\psi}}_{e}(\vec{r})$, and $\hat{\psi}_{\nu}(\vec{r})$ are the field operators for the proton, neutron, electron, and neutrino, respectively. For each field operator $\hat{\psi}$, the corresponding conjugate $\hat{\bar{\psi}} = \hat{\psi}^{\dagger} \gamma^{0}$ is defined using the following Dirac matrix:
\begin{equation}
\label{gamma_0}
\gamma^{0} = \left(
\begin{array}{cc}
I & 0 \\
0& -I \\
\end{array}
\right),
\end{equation}
where $I$ and $0$ are the $2 \times 2$ identity and zero matrices, respectively.
\\
\indent
The matrix element of the interaction potential in Eq. (\ref{inter_ham}) between the initial and final states factorises into purely hadronic ($p, n$ label proton and neutron) and leptonic ($e, \nu$ label electron and neutrino) currents as follows \citep{morresi2018relativistic,taioli2021relativistic}:
\begin{eqnarray}\label{correnti}
\langle f|\hat{V}|i \rangle &=&
\frac{G_{\mathrm F}}{\sqrt{2}}\int d^{3}r \,
\big[ \langle  f_{p,n}|
\hat{\bar{\psi}}_{p}(\vec{r})\gamma^{\mu}(1-x\gamma^{5})
\hat{\psi}_{n}(\vec{r})
|i_{p,n}\rangle \big] \times
\,\nonumber \\
&&\big[
\langle f_{e,\nu}|\hat{\bar{\psi}}_{e}(\vec{r})
\gamma_{\mu}(1-\gamma^{5})
\langle \hat{\psi}_{\nu}(\vec{r})|i_{e,\nu}\rangle
\big] + \mathrm{c.c.} \nonumber \\
&=&\frac{G_{\mathrm F}}{\sqrt{2}}\int d^{3}r \,
\big[ \langle  f_{p}|
\hat{\bar{\psi}}_{p}(\vec{r})|i_p\rangle \gamma^{\mu}(1-x\gamma^{5})\langle f_n|
\hat{\psi}_{n}(\vec{r})
|i_{n}\rangle \big] \times
\,\nonumber \\
&&
\big[
\langle f_{e}|\hat{\bar{\psi}}_{e}(\vec{r})|i_{e}\rangle
\gamma_{\mu}(1-\gamma^{5})
\langle f_{\nu}|\hat{\psi}_{\nu}(\vec{r})|0_{\nu}\rangle
\big] + \mathrm{c.c.} \nonumber \\
&=&\frac{G_\mathrm{F}}{\sqrt{2}}\langle f_{p,n}| J_{\mathrm{had}}^{\mu}| i_{p,n}\rangle \langle f_{e,\nu} | J^{\mathrm{lep}}_\mu | i_{e,\nu} \rangle,
\end{eqnarray}
where $\left|i\right\rangle =\left|i_{p,n}\right\rangle \otimes\left|i_{e,\nu}\right\rangle$ and $\left|f\right\rangle =\left|f_{p,n}\right\rangle \otimes\left|f_{e,\nu}\right\rangle$. The hadronic and leptonic terms can  be further factorised as follows:
\begin{eqnarray}
\left|i_{p,n}\right\rangle &=& \left|i_{p}\right\rangle \otimes \left|i_{n}\right\rangle,~~\left|f_{p,n}\right\rangle = \left|f_{p}\right\rangle \otimes \left|f_{n}\right\rangle, \\
\left|i_{e,\nu}\right\rangle &=&\left|i_{e}\right\rangle \otimes\left|0_{\nu}\right\rangle,~~ \left|f_{e,\nu}\right\rangle =\left|f_{e}\right\rangle \otimes\left|f_{\nu}\right\rangle.
\end{eqnarray}
This is due to the independent-particle nuclear shell model adopted here and the fact that neutrinos do not interact electromagnetically with fermions and can therefore be safely considered free.
In Eq. (\ref{correnti}) we also define the leptonic,
\begin{equation}\label{current_lep}
J_{\mathrm{lep}}^\mu=\hat{\overline{\psi}}_e(\vec{r})\gamma^{\mu}\left(1-\gamma^{5} \right)\hat{\psi}_{\nu}(\vec{r}),
\end{equation}
and hadronic currents,
\begin{equation}\label{current_had}
J_{\mathrm had}^{\mu}=\hat{\overline{\psi}}_p(\vec{r})\gamma^{\mu}\left(1-x\gamma^{5} \right)\hat{\psi}_n(\vec{r}).
\end{equation}
This separation allows the nuclear matrix element
$\langle f_{p,n}|\cdots|i_{p,n}\rangle$ -- which includes the Fermi and
Gamow--Teller (GT) contributions -- to be treated independently of the lepton wave functions $\langle f_{e,\nu}|\cdots|i_{e,\nu}\rangle$.\\
\indent A general expression for the field operators of the initial and final fermionic states, characterised by both a continuum and a discrete spectrum and appearing in the transition operator matrix element (\ref{correnti}), can be written as follows:
\begin{eqnarray}
\label{general_field_incoming}
\hat{\psi}_{i}(\vec{r})&=&\sum_{\phi}\left[\braket{\vec{r}|\phi}\hat{c}_{d,i}+\braket{\phi|\vec{r}}\hat{d}_{d,i}^{\dagger}\right]\nonumber \\ &+&
\int d\boldsymbol{k}\sum_{\boldsymbol{l}}\left[\braket{\vec{r}|\boldsymbol{k},\boldsymbol{l}}\hat{c}_{c,i}+\braket{\boldsymbol{k},\boldsymbol{l}|\vec{r}}\hat{d}_{c,i}^{\dagger}\right],
\end{eqnarray}
where the terms are as follows:
\begin{itemize}
	\item[-]\(\phi\) is a set of wave functions labelled by suitable quantum numbers with a discrete spectrum. For example, total angular momentum (\(j\)) and principal quantum number (\(n\)).
	\item[-]\(\hat{c}_{d,i}\) and \(\hat{d}_{d,i}^{\dagger}\) are the fermion annihilation operator and the antifermion creation operator for the discrete spectrum states, respectively.
	\item[-]\(\boldsymbol{k}\) is a set of continuum quantum numbers (such as energy for a free particle), and \(\boldsymbol{l}\) is a set of discrete spectrum quantum numbers.
	\item[-]\(\hat{c}_{c,i}\) and \(\hat{d}_{c,i}^{\dagger}\)~are the fermion annihilation operator and the antifermion creation operator for the continuum spectrum states, respectively.
\end{itemize}
Similarly, for the final state \(f\), we have the field operator \(\hat{\psi}_{f}^{\dagger}\):
\begin{eqnarray}
\label{general_field_outgoing}
\hat{\psi}_{f}^{\dagger}(\vec{r})&=&\sum_{\phi}\left[\braket{\phi|\vec{r}}\hat{c}_{d,f}^{\dagger}+\braket{\vec{r}|\phi}\hat{d}_{d,f}\right]\nonumber \\ &+&
\int d\boldsymbol{k}\sum_{\boldsymbol{l}}\left[\braket{\boldsymbol{k},\boldsymbol{l}|\vec{r}}\hat{c}_{c,f}^{\dagger}+\braket{\vec{r}|\boldsymbol{k},\boldsymbol{l}}\hat{d}_{c,f}\right].
\end{eqnarray}
Using expressions (\ref{general_field_incoming}) and (\ref{general_field_outgoing}), we can construct the field operators that appear in the hadronic and leptonic currents in Eqs. (\ref{current_lep}) and (\ref{current_had}), respectively. Here, we are interested in EC, AC, and $\beta^+$ decay, which involve (free) neutrinos, continuum electrons and positrons as well as bound protons, neutrons, and electrons.
\paragraph{Electron capture}
In EC, the annihilation of an electron results in the formation of a bound neutron and a neutrino, which interacts very weakly with matter and can therefore be considered free. Thus, only the continuum spectrum part of Eq. (\ref{general_field_outgoing}) needs to be retained. Neutrinos are identified by their relativistic kinetic energy \(W_{\nu}\), their total angular momentum \(K_{\nu}\), and the projection of this angular momentum along a quantisation axis \(M_{\nu}\):
\begin{equation}
\label{neutrino_creat_field_op}
\hat{\psi}_{\nu}^{\dagger}(\vec{r})=\int d W_{\nu}\sum_{K_{\nu},M_{\nu}}\left[\braket{W_{\nu},K_{\nu},M_{\nu}|\vec{r}}\hat{c}_{\nu,c}^{\dagger}+\braket{\vec{r}|W_{\nu},K_{\nu},M_{\nu}}\hat{d}_{\nu,c}\right].
\end{equation}
The decaying Be atom can be partially or fully ionised due to the high BBN temperature. Therefore, for the electron field operator, we had to retain both the discrete ($n,j$, even if only the $s$-symmetry orbitals have finite overlap with the nucleus) and continuum ($W_c$) energy spectrum components:
\begin{eqnarray}
\label{elec_field_op_EC}
&&\hat{\psi}_{e}(\vec{r})=\sum_{n,j}\left[\braket{\vec{r}|n,j}\hat{c}_{d,i}+\braket{n,j|\vec{r}}\hat{d}_{d,i}^{\dagger}\right]\nonumber +\\
&&\int d W_{c}\sum_{K_{c},M_{c}}\left[\braket{\vec{r}|W_{c},K_{c},M_{c}}\hat{c}_{e,c}+\braket{W_{c},K_{c},M_{c}|\vec{r}}\hat{d}_{e,c}^{\dagger}\right].
\end{eqnarray}
In addition, neutrons (\(n\)) and protons (\(p\)) are in bound states within the parent and daughter nuclei, so the continuum spectrum must be excluded. Considering the symmetries of the system, we had to use the quantum numbers associated with \(\hat{J}^{2}\) and \(\hat{J}_{z}\), which we denote as \(j_{p,n}\) and \(M_{p,n}\):
\begin{equation}
\label{hadron_EC_ops}
\begin{aligned}
\hat{\psi}_{p}(\vec{r}) &= \sum_{j_{p}, M_{p}, \xi_{p}} \left[ \braket{\vec{r} | j_{p}, M_{p}, \xi_{p}} \hat{c}_{p} + \braket{j_{p}, M_{p}, \xi_{p} | \vec{r}} \hat{d}_{p}^{\dagger} \right] \\
\hat{\psi}^{\dagger}_{n}(\vec{r}) &= \sum_{j_{n}, M_{n}, \xi_{n}} \left[ \braket{j_{n}, M_{n}, \xi_{n} | \vec{r}} \hat{c}_{n}^{\dagger} + \braket{\vec{r} | j_{n}, M_{n}, \xi_{n}} \hat{d}_{n} \right], \\
\end{aligned}
\end{equation}
where the quantum numbers \(\xi_{p,n}\) define the discrete energy of the state of the nucleon.
\paragraph{\(\overline{\boldsymbol{\nu}}\)-capture}
Field operators are the same as for EC, but both the antineutrino,
\begin{equation}
\label{wave_func1}
\braket{W_{\nu},K_{\nu},M_{\nu}|\vec{r}},
\end{equation}
and positron wave functions,
\begin{equation}
\label{wave_func2}
\braket{\vec{r}|W_{c},K_{c},M_{c}},
\end{equation}
must be negative energy solutions of the Dirac equation.
\paragraph{$\boldsymbol{ \beta^+}$ decay}
For \(\beta^{+}\) decay, the field operators are the same as for EC and AC, but a positron in the continuum spectrum is created. Therefore, the wave function of the positron,
\begin{equation}\label{pos}
\braket{\vec{r}|W_{c},K_{c},M_{c}}
\end{equation}
must also be a negative energy solution of the Dirac equation.
\section{From field operators to wave functions}
In this section, we derive the explicit form of the matrix elements appearing in Eq. (\ref{correnti}), when the field operators have the general form given in Eqs. (\ref{general_field_incoming}) and (\ref{general_field_outgoing}). Specifically, for EC, $\beta^{+}$ decay, and AC, the relevant field operators are those introduced in Eqs. (\ref{neutrino_creat_field_op})-(\ref{pos}). The calculation proceeded in three main steps:
\begin{itemize}
\item We formulated the appropriate Dirac equation and selected a convenient basis of single-particle wave functions (we used a radial-grid representation of the wave functions, but a Gaussian basis set can in principle also be adopted).
\item We solved the Dirac equation using a DHF approach to obtain the antisymmetric many-body wave functions for both nucleons and electrons.
\item The resulting wave functions were substituted into the field operator expressions in Eqs. (\ref{neutrino_creat_field_op})-(\ref{pos}) to compute the required matrix elements.
\end{itemize}
\subsection{Dirac formalism in the context of leptonic and hadronic currents for a central potential}
In the two-component representation, the Dirac spinor is written as
\begin{equation}
\label{spin_repr_dirac}
\psi(\mathbf{x}) =
\begin{pmatrix}
\psi_{u}(\mathbf{x}) \\[2mm]
\psi_{d}(\mathbf{x})
\end{pmatrix},
\end{equation}
where the upper ($u$) and lower ($d$) components correspond to the large and small components in the Dirac representation. Each component contains the spin degrees of freedom, which are included in the variable $\mathbf{x}$ together with the spatial coordinates.
The general form of the time-independent Dirac equation in matrix form is
\begin{equation}
\label{dirac_matrix}
\begin{pmatrix}
m c^{2} + V(\mathbf{x}) & c\,\boldsymbol{\sigma}\!\cdot\!\boldsymbol{p} \\[1mm]
c\,\boldsymbol{\sigma}\!\cdot\!\boldsymbol{p} & -m c^{2} + V(\mathbf{x})
\end{pmatrix}
\begin{pmatrix}
\psi_{u}(\mathbf{x}) \\[1mm]
\psi_{d}(\mathbf{x})
\end{pmatrix}
=
E\begin{pmatrix}
\psi_{u}(\mathbf{x}) \\[1mm]
\psi_{d}(\mathbf{x})
\end{pmatrix},
\end{equation}
where $V(\mathbf{x})$ is, in this case, a central potential; $m$ is the particle's rest mass; $\vec{p}$ is its momentum; and $\boldsymbol{\sigma}$ is the vector of Pauli matrices $[\boldsymbol{\sigma}_x, \boldsymbol{\sigma}_y, \boldsymbol{\sigma}_z]$.\\
\indent Owing to the spherical symmetry of the interaction potential, the states are classified by the quantum numbers $(n, \kappa, j, M)$, where $n$ is the radial (principal) quantum number and $\kappa$ is the Dirac angular quantum number, which is defined through
\begin{equation}
\label{k_quantum_number}
\hat{K}\,\psi(\mathbf{x}) = -\hbar \kappa\,\psi(\mathbf{x}),
\end{equation}
where
\begin{equation}
\label{operator_k}
\hat{K} = \hbar\gamma_0\left(2\hat{\mathbf{S}}\!\cdot\hat{\mathbf{J}} - \frac{1}{2}\right).
\end{equation}
The term $j$ is the eigenvalue of the total angular momentum operator $\vec{\hat{J}}$, and $M$ is its projection onto the arbitrarily chosen quantisation axis since in relativistic quantum mechanics neither the angular momentum $\hat{\mathbf{L}}$ nor the spin $\hat{\mathbf{S}}$ is conserved separately. From the commutation with the Hamiltonian in Eq. (\ref{dirac_matrix}),
$ [\hat{H},\hat{K}] = 0, $
the associated quantum number $\kappa$ acquires two possible values for each total angular momentum $j$, thus determining the orbital angular momentum of the upper and lower components as
\begin{equation}
\label{k_j}
\kappa=\pm\left(j+\frac{1}{2}\right)
\quad\Rightarrow\quad
\begin{aligned}
l_{u} &= j\pm\frac{1}{2},\\
l_{d} &= j\mp\frac{1}{2}.
\end{aligned}
\end{equation}
The spinor in Eq. (\ref{spin_repr_dirac}) can be factorised into radial and angular–spin components:
\begin{align}
\psi_{u}(\mathbf{x}) &= U(r)\,\mathcal{Y}^{j,l_{u}}_{M}(\theta,\phi),\\
\psi_{d}(\mathbf{x}) &= i\,D(r)\,\mathcal{Y}^{j,l_{d}}_{M}(\theta,\phi),
\end{align}
where the angular--spin function is defined as
\begin{equation}
\label{ang_spin_part}
\mathcal{Y}^{j,l}_{M}(\theta,\phi)
=
\sum_{l_{z},s_{z}}
\braket{l\,l_{z},s\,s_{z}\,|\,j\,M}\,
Y_{l}^{l_{z}}(\theta,\phi)\,
\chi_{s}^{\,s_{z}}.
\end{equation}
By introducing the reduced radial functions $u(r)=rU(r)$ and
$d(r)=rD(r)$, we obtained
\begin{equation}
\label{spin_repr_dirac_fact}
\psi(\mathbf{x}) =
\begin{pmatrix}
\dfrac{u(r)}{r}\,\mathcal{Y}^{j,l_{u}}_{M}(\Omega)\\[3mm]
i\,\dfrac{d(r)}{r}\,\mathcal{Y}^{j,l_{d}}_{M}(\Omega)
\end{pmatrix}.
\end{equation}
The Dirac equation for the radial components then becomes
\begin{equation}
\label{matrix_solve_radial_2}
\begin{pmatrix}
V(r)+mc^{2} & -\hbar c\,\dfrac{d}{dr}+\hbar c\,\dfrac{\kappa}{r}\\[1mm]
\hbar c\,\dfrac{d}{dr}+\hbar c\,\dfrac{\kappa}{r} & V(r)-mc^{2}
\end{pmatrix}
\begin{pmatrix}
u_{\kappa}(r)\\
d_{\kappa}(r)
\end{pmatrix} = E
\begin{pmatrix}
u_{\kappa}(r)\\
d_{\kappa}(r)
\end{pmatrix}.
\end{equation}
\subsection{Choice of scalar and vector potentials}
More general potentials (pseudoscalar, axial vector, tensor, etc.) could be introduced in Eq. (\ref{matrix_solve_radial_2}), but for our purposes it is sufficient to consider suitable combinations of scalar and vector potentials when modelling hadronic or leptonic currents. Moreover, we assume a mean-field approximation of the nuclear and electron potentials; that is, we assume an ansatz for the solution of the general Dirac equation (\ref{matrix_solve_radial_2}). In particular, the key assumption is that the many-electron or many-nucleon wave function is represented by a Slater determinant of four-component Dirac spinors, so the total electron or nuclear density \(\rho (\mathbf{x})\), with \(\mathbf{x}=(r,s)\), where \(s\) is the particle spin, is constructed from these relativistic orbitals. Each orbital represents an independent particle moving in an average potential, which must be calculated self-consistently and inherently includes electron-positron (large/small component) or nucleon-antinucleon mixing \citep{morresi2018relativistic}.
Relativistic and quantum approaches have also been exploited in
condensed-matter systems with non-Euclidean geometries and to investigate
effective curved-space phenomena and analogue event horizons \citep{Taioli2016Lobachevsky,Morresi2020EventHorizons}.
\paragraph{Nucleons}
For nucleons, we use the independent-particle model within the relativistic Woods–Saxon phenomenological representation of the nuclear potential \citep{Schwierz2007ParameterizationOT}. In this context, within relativistic mean-field theory, the effective single-particle potential experienced by a nucleon is described as the sum of a Lorentz-scalar attractive field, \( V_{s}(r) \), generated by \(\sigma\)-meson exchange, and a Lorentz-vector repulsive field, \( V_{v}(r) \), associated with \(\omega\)-meson exchange. The Dirac Hamiltonian couples to the combinations \( V_{v}(r) \pm V_{s}(r) \), which determine the dynamics of the upper and lower components of the Dirac spinor.
These combinations reproduce the depth of the central Woods–Saxon potential \citep{morresi2018relativistic}:
\begin{equation}
\label{ws_nuc}
V_{\mathrm{nuc}}(r)
=
-\frac{V}{1+\exp\!\left(\dfrac{r-R_{N}}{a}\right)},
\end{equation}
where \( a \) is the diffuseness parameter, which controls how rapidly the potential falls off at the nuclear surface. Here, \( R_{N} = R_{0}A^{1/3} \) is the nuclear radius, with \( A = N + Z \) the mass number, given by the sum of the atomic number (\( Z \)) and the number of neutrons (\( N \)), and \( R_{0} = 1.25\,\mathrm{fm} \), consistent with empirical nuclear charge radii. The depth
\[
V = V_{0}\left(1 \pm \chi\,\frac{N-Z}{A}\right)
\]
incorporates isospin dependence, with the \( + \) sign for protons and the \( - \) sign for neutrons, where \( V_0 = 45\text{–}55 \) MeV is a constant and \( \chi \) is an isospin-asymmetry coefficient (\( \approx 0.6\text{–}0.9 \)). The mixture of scalar and vector terms also accounts for the spin–orbit splitting
\begin{equation}
\label{ws_so}
V_{\mathrm{SO}}(r)
=
\frac{\tilde{V}}{1+\exp\!\left(\dfrac{r-\tilde{R}_{N}}{a}\right)},
\end{equation}
where $\tilde{V}=\tilde{\lambda}V$ is the spin–orbit strength ($\tilde{\lambda} \approx 20-25$), and $\tilde{R}_{N}=\tilde{R}_{0}A^{1/3}$ is the spin–orbit radius ($\tilde{R}_{0}\approx R_{0}$).
\\
\indent
Thus, the Woods–Saxon nuclear mean-field potential must be regarded as a combination of scalar and vector components,
\begin{equation}
\label{pot_comb}
\begin{aligned}
V_{\mathrm{nuc}}(r)
=V_{v}(r) + V_{s}(r) &= -\dfrac{V}{1 + \exp\!\left(\dfrac{r - R_{N}}{a}\right)},\\
V_{\mathrm{SO}}(r)
=V_{v}(r) - V_{s}(r) &= \dfrac{\tilde{V}}{1 + \exp\!\left(\dfrac{r - \tilde{R}_{N}}{a}\right)},
\end{aligned}
\end{equation}
which leads to the following form of the Dirac matrix in the secular equation for nuclear matter:
\begin{equation}
\label{matrix_solve_radial_vector+scalar}
\begin{pmatrix}
V_{v}(r) + V_{s}(r) + mc^{2} & -\hbar c\,\dfrac{d}{dr} + \hbar c\,\dfrac{\kappa}{r}\\[1mm]
\hbar c\,\dfrac{d}{dr} + \hbar c\,\dfrac{\kappa}{r} & V_{v}(r) - V_{s}(r) - mc^{2}
\end{pmatrix}.
\end{equation}
Here, $m$ is the rest mass of the proton (\(m_{p}\)) or the neutron (\(m_{n}\)).
\paragraph{Leptons}
Electrons move in the Coulomb field of the nucleus and the other electrons, and experience a non-local electron–electron interaction. In the DHF method, this interaction is approximated by the following local potentials \citep{morresi2018relativistic,taioli2021relativistic}:
\begin{eqnarray}
\label{hartree+fock_pot}
V_{\mathrm{Coul}}(r) &=& -\frac{Z}{r},\nonumber \\
V_{\mathrm{Hartree}}(r) &=& \int d^{3}r'\,\frac{\rho(r')}{r_{>}}, \quad r_{>}=\max(r,r'),\nonumber \\
V_{\mathrm{Fock}}(r) &=& \frac{9}{4}\left(\frac{3}{\pi}\rho(r)\right)^{1/3},
\end{eqnarray}
yielding the total electronic potential
\begin{equation}
\label{hf_complete}
V_{\mathrm{el}}(r) =
V_{\mathrm{Coul}}(r) + V_{\mathrm{Hartree}}(r) - V_{\mathrm{Fock}}(r),
\end{equation}
where $V_{\mathrm{Coul}}$ is the electron–nuclear attraction, and $V_{\mathrm{Hartree}}$ and $V_{\mathrm{Fock}}$ are the classical Coulomb and exchange terms representing electron–electron repulsion within the HF approximation to the many-body potential.
This is essentially a local density approximation (LDA) of the electron-electron correlation potential.
Since the electromagnetic interaction is vectorial, the Dirac operator for electrons reduces to
\begin{equation}
\label{matrix_solve_radial_vector_electr}
\begin{pmatrix}
V_{\mathrm{el}}(r) + m_e c^{2} & -\hbar c\,\dfrac{d}{dr} + \hbar c\,\dfrac{\kappa}{r} \\[1mm]
\hbar c\,\dfrac{d}{dr} + \hbar c\,\dfrac{\kappa}{r} & V_{\mathrm{el}}(r) - m_e c^{2}
\end{pmatrix},
\end{equation}
where $m_e$ is the electron rest mass.
\paragraph{Neutrinos}
Neutrinos are treated as free particles. Their wave functions are therefore eigenstates of the Dirac operator without a potential:
\begin{equation}
\label{matrix_solve_radial_vector_neutrino}
\begin{pmatrix}
0 & -\hbar c\,\dfrac{d}{dr}+\hbar c\,\dfrac{\kappa}{r} \\[1mm]
\hbar c\,\dfrac{d}{dr}+\hbar c\,\dfrac{\kappa}{r} & 0
\end{pmatrix}.
\end{equation}
\subsection{Numerical solution of the Dirac equation}
The general Dirac equation (\ref{matrix_solve_radial_2}) can be adapted to the specific cases of EC and AC, provided that the nucleonic and leptonic single-particle wave functions required for the transition matrix elements are calculated by solving Eqs. (\ref{matrix_solve_radial_vector+scalar}), (\ref{matrix_solve_radial_vector_electr}), and (\ref{matrix_solve_radial_vector_neutrino}).
The numerical strategy used to solve Eq. (\ref{matrix_solve_radial_2}) involves finite-difference diagonalisation. These methodologies are described in the appendix \ref{app:dirac_numerics}.
\section{Electron and neutrino capture rates during the NSE epoch}
\subsection{$^7$Be and $^7$Li nuclear shell configurations}
\begin{figure}[hbt!]
\centering
\includegraphics[width=0.35\textwidth]{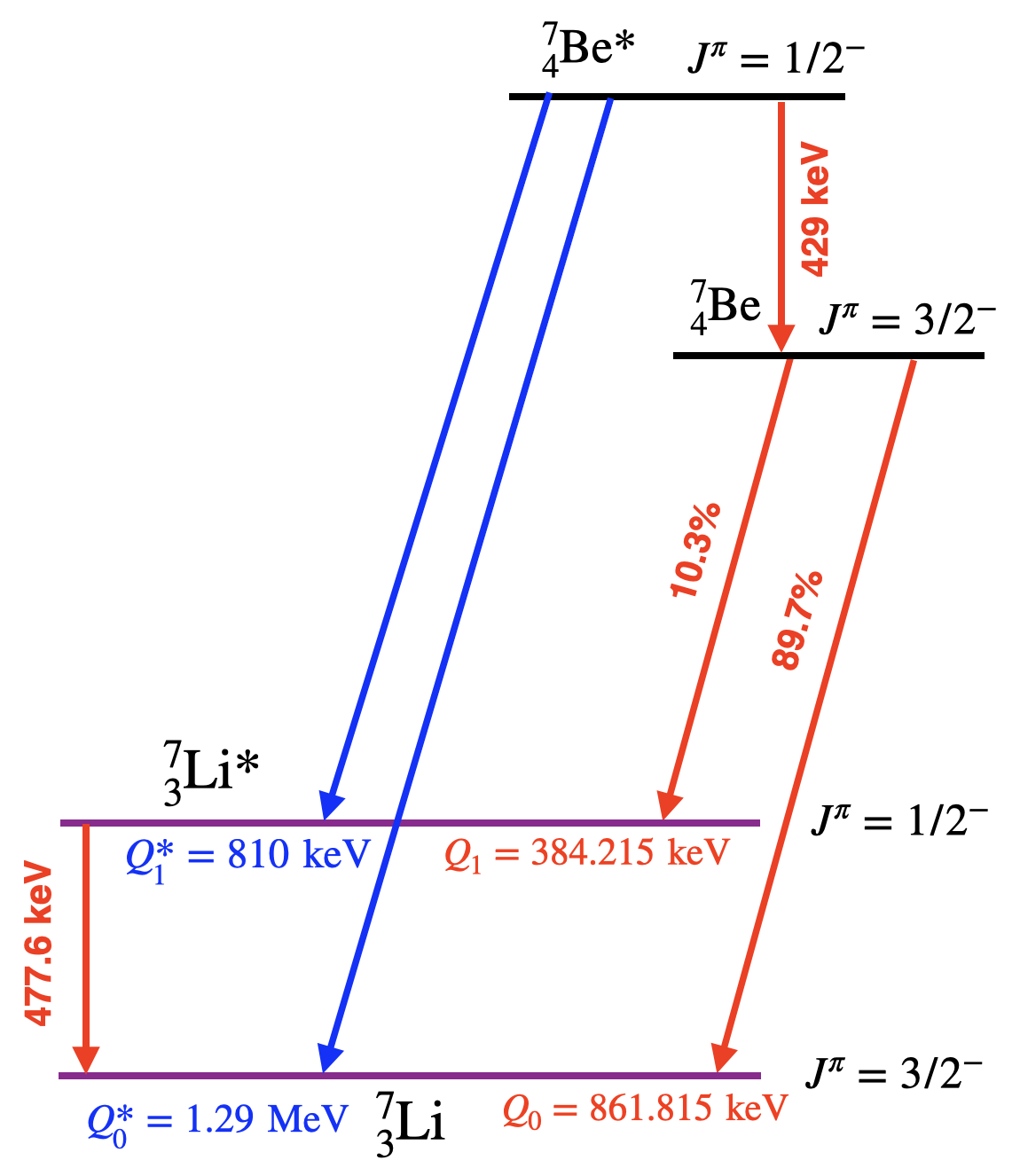}
\caption{$^{7}$Be electron capture decay scheme. The $Q$-values and branching ratios for the four decay channels are also provided (0 = $^{7}$Li nuclear ground state, 1 = $^{7}$Li nuclear excited state; $^*$ indicates decay from the excited state of $^{7}$Be).}
\label{weakrates_scheme}
\end{figure}

\begin{figure}[h]
\centering
\includegraphics[width=0.35\textwidth]{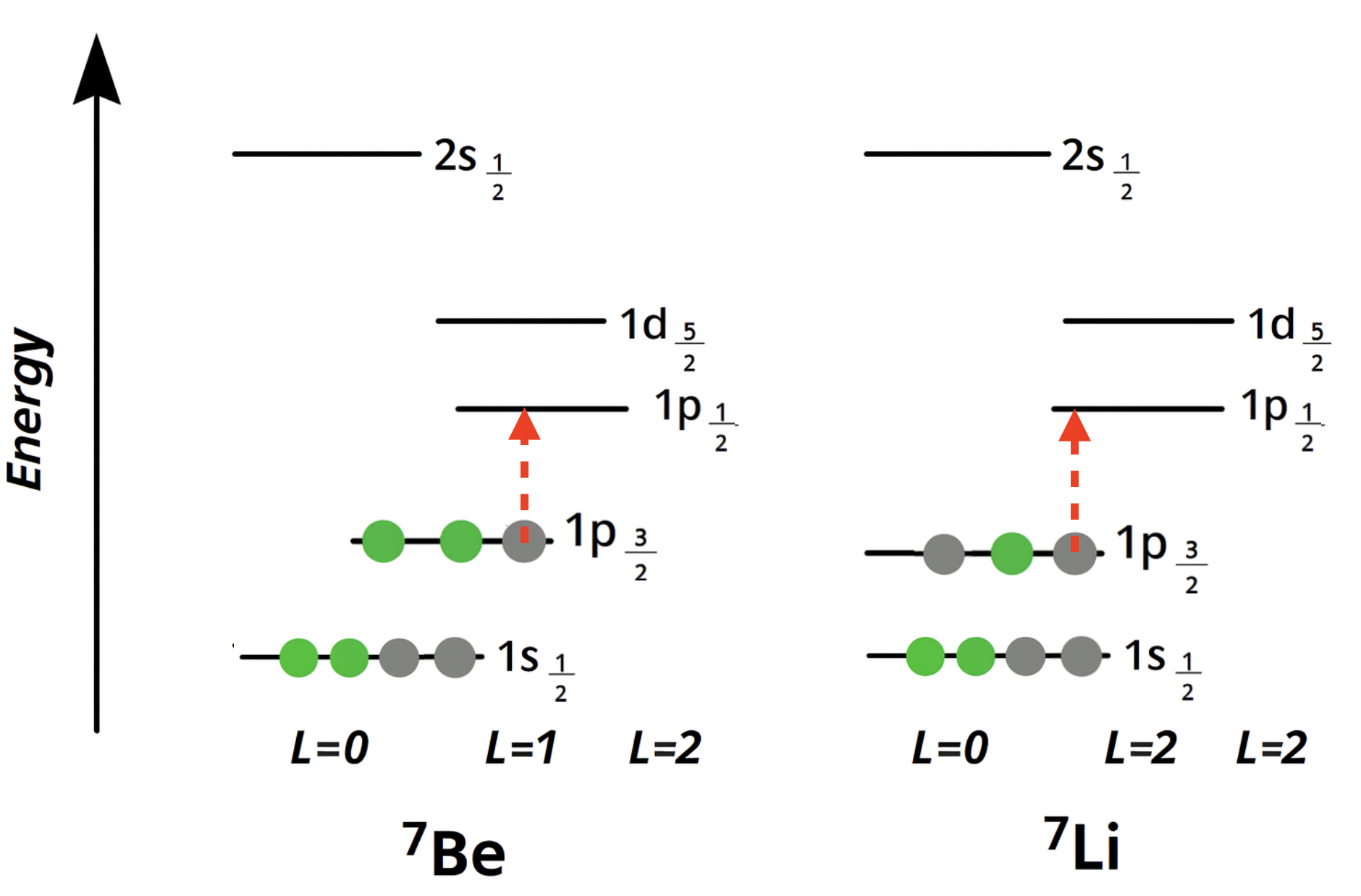}
\caption{
Shell model configurations of the ground state of $^7$Be (left, 3/2$^-$) and $^7$Li (right, 3/2$^-$). Green dots represent protons and grey dots represent neutrons. Also shown are the excited nuclear states of $^7$Be (left, 1/2$^-$) and $^7$Li (right, 1/2$^-$), which result from the promotion of a neutron to the $1p_{1/2}$ nuclear level (red arrows).
}
\label{nucleon level}
\end{figure}

EC on $^{7}$Be occurs through two GT transitions connecting $^{7}$Be to states of $^{7}$Li (we label 0 as the nuclear ground state of Li, 1 as its nuclear excited state, and $^*$ indicates decay from the excited state of Be). Under Earth conditions, $^{7}$Be decays from the ground state (3/2$^-$) via EC (see reaction (\ref{eqn1}) and Fig. (\ref{weakrates_scheme}) for the decay scheme) into the ground state of  $^{7}$Li (3/2$^-$) with a half-life of 53 days in 89.7\% of cases and a $Q$-value $Q_0 = 861.815$ keV.
In 10.3\% of cases ($Q_1 = 384.215$ keV), it decays to the first excited state (1/2$^-$) of \litiosette~(\litiosette$^*$). Both are allowed transitions; the former is a mixed Fermi and GT transition, while the latter is purely GT ($\Delta J = 1$, no parity change). \\
\indent In the astrophysical environment under the NSE conditions considered here, we must also account for the possibility that $^{7}$Be is thermally excited to its first excited state at $E_{^{7}\mathrm{Be}^{\ast}} = 0.429~\mathrm{MeV}$, characterised by the quantum number $J^\pi = 1/2^{-}$ (see Fig. (\ref{weakrates_scheme})). This excited state of $^{7}$Be lies at an excitation energy sufficiently close to the temperature $k_{\rm{B}}T \simeq 0.1~\mathrm{MeV}$, at which this nucleus is synthesised during BBN, to be populated. From this state, the capture can proceed to either the ground state or the first excited state of $^{7}$Li (see Fig. (\ref{weakrates_scheme})).
The transition from $^{7}$Be$^\ast(1/2^{-})$ to the ground state of $^{7}$Li ($J^\pi = 3/2^{-}$) is an allowed GT transition. Due to the additional excitation energy of the parent nucleus, the corresponding $Q$-value increases to $Q^\ast_{0} \approx 1.29~\mathrm{MeV}$. Capture to the first excited state of $^{7}$Li at $0.478~\mathrm{MeV}$ ($J^\pi = 1/2^{-}$) is a superallowed GT transition ($\Delta J = 0$, no parity change), with a reduced $Q$-value $Q^\ast_1 \simeq 0.81~\mathrm{MeV}$. For both these transitions from $^{7}$Be$^*$, the EC decay proceeds entirely via the GT operator, as Fermi transitions are forbidden by isospin selection rules in the $A = 7$ system. The enhancement of the $Q$-value for captures originating from the excited state directly affects the phase space available for the emitted neutrino and thus modifies the total EC rate in a temperature-dependent manner, especially in environments where the population of the $0.429~\mathrm{MeV}$ level becomes non-negligible.
Since $Q^*_{0,1} > m_{e}$, excitation of this nuclear level in $^7$Be may allow positron emission (see reaction (\ref{eqn2})) via this excited state. We also note that the thermal population of the $1/2^{-}$ state is
\begin{equation}\label{Boltzmann}
P_{1} \simeq \frac{2J+1}{\cal{Z}}\, e^{-E^{*}/(k_{\rm{B}}T)}
 \simeq \frac{1}{2}\, e^{-0.4292/(k_{\rm{B}}T)},
\end{equation}
where we approximate the partition function ${\cal{Z}}$ by the ground-state degeneracy. At $k_{\rm{B}}T \simeq 0.1~\mathrm{MeV}$, $P_{1} \sim 1\%$. Furthermore, an antineutrino threshold of $E_{\nu}^{\mathrm{th}} = 0$ applies, although this is less favourable than decay from the ground state. Most importantly, in addition to this population factor, it is necessary to determine how the EC and AC or $\beta^+$ decay rate from $^7$Be$^*$ compares to that of $^7$Be in its ground state, to assess whether all weak processes involving this excited state contribute significantly or only marginally to the overall $^7$Be destruction rate during BBN. Finally, we also note that $\bar{\nu}_{e}$-capture and $e^{-}$-capture transitions to the first excited state of $^7$Li are allowed, but their $Q$-values are less favourable.
To calculate the matrix element in Eq. (\ref{correnti}), we must specify the initial and final configurations on which the weak interaction potential $\hat{V}$ acts. This requires knowledge of the wave functions of the states involved in the electro-nuclear transition, namely, the nuclear and atomic shell configurations of both $^7$Be and $^7$Li. Within the nuclear shell model, the low-lying level structure of $^7$Be and $^7$Li is described as a closed $1s_{1/2}$ shell with three valence nucleons in the $1p_{3/2}$ shell (see Fig.(~\ref{nucleon level})). The corresponding shell model configurations of the two nuclei can be described as follows:
\begin{itemize}
\item $^7$Be (left side of Fig. (\ref{nucleon level})).
The nucleus contains four protons and three neutrons. Two protons and two neutrons fill the $1s_{1/2}$ closed shell with total spin-parity $J = 0^+$. In the incomplete $1p_{3/2}$ subshell, the remaining two protons tend to pair off to total angular momentum $J_{p} = 0$, while the remaining unpaired neutron carries $j_n = 3/2$. As the $p$-shell has negative parity, the ground state is $J^{\pi} = 3/2^-$. In our simulations, we therefore assumed a core with $J_{C} = 0$ that couples geometrically with the remaining neutron and essentially acts as a spectator for the transition.
\item $^7$Li (right side of Fig. (\ref{nucleon level})).
The level scheme is analogous to that of $^7$Be, with the only difference being that the unpaired nucleon in the $1p_{3/2}$ orbital is a proton. Consequently, the ground state is also $J^{\pi} = 3/2^{-}$.
\end{itemize}
Similar considerations apply to determining the symmetry of the excited nuclear levels of $^7$Be and $^7$Li, which result from the promotion of a neutron to the $1p_{1/2}$ nuclear level (see red arrows in Fig.~(\ref{nucleon level})).
These configurations are consistent with the observed level scheme and standard independent-particle shell model predictions, resulting in the ground and first excited states (both 1/2$^-$) shown in Fig.~(\ref{weakrates_scheme}). Unpaired nucleons of opposite isospin tend to couple to the maximum total angular momentum; in this context, $J_C = 3$ is the physically realised value, which we denote as the core-coupling angular momentum.
At even higher excitation energies, the pure shell model picture becomes less reliable, and significant configuration mixing is expected. Nevertheless, for the weak processes of interest, the relevant matrix elements are dominated by transitions involving the valence nucleons in the $1p$ shell.
\subsection{Electron and antineutrino capture rate}
For EC or AC on $^7$Be, four single--particle transitions may be considered:
\begin{itemize}
\item The $^7$Be ground state to $^7$Li ground state:
$p(1p_{3/2}) \rightarrow n(1p_{3/2})$;
\item The $^7$Be excited state to $^7$Li ground state:
$p(1p_{1/2}) \rightarrow n(1p_{3/2})$;
\item The $^7$Be excited state to $^7$Li excited state:
$p(1p_{1/2}) \rightarrow n(1p_{1/2})$,
\item The $^7$Be ground state to $^7$Li excited state:
$p(1p_{3/2}) \rightarrow n(1p_{1/2})$.
\end{itemize}
In all cases, a proton is converted into a neutron, while two nucleons (one proton and one neutron) remain as spectators in the $p$ shell. \\
\indent Within our ab initio framework, the decay rate per $^7$Be nucleus is obtained using Eq. (\ref{FGR}), where DHF calculations of the initial and final state nuclear and electronic wave functions are performed and substituted into the weak leptonic and hadronic currents in Eq. (\ref{correnti}). Additionally, this rate must be multiplied by the initial and final population factors and projectile densities, also taking into account the temperature through the chemical potentials. In particular, the electron, positron, and relic antineutrino population factors can be obtained via Fermi–Dirac distributions, each with its own temperature, as follows:
\begin{equation}
\label{pop_fact_FD}
\begin{aligned}
&f_{e^{\pm}}(E,T,\mu_e)=\frac{1}{1+e^{\frac{E\pm\mu_{e}}{k_{\rm{B}}T}}},\\
&f_{\nu}(E,T_{\nu})=\frac{1}{1+e^{\frac{E}{k_{\rm{B}}T_{\nu}}}},\\
\end{aligned}
\end{equation}
where \(T_{\nu}\) is the relic (anti-)neutrino temperature calculated in Eq. (\ref{neutrino_photon_temp}) of the Appendix (recall that we assume \(\mu_{\nu} = \mu_{\bar{\nu}} = 0\); see section (\ref{neu_chem}) of the Appendix), while electrons are in thermal equilibrium with photons, so \(T = T_{\gamma}\) is the usual photon temperature, and \(\mu_e\) is the electron chemical potential. Moreover, after the transition, the final neutrino must occupy a free state. Therefore, the required neutrino population factor is
\begin{equation}
\label{neutrino_pop}
g_{\nu}(E,T_{\nu})=1-f_{\nu}(E,T_{\nu})=1-\frac{1}{1+e^{\frac{E}{k_{\rm{B}}T_{\nu}}}}.
\end{equation}
The chemical potential of the electrons can be obtained by inverting Eq. (\ref{electron_density_integral_chemic_pot}) for a given proton density, assuming charge neutrality of the Universe and following the procedure explained in Appendix \ref{lep_bar_neu}.4.\\
\index The total rate for EC is obtained by integrating over all energies:
\begin{equation}
\label{total_EC_rate}
\lambda_{\mathrm{EC}}(T)=\int_{0}^{\infty}\Gamma(E)f_{e^-}(E,T,\mu_e)g_{\nu}(E,T_{\nu})dE.
\end{equation}
Moreover, the AC rate is
\begin{equation}
\label{total_ancap_rate}
\lambda_{\rm AC}(T)=\int_{-\infty}^{0}\Gamma(E)\left[1-f_{e^+}(E,T,\mu_e)\right]f_{\nu}(E,T_{\nu})\,dE.
\end{equation}
The integrals (\ref{total_EC_rate}) and (\ref{total_ancap_rate}) have been evaluated over the temperature range relevant after NSE freeze-out ($1~\mathrm{keV} < k_{\rm B}T < 100~\mathrm{keV}$) and beyond, up to $275~\mathrm{keV}$.
Finally, the total weak capture rate of ${}^{7}$Be is
\begin{equation}
\lambda_{\rm weak}(T) =
\lambda_{\rm EC}(T) + \lambda_{\rm AC}(T),
\end{equation}
which is evaluated numerically as a function of temperature (or cosmic time).
\begin{figure*}[hbt!]
\centering
\includegraphics[width=0.43\textwidth]{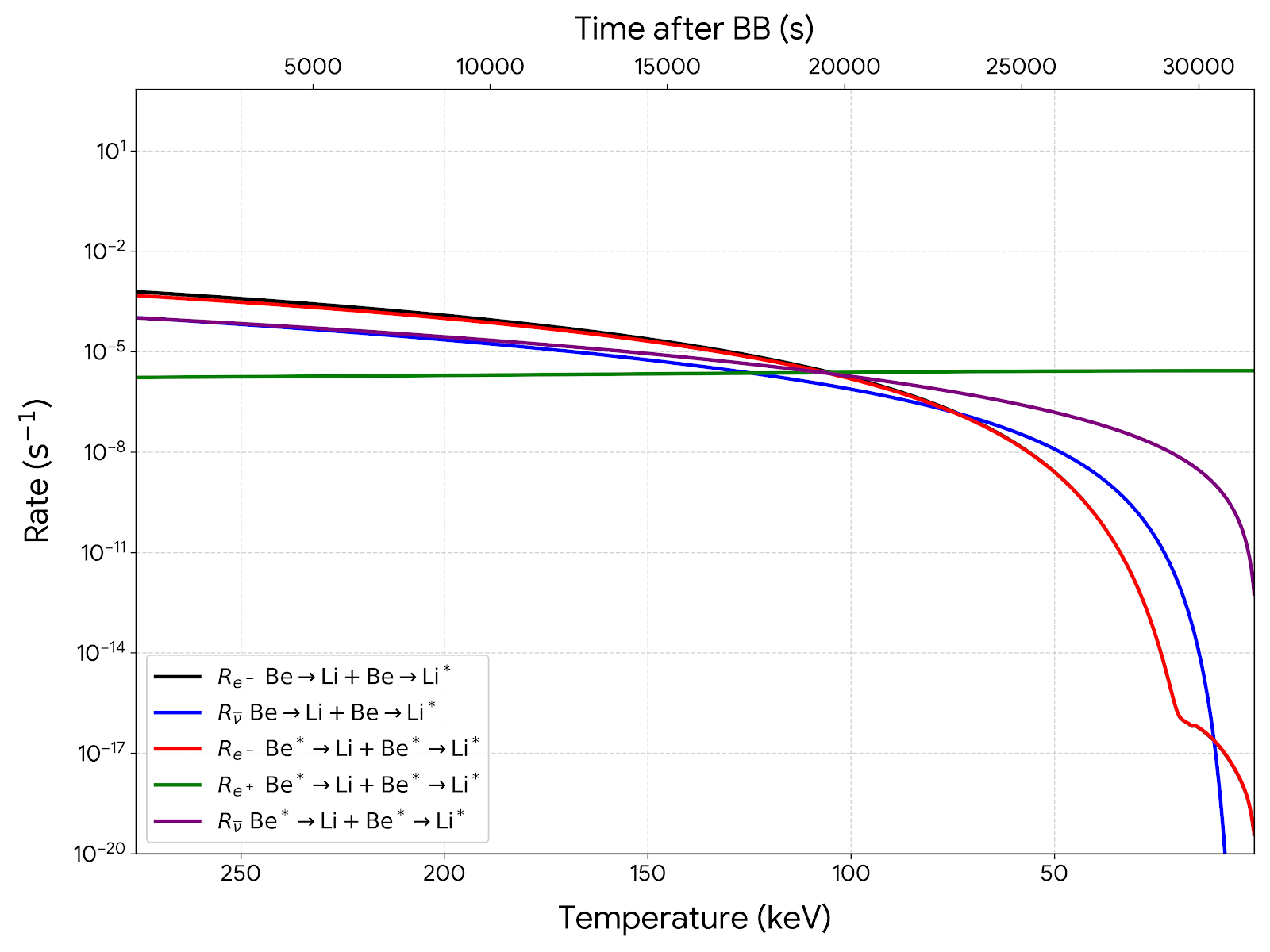}
\includegraphics[width=0.47\textwidth]{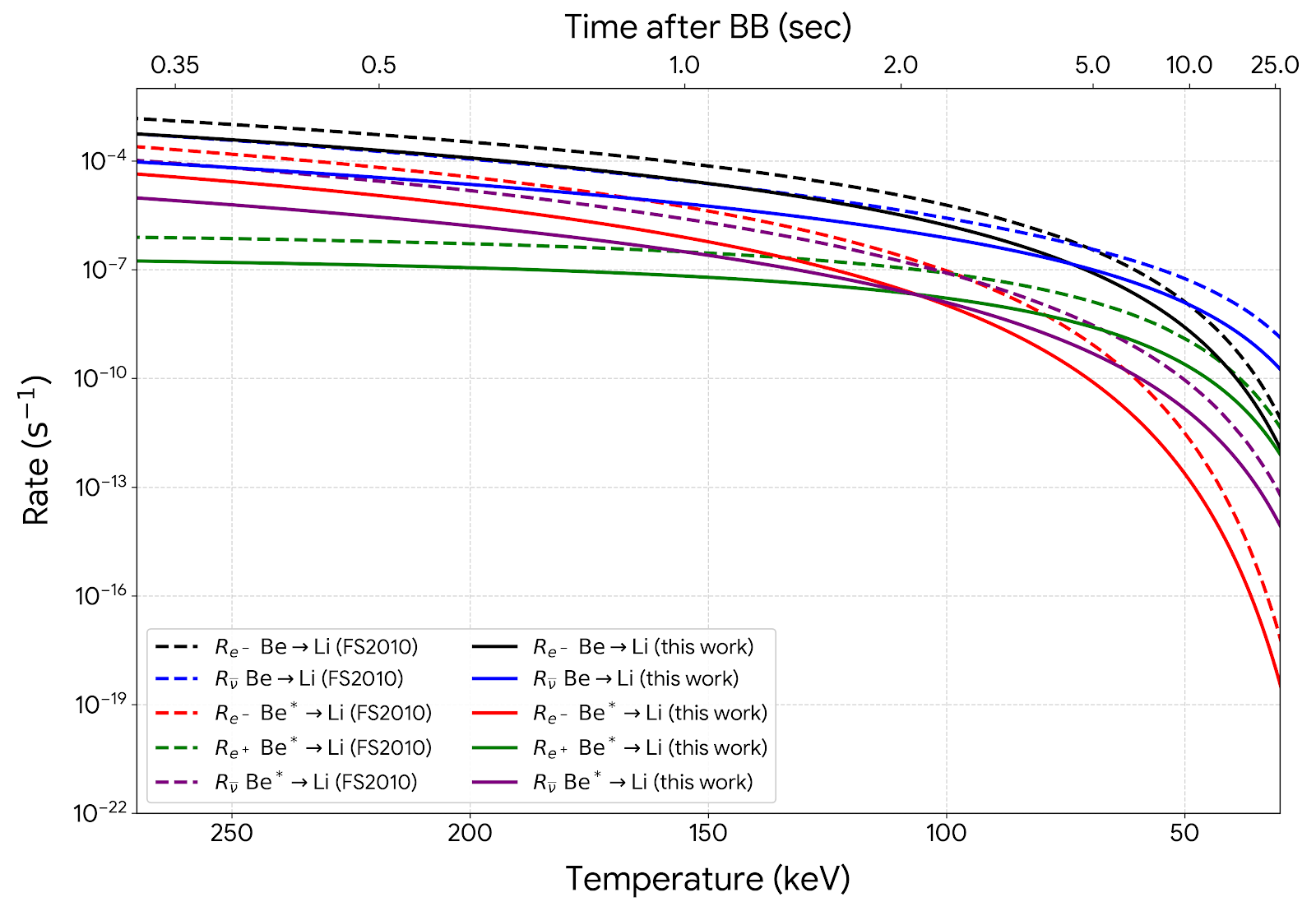}
\includegraphics[width=0.45\textwidth]{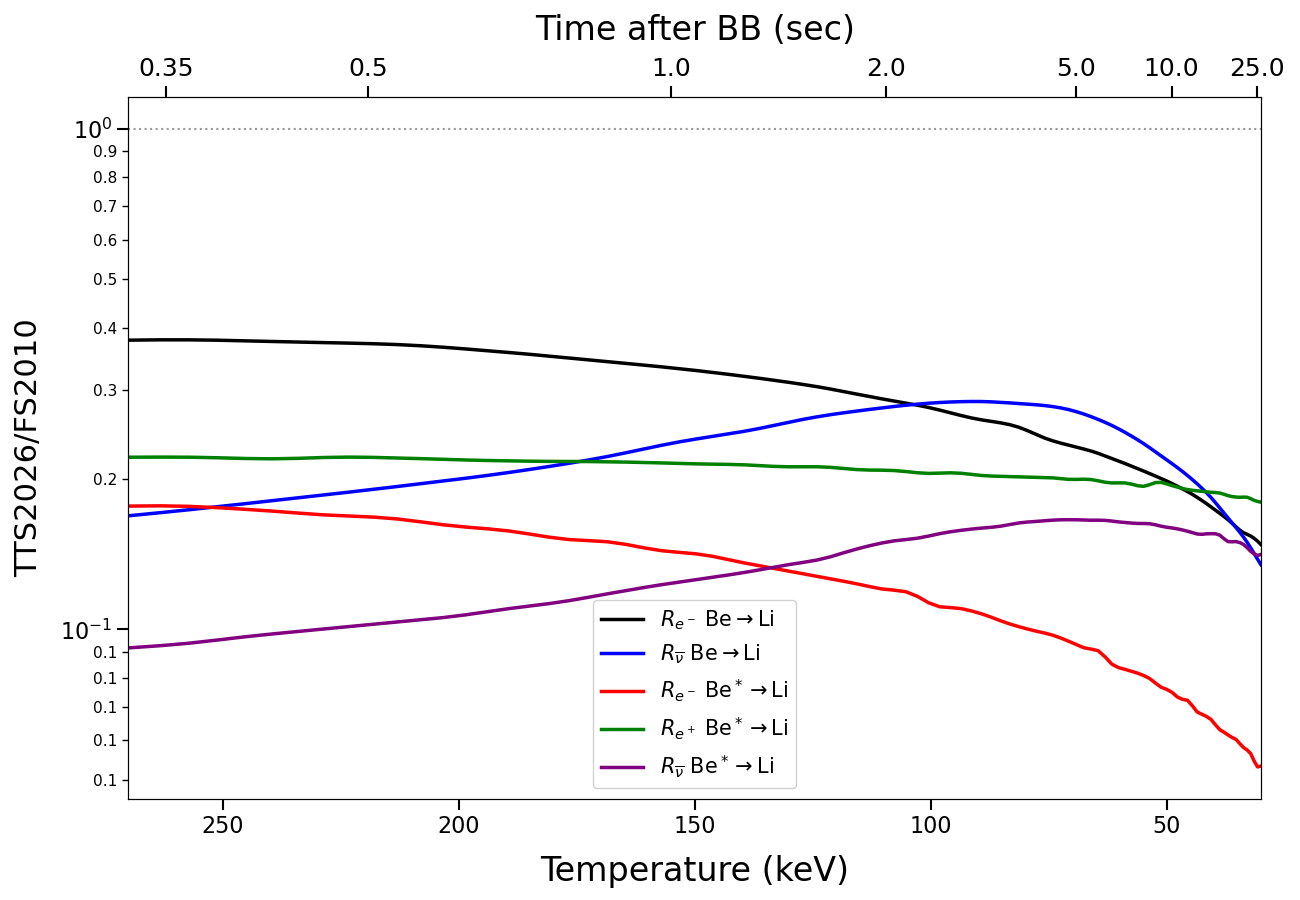}
\includegraphics[width=0.44\textwidth]{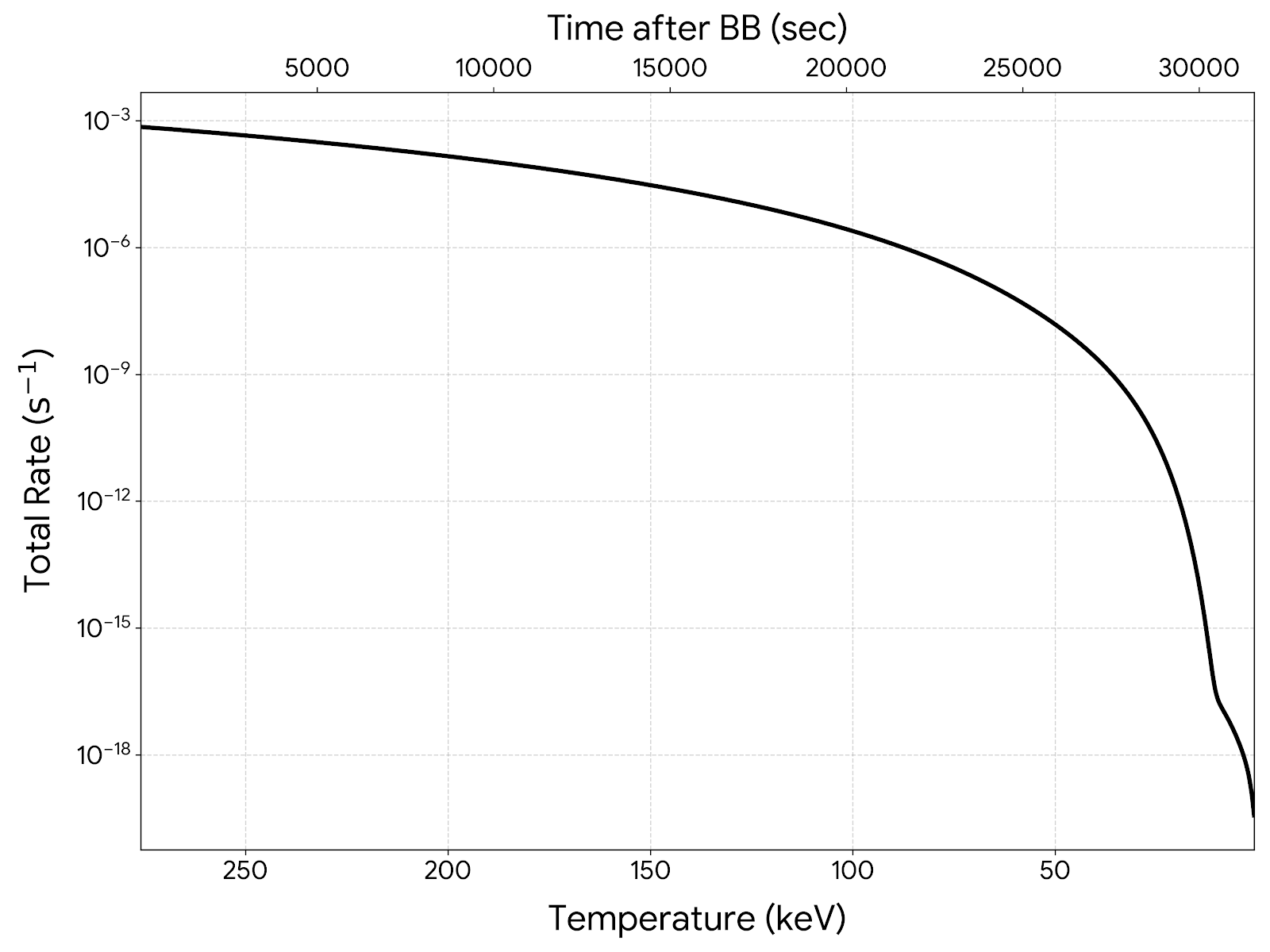}
\caption{Top left: Electron ($R_{e^-}$) and antineutrino ($R_{\overline{\nu}}$) capture rates as well as positron-emission ($R_{e^+}$) decay rates for the $^7$Be$\rightarrow{}^7$Li transitions shown as functions of temperature (lower $x$-axis) and indicative time (seconds) elapsed since the Big Bang (upper $x$-axis), according to the time--temperature parametrization introduced above. The decay rate for each individual transition is calculated assuming it is the only populated state (no population or degeneracy-factor correction). Be$^*$ and Li$^*$ denote the excited states of Be and Li, respectively.
Top right: Comparison between our new rates (solid
coloured lines) and those (dashed coloured lines) obtained in
Ref. \citep{fuller_smith}, including degeneracy and thermal
population factors as in Eq. (\ref{Boltzmann}).
{ Bottom left: Ratio between our new rates (TTS2026) and those obtained in Ref. \citep{fuller_smith} (FS2010), including degeneracy and thermal population factors as in Eq. (\ref{Boltzmann}).}
Bottom right:
Total rate obtained as
the sum of the rates of the different decay channels weighted by the respective
degeneracy and thermal population factors, as in Eq. (\ref{Boltzmann}).}
\label{fig:Be7_rates}
\end{figure*}
A similar concept applies when determining the decay rates of ${^7}$Be excited nuclear states, which opens the positron emission channel. For positron emission, similar calculations are required, but with positron kinematics replacing those of the captured electron or antineutrino. Finally, the total rate is the sum of the independent rates for the ground and excited nuclear states for all possible weak reactions, where the relevant population and state degeneracy factors follow a Boltzmann distribution as in Eq. (\ref{Boltzmann}).
\begin{table*}[hbt!]
\caption{Weak-interaction rates for the $^7$Be$\rightarrow{}^7$Li transitions.}
\label{tab:Be7_rates_full}
\centering
\begin{tabular}{c c c c c c c c}
\hline\hline
$t$  &
$k_{\rm{B}}T$  &
$R_{e^-}$  &
$R_{\overline{\nu}}$ &
$R_{e^-}$  &
$R_{e^+}$ &
$R_{\overline{\nu}}$ & $R_{\mathrm{tot}}$ \\
(s) & (keV) & Be$\rightarrow$(Li + Li$^*$)  & Be$\rightarrow$(Li + Li$^*$) &
Be$^*$$\rightarrow$(Li + Li$^*$) &
Be$^*$$\rightarrow$(Li + Li$^*$) & Be$^*$$\rightarrow$(Li + Li$^*$) &
\\
\hline
3.156e-01 & 2.759e+02 & 6.176e-04 & 1.035e-04 & 4.713e-04 & 1.682e-06 & 1.018e-04 & 7.072e-04 \\
5.049e-01 & 2.181e+02 & 1.917e-04 & 3.403e-05 & 1.532e-04 & 1.867e-06 & 3.926e-05 & 2.237e-04 \\
8.078e-01 & 1.724e+02 & 5.364e-05 & 1.105e-05 & 4.494e-05 & 2.048e-06 & 1.517e-05 & 6.459e-05 \\
1.355e+00 & 1.332e+02 & 1.154e-05 & 3.158e-06 & 1.014e-05 & 2.226e-06 & 5.449e-06 & 1.476e-05 \\
2.167e+00 & 1.053e+02 & 2.410e-06 & 9.749e-07 & 2.194e-06 & 2.360e-06 & 2.206e-06 & 3.413e-06 \\
3.462e+00 & 8.329e+01 & 4.045e-07 & 2.824e-07 & 3.791e-07 & 2.464e-06 & 9.219e-07 & 6.957e-07 \\
5.500e+00 & 6.608e+01 & 5.165e-08 & 7.476e-08 & 4.951e-08 & 2.539e-06 & 4.017e-07 & 1.286e-07 \\
9.151e+00 & 5.123e+01 & 3.427e-09 & 1.453e-08 & 3.349e-09 & 2.593e-06 & 1.667e-07 & 1.828e-08 \\
1.454e+01 & 4.065e+01 & 1.708e-10 & 2.633e-09 & 1.692e-10 & 2.624e-06 & 7.696e-08 & 2.839e-09 \\
2.419e+01 & 3.151e+01 & 2.941e-12 & 2.902e-10 & 2.946e-12 & 2.643e-06 & 3.357e-08 & 2.948e-10 \\
3.842e+01 & 2.500e+01 & 3.017e-14 & 2.651e-11 & 3.046e-14 & 2.652e-06 & 1.600e-08 & 2.658e-11 \\
6.103e+01 & 1.984e+01 & 1.783e-16 & 1.498e-12 & 1.811e-16 & 2.657e-06 & 7.695e-09 & 1.499e-12 \\
1.015e+02 & 1.538e+01 & 6.567e-17 & 3.110e-14 & 6.704e-17 & 2.660e-06 & 3.468e-09 & 3.117e-14 \\
1.613e+02 & 1.220e+01 & 3.480e-17 & 3.997e-16 & 3.566e-17 & 2.661e-06 & 1.691e-09 & 4.345e-16 \\
2.563e+02 & 9.681e+00 & 1.695e-17 & 1.870e-18 & 1.741e-17 & 2.662e-06 & 8.283e-10 & 1.882e-17 \\
4.264e+02 & 7.505e+00 & 7.629e-18 & 1.155e-21 & 7.856e-18 & 2.662e-06 & 3.795e-10 & 7.630e-18 \\
6.773e+02 & 5.955e+00 & 3.853e-18 & 2.498e-25 & 3.974e-18 & 2.663e-06 & 1.873e-10 & 3.853e-18 \\
1.076e+03 & 4.724e+00 & 2.057e-18 & 6.731e-30 & 2.124e-18 & 2.663e-06 & 9.262e-11 & 2.057e-18 \\
1.790e+03 & 3.663e+00 & 1.091e-18 & 2.977e-36 & 1.128e-18 & 2.663e-06 & 4.280e-11 & 1.091e-18 \\
2.844e+03 & 2.906e+00 & 6.300e-19 & 1.457e-43 & 6.518e-19 & 2.663e-06 & 2.125e-11 & 6.300e-19 \\
4.732e+03 & 2.253e+00 & 3.431e-19 & 9.190e-54 & 3.552e-19 & 2.663e-06 & 9.849e-12 & 3.431e-19 \\
7.517e+03 & 1.787e+00 & 1.942e-19 & 1.578e-65 & 2.012e-19 & 2.663e-06 & 4.901e-12 & 1.942e-19 \\
1.194e+04 & 1.418e+00 & 1.101e-19 & 2.629e-80 & 1.140e-19 & 2.663e-06 & 2.441e-12 & 1.101e-19 \\
1.987e+04 & 1.099e+00 & 6.094e-20 & 5.528e-101 & 6.316e-20 & 2.663e-06 & 1.135e-12 & 6.094e-20 \\
3.156e+04 & 8.723e-01 & 3.725e-20 & 6.943e-125 & 3.862e-20 & 2.663e-06 & 5.665e-13 & 3.725e-20 \\
\hline
\end{tabular}
	\tablefoot{
    First and second columns: Indicative time (seconds) elapsed since the Big Bang and the corresponding plasma temperature (keV), according to the time--temperature parametrization introduced above. 3rd and 4th columns: electron ($R_{e^-}$) and antineutrino ($R_{\bar{\nu}}$) capture rates from $^7$Be (ground) to $^7$Li (ground) and $^7$Li$^*$ (excited nuclear state). 5th, 6th, and 7th columns: electron ($R_{e^-}$), positron ($R_{e^+}$), and antineutrino ($R_{\bar{\nu}}$) capture rates from $^7$Be$^*$ (excited nuclear state) to $^7$Li (ground) and $^7$Li$^*$ (excited nuclear state). Last column: total rate ($R_{\mathrm{tot}}$). Rates are given in units of s$^{-1}$. Only the total rate (last column) includes the population and degeneracy factors of the nuclear states from Eq. (\ref{Boltzmann}). The complete table, containing all calculated temperatures, is available as Supporting Material in the online version of the paper.}
\end{table*}
The results are listed in Tab. (\ref{tab:Be7_rates_full}) and shown in Fig.~(\ref{fig:Be7_rates}). Within the same [1–275] keV range, we have also provided a table in a supporting information file with 250 points, which should be sufficient for interpolation. Additionally, Tables~(\ref{tab:rate_Be7_ee}), (\ref{tab:rate_Be7_eg}), (\ref{tab:rate_Be7_gg}), and (\ref{tab:rate_Be7_ge}) in the Appendix present the results specific to each transition type (EC, positron decay, and AC) and nuclear level (ground or excited states of Be and Li). \\
\indent
We note that only in the last column of Tab. (\ref{tab:Be7_rates_full}) are the decay rates weighted using the population and degeneracy factors of the nuclear states from Eq. (\ref{Boltzmann}). This is because we assume that the decay characterising each individual transition shown in columns three to seven occurs as if the occupied parent nuclear state were the only populated state.
These individual transition rates, without including degeneracy or thermal occupation, are reported in the top left panel of Fig. (\ref{fig:Be7_rates}). We observe that the contribution of \ancap-capture becomes more significant compared to EC as temperature decreases (or time increases), while positron emission, which occurs only from the \berilliosette~excited nuclear state, remains roughly constant with temperature. { In the latter case, two phenomena occur: on the one hand, the Be nuclear excited state becomes exponentially depleted, while, as \(e^+ e^-\) annihilation proceeds, the positron Fermi sea becomes progressively less occupied, enhancing the probability of this decay channel.} 
Moreover, in the top-right panel of Fig. (\ref{fig:Be7_rates}), our rates (solid lines), weighted with the appropriate degeneracy and thermal population factors as in Eq. (\ref{Boltzmann}), are compared with those from Ref. \citep{fuller_smith} (dashed lines), showing that our rates are consistently lower at all temperatures.
Our self-consistent relativistic ab initio calculations provide a systematic revision of the weak transition rates compared with earlier work \citep{fuller_smith}. However, rather than a uniform shift, the discrepancy is strongly temperature-dependent, even though all rates decay exponentially as the temperature decreases. EC to the ground state is the dominant process at high energy (see black solid and dashed lines in the top-right panel of Fig. (\ref{fig:Be7_rates})), while the ground-state AC rate (see blue solid and dashed lines in the top-right panel of Fig. (\ref{fig:Be7_rates})) becomes comparable to EC at a higher temperature (corresponding to shorter times after the Big Bang) than in Ref. \citep{fuller_smith}.
Overall, the previously parametrized rates, based on $\log(ft)$ analysis, tend to overestimate the transition channels in the high-energy window (250 to 100 keV), with the EC rate from Be$^*$ higher than our estimates. In particular, at the temperatures relevant for BBN, when the abundance of $^7$Be builds up ($T \simeq 0.1$ MeV), the total weak destruction rate is dominated by EC and AC through ground-state to ground-state and ground-state to first-excited-state transitions. Around this epoch, the contributions from the two processes are of comparable magnitude in our assessment.
As the Universe cools, however, the electron-positron plasma rapidly annihilates, causing the electron density to decrease exponentially below $T \simeq 80$ keV. Consequently, the EC rate becomes strongly suppressed, whereas AC remains comparatively more efficient owing to the persistent relic antineutrino background. Although electron-positron annihilation heats the photon bath relative to the decoupled neutrino background, the electron density decreases far more rapidly than the high-energy tail of the relic antineutrino distribution that is relevant for the capture process.
Consequently, AC becomes the dominant weak destruction channel of $^7$Be at low temperatures.\\
\indent
The first excited state of $^7$Be lies at an excitation energy of 429.2 keV, substantially above the characteristic temperature ($T\simeq 0.1$ MeV) at which $^7$Be is synthesized during BBN. Consequently, its thermal population is strongly suppressed, reaching only about 1\% at $T\simeq 0.1$ MeV. As a result, weak processes involving this state provide only a minor contribution to the overall destruction rate of $^7$Be during its production epoch.\\
\indent
The rates associated with the thermally populated first excited state are also shown in the top-right panel of Fig. (\ref{fig:Be7_rates}). At $T \simeq 0.1$ MeV, the AC, EC, and positron decay rates are of comparable magnitude and together account for only a few percent of the total weak destruction rate. As the temperature decreases, however, an interesting behaviour emerges: Despite the exponentially decreasing population of the excited state, the positron decay channel remains sufficiently efficient to become comparable to the EC rate from the ground state. This reflects the different temperature dependencies governing the competing weak processes and highlights the non-negligible role that positron decay from the excited $^7$Be nuclear state may play in the late stages of BBN. The same trend is visible in the previous simulations in Ref. \citep{fuller_smith}, although our results show localised differences that reflect the sensitivity of the Gamow–Teller matrix elements to the valence-shell configurations and nuclear correlations considered here.\\
\indent { Moreover, the ratio (TTS2026/FS2010) between our first-principles rates (TTS2026) and those obtained in Ref. \citep{fuller_smith} (FS2010) is shown in the bottom-left panel of Fig. \ref{fig:Be7_rates}. The ratios typically lie between $\simeq$0.1 and 0.4 over the BBN temperature range, corresponding to a reduction of the present rates by a factor of $\simeq$2.5–10 relative to the semi-empirical $\log(ft)$-based estimates. As the thermal population and degeneracy factors are applied consistently to both TTS2026 (this work) and FS2010 when constructing the ratio, we attribute the dominant, nearly channel-independent suppression of the rates to the different treatments of the hadronic and leptonic currents and their channel-dependent relativistic phase space. FS2010 relies on semi-empirical $\log(ft)$ systematics, together with phenomenological shape and Fermi functions, to deal with the weak transition matrix elements, whereas in the present work both the hadronic and leptonic currents entering the weak-interaction matrix element are evaluated explicitly from relativistic single-particle wave functions. In particular, our treatment includes the spatial structure of the leptonic current, continuum and bound electronic states, non-orthogonality effects, and the corresponding relativistic phase-space integrations, rather than absorbing these effects into a semi-empirical $\log(ft)$ value.
The systematic reduction of the present rates relative to the semi-empirical compilation of Ref. \citep{fuller_smith} is consistent with our previous investigations of stellar weak-interaction rates \citep{taioli2022theoretical}. In that work, the same first-principles formalism was benchmarked against the widely used Takahashi \& Yokoi prescription \citep{TAKAHASHI1983578,TAKAHASHI1987375}, which is likewise based on $\log(ft)$ systematics. By selectively including nuclear excited states and electronic degrees of freedom, it was shown that these effects can substantially modify the stellar rates, but do not eliminate the systematic differences between the semi-empirical and microscopic approaches. Even after consistently accounting for thermal population of excited nuclear states and electronic excitations, the fully microscopic calculations generally yielded smaller weak rates than the corresponding $\log(ft)$ estimates, with deviations ranging from factors of a few up to nearly two orders of magnitude depending on the isotope and plasma conditions. The present comparison with FS2010 therefore follows the same overall trend, suggesting that the observed reduction primarily reflects the replacement of phenomenological $\log(ft)$ strengths by an explicit evaluation of the hadronic and leptonic weak currents together with the corresponding relativistic phase-space integrals. Although the quantitative differences are nucleus dependent, these previous benchmarks indicate that the discrepancy observed here is not an isolated feature of the present calculation, but rather a systematic consequence of replacing semi-empirical weak-transition strengths with a fully microscopic treatment of the weak-interaction matrix elements.}\\
\indent Finally, the bottom-right panel of Fig. \ref{fig:Be7_rates} shows that the total decay rate, obtained as the sum of the rates of the different processes weighted by the respective degeneracy and thermal population factors as in Eq. (\ref{Boltzmann}), decreases sharply as the temperature falls.
\\
\indent We note that, despite discrepancies between ab initio and phenomenological analyses, which leave the global cosmological trend of primordial $^7$Be processing qualitatively unchanged, the structural robustness of our updated BBN network is confirmed.
Even $\approx$ 2 s after the Big Bang, corresponding to a temperature of 105 keV, the EC half-life of \berilliosette~is approximately 2 days, which is too long compared to the synthesis of primordial \litiosette, believed to have occurred within the first 1000 s of the Universe’s existence.
\section{Discussion and conclusions}
\label{sec:discussion}
In this work, we have reassessed several weak and radiative processes involving $^7$Be and $^7$Li under BBN conditions with the aim of quantifying their possible impact on the long-standing cosmological lithium problem. In particular, we focused on (i) EC and AC on $^7$Be, (ii) positron emission from the nuclear excited state of $^7$Be, and (iii) PC via both the standard radiative channel $^7$Be$(p,\gamma)^8$B, including the possibility of SE from the dense photon background, and an Auger-like three-body variant, in which the capture energy is transferred to a continuum electron (reported in Appendix A).\\
\indent
Our calculations were carried out within a first-principles framework that treats leptonic and hadronic currents consistently and includes temperature- and density-dependent effects appropriate to the post-NSE epoch ($10\,\mathrm{keV} < k_{\rm B}T < 100\,\mathrm{keV}$) and to earlier times, approximately $0.3\,\mathrm{sec}$ after the Big Bang (corresponding to a photon temperature of $275\,\mathrm{keV}$). In contrast to previous estimates based on semi-empirical log$(ft)$ systematics, our approach explicitly accounts for ionisation, excitation, relativistic effects, and the actual populations of bound and continuum leptonic states in the primordial plasma, and it calculates the nuclear wave functions of ground and excited states within an ab initio framework.\\
\indent
We find that the in situ EC rate of $^7$Be decreases rapidly as
the Universe expands and cools, reflecting the dilution of the electron density
and the progressive departure from pair equilibrium. The inclusion of the
AC channel enhances the total weak interaction rate at early
times; however, its contribution becomes negligible as the neutrino background
dilutes after weak freeze-out. Overall, the combined effect of EC and AC and $\beta^+$ decay led to significant corrections in the $^7$Be destruction channels during BBN, resulting in a half-life of approximately 2 days. However, this is still not entirely sufficient to reconcile the observations with the BBN model in the very early stages of the Universe. \\
\indent
For PC (whose cross section formulae and calculations are given in Appendix A), we confirm that plasma screening only results in a modest enhancement (of the order of a few percent) of the thermally averaged $^7$Be$(p,\gamma)^8$B rate in the temperature range considered.
The Auger-like
three-body capture mechanism, although conceptually interesting and analogous
to internal conversion processes in atomic physics, was found to be subdominant.
At $k_{\rm{B}}T \simeq 100$~keV, its cross section is approximately 0.004\% of the radiative PC of the radiative-only channel, but it decreases steeply with temperature and becomes negligible ($\sim 10^{-10}$ of the radiative rate) by $k_{\rm{B}}T \simeq 10$~keV. As a
result, this channel does not provide a significant additional pathway for
$^7$Be depletion during the relevant BBN epoch. This picture does not change significantly if we also consider adding a novel mechanism to the standard PC mechanism due to the presence of a high-density $\gamma$-photon background, which is conceptually similar to SE caused by lasers in cavities. The enhancement of the cross section due to this background is less than 5\%.\\
\indent
Finally, it is worth commenting on the potential implications of the weak interaction rates discussed here for solar neutrino physics. In the solar core, EC on
\({}^{7}\mathrm{Be}\) competes with PC,
\({}^{7}\mathrm{Be}(p,\gamma){}^{8}\mathrm{B}\),
in determining the branching ratio between the
\({}^{7}\mathrm{Be}\) and
\({}^{8}\mathrm{B}\) neutrino production channels.
However, the thermodynamic conditions in the solar core
(\(T\simeq1.3\,\mathrm{keV}\) and high baryon density)
differ substantially from those encountered during BBN
(\(T\simeq30\)--\(280\,\mathrm{keV}\)).
At solar temperatures, the first excited state of
\({}^{7}\mathrm{Be}\), located at an excitation energy of
429.2 keV, is completely depopulated since the corresponding
Boltzmann factor is
\[
\exp\!\left(-\frac{429.2}{1.3}\right)\approx10^{-144}.
\]
Consequently, all weak channels involving
\({}^{7}\mathrm{Be}^{*}\), which play an important role in the
BBN lithium problem discussed in the present work,
are physically irrelevant under solar conditions.
Likewise, positron decay and AC channels are
negligible in the solar plasma, so the only weak process
directly affecting solar neutrino production is the ground-state
EC reaction,
\[
{}^{7}\mathrm{Be}+e^{-}\rightarrow{}^{7}\mathrm{Li}+\nu_{e}.
\]
Nevertheless, some qualitative conclusions regarding the impact
of an improved treatment of weak interaction rates on solar
neutrino production can be drawn from previous investigations.
As shown in Ref.~\citep{vescovi2019effects}, adopting a revised
\({}^{7}\mathrm{Be}\) EC rate that incorporates a
more accurate description of plasma effects reduces the
ground-state EC rate by approximately \(4\%\).
This leaves a larger fraction of
\({}^{7}\mathrm{Be}\) available for PC,
thereby increasing the predicted
\({}^{8}\mathrm{B}\) neutrino flux by up to \(2.7\%\),
while leaving the overall solar structure essentially
unchanged. Interestingly, this revised prediction leads to
slightly improved agreement with the Sudbury Neutrino
Observatory (SNO) measurements of the neutral-current
\({}^{8}\mathrm{B}\) neutrino flux.
Although the multichannel formalism developed in the present work has been specifically derived for the thermodynamic conditions of BBN, its extension to the physical conditions of the solar core, including the appropriate treatment of plasma screening and electron distributions, could provide further refinements to Standard Solar Model calculations and represents an interesting direction for future work.\\
\indent
Taken together, our results indicate that neither enhanced weak interaction
rates nor Auger-like PC processes are capable on their own of
resolving the discrepancy between the primordial $^7$Li abundance predicted by
standard BBN and that inferred from observations of metal-poor halo stars.
Nevertheless, our study highlights the importance of computing nuclear and
electroweak reaction rates under cosmological conditions using accurate
first-principles simulations before considering unexplored physics beyond the
Standard Model. Future work should refine the three-body matrix element beyond
the plane-wave approximation, reassess screening using kinetic plasma models,
and propagate the updated rates through a complete BBN network. Moreover, we
believe the conditions under which BBN models predict nucleus formation should
be revised to account for possible earlier formation, even before nuclear
statistical equilibrium, under conditions different from those typically
considered. This would further reduce the decay half-life of the relevant
isotopes and result in closer agreement with observations.

\section*{Data availability}

Tables 1 and A.1 with 250
points, which should be sufficient for interpolation, are also available in electronic form at the CDS via
anonymous ftp to cdsarc.u-strasbg.fr (130.79.128.5) or via
\url{http://cdsweb.u-strasbg.fr/cgi-bin/qcat?J/A+A/}.

\begin{acknowledgements}
This work has received funding from the European Union under the Mimosa grant agreement No. 10104665 and from INFN through the project PANDORA Gr3, funded by the 3rd National Scientific Commission. The authors acknowledge fruitful discussions with Dr. Martino Silvetti (University of Modena and Reggio Emilia).
\end{acknowledgements}

\begin{appendix}

\section{Proton capture on ${}^{7}$Be}
\begin{figure}[hbt!]
\centering
\includegraphics[width=0.3\textwidth]{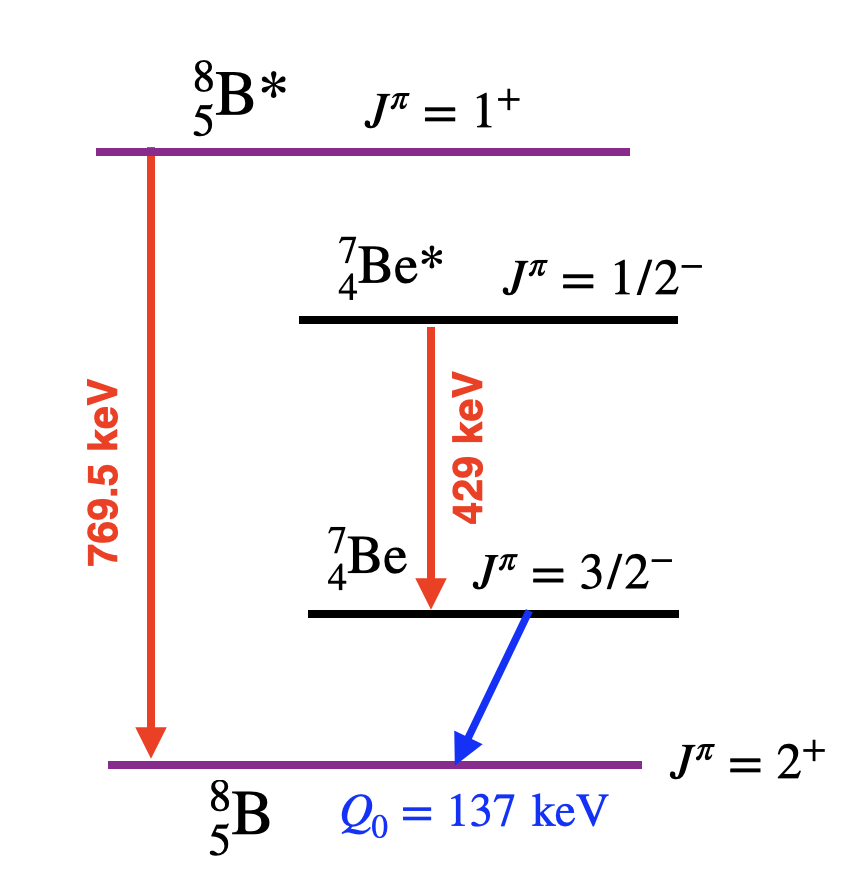}
\caption{$^{7}$Be proton capture decay scheme. The $Q$-value for the only decay channel considered here (ground-to-ground state decay) is also provided (0 = nuclear ground state, $^*$ indicates the excited states of $^{7}$Be and $^{8}$B).}
\label{BetoB}
\end{figure}

The reaction $^{7}\mathrm{Be}(p,\gamma)^{8}\mathrm{B}$ at BBN energies is dominated by non-resonant direct capture, primarily populating the $^{8}\mathrm{B}$ ground state ($J^\pi=2^+$) with a $Q$-value of 137 keV. Higher-lying resonant states (at approximately 769.5 keV) are at too high in energy for proton kinetic energies below NSE and therefore contribute only subdominantly. A diagram of the nuclear levels involved in this reaction is shown in Fig. (\ref{BetoB}).
\\
\indent
In this context, we note that the photon carries away angular momentum 1. At the low energies relevant for BBN, this reaction proceeds predominantly via electric E1 dipole transitions from incoming $s$-wave protons, due to the absence of a centrifugal barrier for $l=0$, with a subleading contribution from $d$-wave capture to the bound $p_{3/2}$ configuration of $^{8}\mathrm{B}$. The $s$-wave to $p$-shell bound state transition, together with the selection rules (parity change $\Delta\pi = -1$ and $\Delta l = \pm 1$), leads to a $2^+$ nuclear ground state of $^{8}\mathrm{B}$.
\subsection{Radiative capture ${}^{7}\mathrm{Be}(p,\gamma){}^{8}\mathrm{B}$: Classical approach}
We have analysed two types of PC, as reported in (\ref{Be_to_B}) and (\ref{radiative_b}).
Because the first excited state of $^{8}\mathrm{B}$ lies at $770~\mathrm{keV}$, well above the reaction $Q$-value ($Q \simeq 137~\mathrm{keV}$), PC at astrophysical energies can populate only the ground state of $^{8}\mathrm{B}$. Thus, in these reactions, almost the entire $Q$-value (137 keV) is carried away by the emitted photon, while the recoil energy of the $^{8}\mathrm{B}$ nucleus is negligible ($<1~\mathrm{keV}$).\\
\indent
For these reactions, we must determine the corresponding cross sections, from which we can, in principle, obtain the nuclear decay rate of ${}^{7}\mathrm{Be}$ by multiplying by the proton flux (number of protons per unit area per unit time).\\
\indent The thermally averaged reaction rate per particle pair per unit volume can be determined by considering protons and Be nuclei as classical particles, that is their velocities $v_{p}, v_{\mathrm{Be}}$ are distributed according to the following Maxwell-Boltzmann distribution
\begin{equation}
f(v_i) = 4\pi \left(\frac{m_i}{2\pi k_{\rm{B}}T}\right)^{3/2}
v_i^{2}\exp^{\left(-\frac{m_i v_i^{2}}{2k_{\rm{B}}T}\right)},
\end{equation}
where $v_i = v_{\mathrm{Be}},~m_i = m_{\mathrm{Be}}$ or $v_i = v_p, ~m_i = m_p$ are the velocities and masses of Be and $p$, respectively, and $T$ is the plasma temperature.
As each proton and Be nucleus has a different momentum, we must integrate over all possible momenta. \\
\indent The PC rate ($R_{\mathrm{PC}}$) per target particle of Be per unit volume is therefore given by
\begin{eqnarray}\label{rate_PBe}
&R_{\mathrm{PC}}  =n_{p}^{\mathrm{free}}(T)\int\frac{m_{p}^{3}d^{3}v_{p}}{\left(2\pi m_{p}k_{\rm{B}}T\right)^{\frac{3}{2}}}\int\frac{m_{\mathrm{Be}}^{3}d^{3}v_{\mathrm{Be}}}{\left(2\pi m_{\mathrm{Be}}k_{\rm{B}}T\right)^{\frac{3}{2}}}e^{-\frac{m_{p}v_{p}^{2}}{2k_{\rm{B}}T}-\frac{m_{\mathrm{Be}}v_{\mathrm{Be}}^{2}}{2k_{\rm{B}}T}}v_{\mathrm{rel}}\sigma\left(v_{\mathrm{rel}}\right)\nonumber\\
& =n_{p}^{\mathrm{free}}(T)\frac{\left(m_{p}m_{\mathrm{Be}}\right)^{\frac{3}{2}}}{\left(2\pi k_{\rm{B}}T\right)^{3}}\int d^{3}v_{\mathrm{rel}}\int d^{3}V_{\mathrm{CM}}e^{-\frac{MV_{\mathrm{CM}}^{2}}{2k_{\rm{B}}T}-\frac{\mu v_{\mathrm{rel}}^{2}}{2k_{\rm{B}}T}}v_{\mathrm{rel}}\sigma\left(v_{\mathrm{rel}}\right)\nonumber\\
& =n_{p}^{\mathrm{free}}(T)\left(\frac{m_{p}m_{\mathrm{Be}}}{m_{p}+m_{\mathrm{Be}}}\right)^{\frac{3}{2}}\frac{1}{\left(2\pi k_{\rm{B}}T\right)^{\frac{3}{2}}}\int d^{3}v_{\mathrm{rel}}e^{-\frac{\mu v_{\mathrm{rel}}^{2}}{2k_{\rm{B}}T}}v_{\mathrm{rel}}\sigma\left(v_{\mathrm{rel}}\right)\nonumber\\
&=n_{p}^{\mathrm{free}}(T)\left(\frac{\mu}{2\pi k_{\rm{B}}T}\right)^{\frac{3}{2}}\int d^{3}v_{\mathrm{rel}}e^{-\frac{\mu v_{\mathrm{rel}}^{2}}{2k_{\rm{B}}T}}v_{\mathrm{rel}}\sigma\left(v{_\mathrm{rel}}\right),
\end{eqnarray}
where $\sigma({v_{\rm rel}})$ is the reaction cross section calculated at the relative velocity ${v_{\rm rel}}$ of the Be target and the incident proton, $V_{\mathrm{CM}}$ is the speed of the centre of mass, $n_{p}^{\mathrm{free}}(T)$ is the free proton number density at the plasma temperature $T$, and $\mu = \frac{m_{p} m_{Be}}{m_{p} + m_{Be}}$ is the reduced mass of the two colliding bodies.\\
\indent The expression for the cross section of the strong force-driven reaction (\ref{Be_to_B}) as a function of the photon temperature in the range 1 keV < $k_{\rm{B}}T$ < 275 keV can be given in atomic units as \citep{PhysRevC.67.065805,filippone1,filippone2,PhysRevC.68.065803}
\begin{equation}\label{cs_proton}
\sigma\left(v_\mathrm{rel}\right)=\frac{1}{e^{2\pi\eta}-1}\left(\frac{2S_{17}(0)}{\mu v_\mathrm{rel}^{2}}\right),
\end{equation}
where $\eta = \frac{Z_p Z_\mathrm{Be}}{v_\mathrm{rel}}$ is the Sommerfeld parameter in a.u., $Z_p$ and $Z_\mathrm{Be}$ are the atomic numbers of the proton and Be, and $S_{17}(0) \simeq 21.2~\mathrm{eV\,b} = 2.7822 \times 10^{-8}~\mathrm{a.u.}$ is the experimental value of the $S$-factor, which describes quantum tunnelling through the Coulomb barrier given by the Gamow energy
\begin{equation}
\label{gamow_energy}
E_{G}=\frac{(2\pi\alpha Z_pZ_{\mathrm{Be}})^{2}\mu c^{2}}{2}=13,798.8~\rm{keV}.
\end{equation}
In Eq. (\ref{cs_proton}), which describes $s$-wave scattering of a Coulomb-distorted plane wave on the point-like Be nucleus, we have assumed that the S-factor has a weak energy dependence, represented by a linear expansion $S_{17}(E) = S_{17}(0) + S'_{17} E$, and further that $S_{17}(E) = S_{17}(0)$ is constant at all energies.\\
\indent
Using the cross section in Eq. (\ref{cs_proton}) to calculate $R_{\mathrm{PC}}$ in Eq. (\ref{rate_PBe}), the thermally averaged rate per particle pair is therefore
\begin{eqnarray}\label{final_rate}
&&R_{\mathrm{PC}} =n_{p}^{\mathrm{free}}(T)\left(\frac{\mu}{2\pi k_{\rm{B}}T}\right)^{\frac{3}{2}}\times \nonumber \\ &&\int_{0}^{\infty}4\pi v_{rel}^{2}dv_{rel}v_{rel}\frac{1}{e^{2\pi\eta}-1}\left(\frac{2S_{17}(0)}{\mu v_{rel}^{2}}\right)e^{-\frac{\mu v_{rel}^{2}}{2k_{\rm{B}}T}}.
\end{eqnarray}
The net number density of free protons, $n_{p}^{\mathrm{free}}$, can be obtained by following the procedure explained in Appendix \ref{baden} (see, in particular, Eq. (\ref{free_proton_density})).
We note that plasma screening could be taken into account by multiplying the integrand by the Debye static enhancement factor
\begin{equation}\label{debye}
f_{\rm scr} = \exp\!\left(\frac{Z_{p} Z_{\rm Be} e^{2}}{\lambda_{D}k_{\rm{B}}T}\right),
\end{equation}
where $\lambda_{D}$ is the Debye radius at the electron density at temperature $T$. In addition, a moving proton within the plasma modifies the local electrostatic potential. While it does not alter the Debye length itself, which remains a global collective property of the plasma, it changes the dynamic screening. This can, in turn, modify the local density of positive and negative charges around the beryllium nucleus and thus indirectly affect the Debye length.
\begin{table*}
\caption{Proton-capture rates on $^7$Be during Big Bang nucleosynthesis.}
\label{tab:Be7_proton_capture}
\centering
\begin{tabular}{cccccc}
\hline\hline
Time after &
$k_{\rm B}T$ &
$n_{\rm p}$ &
$R_{\rm PC}$ &
$R_{\rm PC+SE}$ & $f_{\rm scr}$\\
Big Bang &
(keV) &
($a_0^{-3}$) &
(s$^{-1}$) &
(s$^{-1}$) & \\
\hline
$3.156\times10^{-1}$ & 275.854 & $5.316\times10^{-5}$ & $6.669\times10^{-2}$ & $7.085\times10^{-2}$ & 1.032 \\
$4.817\times10^{-1}$ & 223.276 & $2.819\times10^{-5}$ & $2.426\times10^{-2}$ & $2.541\times10^{-2}$ & 1.004 \\
$7.016\times10^{-1}$ & 185.016 & $1.604\times10^{-5}$ & $9.585\times10^{-3}$ & $9.930\times10^{-3}$ &  \\
$1.071\times10^{0}$  & 149.752 & $8.506\times10^{-6}$ & $3.251\times10^{-3}$ & $3.334\times10^{-3}$ & \\
$1.560\times10^{0}$  & 124.091 & $4.840\times10^{-6}$ & $1.201\times10^{-3}$ & $1.223\times10^{-3}$ & \\
$2.381\times10^{0}$  & 100.439 & $2.566\times10^{-6}$ & $3.759\times10^{-4}$ & $3.802\times10^{-4}$ & \\
$3.462\times10^{0}$  & 83.287  & $1.463\times10^{-6}$ & $1.291\times10^{-4}$ & $1.300\times10^{-4}$ & \\
$5.251\times10^{0}$  & 67.628  & $7.834\times10^{-7}$ & $3.751\times10^{-5}$ & $3.767\times10^{-5}$ & \\
$7.964\times10^{0}$  & 54.913  & $4.194\times10^{-7}$ & $1.034\times10^{-5}$ & $1.036\times10^{-5}$ & \\
$1.208\times10^{1}$  & 44.588  & $2.245\times10^{-7}$ & $2.690\times10^{-6}$ & $2.693\times10^{-6}$ & \\
$1.832\times10^{1}$  & 36.205  & $1.202\times10^{-7}$ & $6.585\times10^{-7}$ & $6.587\times10^{-7}$ & \\
$2.779\times10^{1}$  & 29.398  & $6.435\times10^{-8}$ & $1.509\times10^{-7}$ & $1.509\times10^{-7}$ & \\
$4.215\times10^{1}$  & 23.871  & $3.445\times10^{-8}$ & $3.222\times10^{-8}$ & $3.223\times10^{-8}$ & \\
$6.392\times10^{1}$  & 19.383  & $1.844\times10^{-8}$ & $6.379\times10^{-9}$ & $6.379\times10^{-9}$ & \\
$9.696\times10^{1}$  & 15.738  & $9.873\times10^{-9}$ & $1.164\times10^{-9}$ & $1.164\times10^{-9}$ & \\
$1.471\times10^{2}$  & 12.779  & $5.286\times10^{-9}$ & $1.948\times10^{-10}$ & $1.948\times10^{-10}$ & \\
$2.230\times10^{2}$  & 10.377  & $2.830\times10^{-9}$ & $2.968\times10^{-11}$ & $2.968\times10^{-11}$ & \\
$3.383\times10^{2}$  & 8.426   & $1.515\times10^{-9}$ & $4.087\times10^{-12}$ & $4.087\times10^{-12}$ & \\
$5.131\times10^{2}$  & 6.842   & $8.110\times10^{-10}$ & $5.071\times10^{-13}$ & $5.071\times10^{-13}$ & \\
$7.782\times10^{2}$  & 5.555   & $4.342\times10^{-10}$ & $5.587\times10^{-14}$ & $5.587\times10^{-14}$ & \\
$1.180\times10^{3}$  & 4.511   & $2.324\times10^{-10}$ & $5.452\times10^{-15}$ & $5.452\times10^{-15}$ & \\
$1.790\times10^{3}$  & 3.663   & $1.244\times10^{-10}$ & $4.663\times10^{-16}$ & $4.663\times10^{-16}$ & \\
$2.715\times10^{3}$  & 2.974   & $6.662\times10^{-11}$ & $3.456\times10^{-17}$ & $3.456\times10^{-17}$ & \\
$6.542\times10^{3}$  & 1.916   & $1.781\times10^{-11}$ & $8.533\times10^{-20}$ & $8.533\times10^{-20}$ & \\
$3.156\times10^{4}$  & 0.8723  & $1.681\times10^{-12}$ & $2.393\times10^{-25}$ & $2.393\times10^{-25}$ & \\
\hline
\end{tabular}
\tablefoot{Proton-capture rates on $^7$Be during Big Bang nucleosynthesis, averaged over the relative proton velocity according to Eq. (\ref{final_rate}). The first column shows the indicative time elapsed since the Big Bang, according to the time--temperature parametrization introduced above, while the second gives the corresponding plasma temperature. The third column lists the proton number density adopted in the calculations. The fourth column reports the one-body proton-capture rate, $R_{\mathrm{PC}}$, whereas the fifth includes the contribution from nuclear SE, $R_{\mathrm{PC+SE}}$. The complete table, containing all calculated temperatures, is available as Supporting Material in the online version of the paper. In the last column we report the Debye screening factor from Eq. (\ref{debye}).}
\end{table*}
The results of the PC rate are presented in Tab. (\ref{tab:Be7_proton_capture}) for various temperatures below and above NSE according to Eq. (\ref{final_rate}).
In the last column of Tab. (\ref{tab:Be7_proton_capture}), we also show the effect of adding Debye screening to the Coulomb potential barrier that the proton must overcome to be captured by beryllium. This effect is small and only slightly alters the barrier, so it can be safely neglected.
\subsection{Theory of proton capture with photon stimulated emission: Quantum approach}
The approach described above relies on a classical approximation of two-body scattering. However, the presence of a highly energetic photon background below NSE, further enhanced by the release of $\gamma$-radiation upon PC on $^7$Be, may generate SE from nuclear excited state levels, increasing the PC cross section. In principle, this mechanism is similar to laser SE for electrons, where the enhancement of the radiative transition rate is induced by the ambient photon field. The effect is described by a multiplicative factor $1+\langle n\rangle$, where $\langle n\rangle$ is given by the Bose–Einstein distribution. This enhancement becomes significant at low photon energies and high radiation densities. Here, we aim to include an analogous mechanism, but involving nuclear rather than electronic SE. \\
\indent In quantum field theory, each application of the photon creation operator introduces a prefactor when it acts on a single-mode Fock state \(|n\rangle\), which contains \(n\) photons. This square root factor, \(\sqrt{n+1}\), increases with the number of photons and can lead to SE. This effect has not been considered previously, and here we aim to estimate its impact on the PC rate.  \\
\indent
To account for this effect, we must extend the classical cross section in Eq. (\ref{final_rate}) and calculate the differential transition rate for a proton with relative momentum $\vec{p}$ scattered by a Coulomb potential, emitting a photon with momentum $\vec{k}_{\rm{ph}}$, polarisation $\hat{\vec{\epsilon}}$, and energy $c\vec{k}_{\rm{ph}}$ as follows:
\begin{eqnarray}\label{pro_fermi}
&&dP_{i\rightarrow f}
=2\pi\left|
\left\langle
\psi_{\alpha,\vec{p}}^{+};n,\hat{\vec{\epsilon}},\vec{k}_{\rm{ph}}\left|H_{\mathrm{int}}\right|\psi_{\beta};n+1,\hat{\vec{\epsilon}},\vec{k}_{\rm{ph}}
\right\rangle
\right|^{2}\times\nonumber \\
&&\delta\left(E_{\alpha}+\frac{p^{2}}{2\mu}-E_{\beta}-ck_{\rm{ph}}\right)
\frac{d^{3}\vec{k}_{\rm{ph}}}{(2\pi)^{3}},
\end{eqnarray}
where $|\psi_{\alpha,\vec{p}}^{+};n,\hat{\vec{\epsilon}},\vec{k}_{\rm{ph}}\rangle$ is the tensor product of the quantum state with $n$ photons, $|n,\hat{\vec{\epsilon}},\vec{k}_{\rm{ph}}\rangle$, and the Coulomb-distorted wave function $|\psi_{\alpha,\vec{p}}^{+}\rangle$ of the system formed by a proton with relative momentum $\vec{p}$ ($^{+}$ denotes outgoing wave boundary conditions) impinging on the $^7$Be atom in the quantum state denoted by $\alpha$, which can be specified, for example, by the quantum numbers ($n, l, m$); $\psi_{\beta}$ is the $^8$B wave function; $E_{\alpha}$ and $E_{\beta}$ are the total energies of the $^7$Be and $^8$B atoms in the states $\alpha$ and $\beta$, respectively; $H_{\mathrm{int}} = -i\,\vec{A} \cdot \vec{\nabla}/\mu$, where $\vec{A}$ is the photon annihilation operator (or the vector potential of the electromagnetic field) and $\vec{\nabla}$ is the gradient (or momentum operator). In this context, we note that
\begin{eqnarray}
&&(n+1)\left|
\left\langle
\psi_{\alpha,\vec{p}}^{+};0,\hat{\vec{\epsilon}},\vec{k}_{\rm{ph}}\left|H_{\mathrm{int}}\right|\psi_{\beta};1,\hat{\vec{\epsilon}},\vec{k}_{\rm{ph}}
\right\rangle
\right|^{2}=\nonumber \\
&&\left|
\left\langle
\psi_{\alpha,\vec{p}}^{+};n,\hat{\vec{\epsilon}},\vec{k}_{\rm{ph}}\left|H_{\mathrm{int}}\right|\psi_{\beta};n+1,\hat{\vec{\epsilon}},\vec{k}_{\rm{ph}}
\right\rangle
\right|^{2},
\end{eqnarray}
where the square of the matrix element on the left-hand side corresponds to what is measured in experiments.\\
\indent
To obtain the cross section without photon background ($n=0$ in Eq. (\ref{pro_fermi})), we integrate over all photon directions and polarisations, assuming the long-wavelength limit approximation; that is, the photon wavelength is much larger than the spatial extent of the final proton wave function, which is confined within the $^8$B potential well after capture. This reads per Be atom
\begin{eqnarray}\label{cs_01}
&&\sigma_{0}  =\frac{1}{\mu p}\int\frac{d^{3}\vec{k}_{\rm{ph}}}{\left(2\pi\right)^{3}}\sum_{\vec{\hat{{\epsilon}}}}2\pi\left|\hat{\epsilon}\cdot\left\langle \psi_{\alpha,\vec{p}}^{+}\left|\vec{\nabla}\right|\psi_{\beta}\right\rangle \right|^{2}\times\nonumber\\ && \delta\left(E_{\alpha}+\frac{p^{2}}{2\mu}-E_{\beta}-ck_{\rm{ph}}\right)\nonumber \\
&& =\frac{1}{\mu p}\int\frac{k^{2}_{\rm{ph}}dk_{\rm{ph}}d\Omega_{\vec{k}_{\rm{ph}}}}{\left(2\pi\right)^{3}}\sum_{\vec{\hat{{\epsilon}}}}2\pi\left|\hat{\epsilon}\cdot\left\langle \psi_{\alpha,\vec{p}}^{+}\left|\vec{\nabla}\right|\psi_{\beta}\right\rangle \right|^{2}\times\nonumber \\ && \delta\left(E_{\alpha}+\frac{p^{2}}{2\mu}-E_{\beta}-ck_{\rm{ph}}\right)\nonumber \\
&&=\frac{1}{\mu p }\int\frac{d\Omega_{\vec{k}_{\rm{ph}}}}{\left(2\pi\right)^{3}}\frac{\left(E_{\alpha}+\frac{p^{2}}{2\mu}-E_{\beta}\right)^{2}}{c^{3}}\sum_{\vec{\hat{\epsilon}}}2\pi\left|\vec{\hat{\epsilon}}\cdot\left\langle \psi_{\alpha,\vec{p}}^{+}\left|\vec{\nabla}\right|\psi_{\beta}\right\rangle \right|^{2}.\nonumber \\
\end{eqnarray}
Due to the orthogonality between the photon polarisation vector $\vec{\hat{\epsilon}}$ and the wave vector $\vec{k}_{\rm{ph}}$, the sum over polarisations can be written as
\begin{equation}\label{epsilon_eq}
\sum_{\vec{\hat{\epsilon}}}
\left|
\vec{\hat{\epsilon}}\cdot
\left\langle \psi_{\alpha,\vec{p}} \left| \vec{\nabla} \right| \psi_{\beta} \right\rangle
\right|^{2}
=
\left\|
\left\langle \psi_{\alpha,\vec{p}} \left| \vec{\nabla} \right| \psi_{\beta} \right\rangle
\right\|^{2}
-
\left|
\vec{\hat{k}}_{\rm{ph}}\cdot
\left\langle \psi_{\alpha,\vec{p}} \left| \vec{\nabla} \right| \psi_{\beta} \right\rangle
\right|^{2}.
\end{equation}
Defining $\theta$ as the angle between $\vec{\hat{k}}_{\rm{ph}}$ and $\left\langle \psi_{\alpha,\vec{p}}^{+}| \vec{\nabla}|\psi_{\beta} \right\rangle$, the expression (\ref{epsilon_eq}) reduces to
\begin{equation}
\sum_{\hat{\epsilon}}
\left|
\hat{\epsilon}\cdot
\left\langle \psi_{\alpha,\vec{p}}^{+} \left| \vec{\nabla} \right| \psi_{\beta} \right\rangle
\right|^{2}
=
\left\|
\left\langle \psi_{\alpha,\vec{p}}^{+} \left| \vec{\nabla} \right| \psi_{\beta} \right\rangle
\right\|^{2}
\left(1-\cos^{2}\theta\right).
\end{equation}
We therefore obtained from Eqs. (\ref{cs_01}) and (\ref{epsilon_eq})
\begin{eqnarray}\label{cs_02}
&&\sigma_{0}  =
\frac{1}{\mu p}
\int_{0}^{\pi}
\frac{2\pi \sin\theta\, d\theta}{(2\pi)^{3}}
\frac{\left(E_{\alpha}+\frac{p^{2}}{2\mu}-E_{\beta}\right)^{2}}{c^{3}} \times \nonumber \\
&&\,2\pi
\left\|
\left\langle \psi_{\alpha,\vec{p}}^{+} \left| \vec{\nabla} \right| \psi_{\beta} \right\rangle
\right\|^{2}
\left(1-\cos^{2}\theta\right)\nonumber
\\
&&=
\frac{1}{\mu p}
\frac{2}{3\pi}
\frac{\left(E_{\alpha}+\frac{p^{2}}{2\mu}-E_{\beta}\right)^{2}}{c^{3}}
\left\|
\left\langle \psi_{\alpha,\vec{p}}^{+} \left| \vec{\nabla} \right| \psi_{\beta} \right\rangle
\right\|^{2}.
\end{eqnarray}
Assuming that the squared matrix element can be factorized as a constant
coefficient multiplying the Sommerfeld factor, we write
\begin{equation}\label{assum}
\left(E_{\alpha}+\frac{p^{2}}{2\mu}-E_{\beta}\right)^{2}
\left\|
\left\langle \psi_{\alpha,\vec{p}}^{+} \left| \vec{\nabla} \right| \psi_{\beta} \right\rangle
\right\|^{2}
\simeq
K\,\frac{2\pi\eta}{e^{2\pi\eta}-1},
\end{equation}
where $\eta$ is the Sommerfeld parameter defined previously. This assumption is supported by experimental data \citep{PhysRevC.68.065803,PhysRevC.67.065805} and is based on the idea that the matrix element on the left is strongly confined within the $^8$B nuclear region due to the capture. Essentially, the proton wave function outside the nucleus before capture may alter the amplitude, but not the shape, of the proton wave function inside after capture, as the latter is mainly determined by the narrow and deep potential well of the nucleus, rather than by its long tail far from the scattering region. Additionally, the kinetic energy outside the nucleus is of the order of keV, while inside the nucleus it is of the order of MeV. The prefactor $K$ in Eq. (\ref{assum}) essentially contains information about the nucleus, while $\eta$ incorporates extra-nuclear information. This factorisation also appears in the phenomenological cross section in Eq. (\ref{cs_proton}).\\
\indent
The resulting cross section then reads
\begin{equation}\label{cs_03}
\sigma_0\equiv \sigma(\epsilon_{p})
=
\frac{2}{3\pi}
\frac{\mu Z_{p} Z_{\rm{Be}}}{\mu p^{2}}
\frac{2\pi}{c^{3}}
\frac{K}{e^{2\pi\eta}-1}=
\frac{S(0)}{\epsilon_{p}}
\frac{1}{e^{2\pi\eta}-1},
\end{equation}
with
\begin{equation}\label{kappa}
S(0)=\frac{2Z_{p} Z_{\rm{Be}}}{3\mu c^{3}}\,K, ~~\epsilon_{p} =\frac{p^2}{2\mu}.
\end{equation}
In atomic units, $S(0)\simeq 2.7822\times 10^{-8}$, giving
$K\simeq 1.1 \times 10^{-5}$.
Finally, the cross section in the presence of a background of $n$ photons can be expressed as
\begin{equation}
\sigma_n = (1 + \langle n \rangle)\sigma_0=\left(1+\frac{1}{e^{\epsilon_{\mathrm{ph}}/kT}-1}\right)\sigma_0,
\end{equation}
\begin{figure}[hbt!]
\centering
\includegraphics[width=0.75\columnwidth]{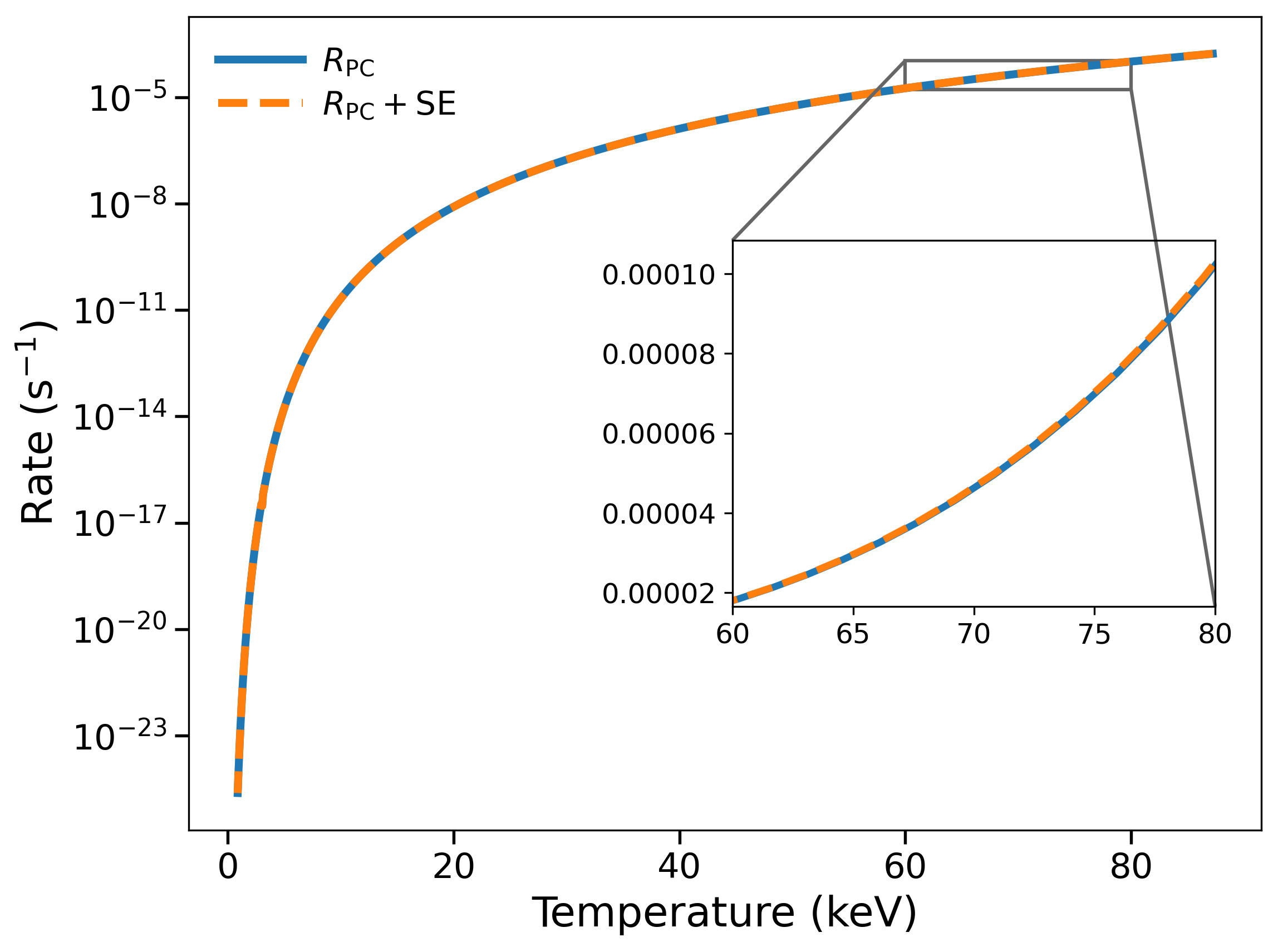}
\caption{
Proton capture rate on $^7$Be as a function of temperature.
The solid line shows $R_{\mathrm{PC}}$,
while the dashed line shows $R_{\mathrm{PC}}+\mathrm{SE}$,
where $\mathrm{SE}$ denotes SE.
The inset zooms into the $60$--$80$~keV range (box in the main panel).
}\label{fig:rate_cattura_protonica}
\end{figure}
where the average value $\langle n \rangle$ is calculated by assuming a statistical mixture of photons distributed according to Bose-Einstein statistics at an energy corresponding to the photon emission, which of course depends on the initial kinetic energy of the proton.
The results for the PC rate with SE are presented in the third column of Tab. (\ref{tab:Be7_proton_capture}) for various temperatures below and above NSE. As the temperature decreases, the effect of SE disappears due to the absence of the photon background at low temperatures. Under these conditions the rates with SE (third column in Tab. (\ref{tab:Be7_proton_capture})) and without SE (second column in Tab. (\ref{tab:Be7_proton_capture})) become similar. These data are also shown in Fig. (\ref{fig:rate_cattura_protonica}).
\subsection{Theory of proton capture with energy transfer to a continuum electron: Auger-like process}
Finally, we have studied a different PC process in which an excited \borootto~nucleus, after PC on $^7$Be, decays by emitting a virtual \(\gamma\)-photon that transfers energy to a nearby continuum electron (see Eq. (\ref{radiative_b})). This is a three-body process, similar to an Auger process, which typically occurs in atoms, molecules, and solids with an inner-shell vacancy that de-excite by ejecting valence electrons from bound levels with characteristic energies, appearing as peaks in the energy spectrum \citep{taioli2010electron, taioli2021relativistic, taioli2024advancements}.\\
\indent
Typically, non-radiative Auger decay in atoms competes with radiative decay under ambient conditions, that is, with photon emission, depending on factors such as atomic number and ionisation state. However, under BBN conditions, the \berilliosette~nucleus is usually fully ionised and surrounded by a plasma of free protons and electrons. Auger conversion under these conditions may differ from the bound-to-continuum emission process described above and may instead be a continuum-to-continuum mechanism, in which a virtual photon emitted by the excited nucleus scatters a continuum electron and changes its momentum. Our aim is to determine whether this non-radiative Auger conversion channel could compete with the radiative PC process studied in the previous section.\\
\indent
The differential transition rate for the Auger-like capture process, in atomic units, is
\begin{eqnarray}
&&dP_{i\rightarrow f}=2\pi\left|\left\langle \psi_{\alpha,\vec{p}}^{+};\varphi_{\vec{k}}\left|H_{int}\right|\psi_{\beta};\varphi_{\vec{k}^{\prime}}\right\rangle \right|^{2}\times \nonumber \\ &&\delta\left(E_{\alpha}+\frac{p^{2}}{2\mu}+c\sqrt{k^{2}+c^{2}}-E_{\beta}-c\sqrt{k^{\prime2}+c^{2}}\right)\frac{d^{3}\vec{k}}{\left(2\pi\right)^{3}}\frac{d^{3}\vec{k^{\prime}}}{\left(2\pi\right)^{3}},\nonumber \\
\end{eqnarray}
where $H_{\mathrm{int}} = 1/|\vec{r}_{p} - \vec{r}_{e}|$ is the Coulomb potential describing the interaction between an electron at position $\vec{r}_{e}$ and a proton at $\vec{r}_{p}$, connecting the initial state ($|\psi_{\alpha,\vec{p}}^{+};\varphi_{\vec{k}}\rangle$) and the final state ($|\psi_{\beta};\varphi_{\vec{k}^{\prime}}\rangle$), in which an electron $\varphi_{\vec{k}}$ with momentum $\vec{k}$ in the energy continuum is scattered out with momentum $\vec{k}'$ upon absorption of a virtual $\gamma$-photon produced by the nuclear transition from state $\alpha$ to state $\beta$ induced by capture of a proton with momentum $\vec{p}$.
This transition rate must be averaged over all initial electron states and integrated over all final states to obtain the corresponding cross section per Be atom as follows:
\begin{eqnarray}
&&\sigma  =\frac{\mu}{p}\int\int2\pi\frac{d^{3}\vec{k}}{\left(2\pi\right)^{3}}\frac{d^{3}\vec{k^{\prime}}}{\left(2\pi\right)^{3}}\left|\left\langle \psi_{\alpha,\vec{p}}^{+};\varphi_{\vec{k}}\left|H_{int}\right|\psi_{\beta};\varphi_{\vec{k}^{\prime}}\right\rangle \right|^{2}\times\nonumber \\
 && \delta\left(E_{\alpha}+\frac{p^{2}}{2\mu}+c\sqrt{k^{2}+c^{2}}-E_{\beta}-c\sqrt{k^{\prime2}+c^{2}}\right)\rho\left(\vec{k}\right),
\end{eqnarray}
where $\rho(\vec{k})$ is the electron density at momentum $\vec{k}$.
Approximating electron wave functions by plane waves near the $^7$Be nucleus ($\vec{r}_{e} \approx \vec{r}_{p}$)
\begin{eqnarray}
&&\sigma  =\frac{\mu}{p}\int\int2\pi\frac{d^{3}\vec{k}}{\left(2\pi\right)^{3}}\frac{d^{3}\vec{k^{\prime}}}{\left(2\pi\right)^{3}}\left|\left\langle \psi_{\alpha,\vec{p}}^{+}\left|e^{i\left(\vec{k}-\vec{k'}\right)\cdot\vec{r_{p}}}\right|\psi_{\beta}\right\rangle \right|^{2}\times\nonumber \\
 && \frac{\left(4\pi\right)^{2}\rho\left(\vec{k}\right)}{\left|\vec{k}^{\prime}-\vec{k}\right|^{4}}\delta\left(E_{\alpha}+\frac{p^{2}}{2\mu}+c\sqrt{k^{2}+c^{2}}-E_{\beta}-c\sqrt{k^{\prime2}+c^{2}}\right).\nonumber \\
\end{eqnarray}
Assuming orthogonality of $\psi_{\alpha,\vec{p}}^{+}$ and $\psi_{\beta}$, and expanding the exponential for small arguments within the nuclear volume ($|e^{i(\vec{k}-\vec{k'})\cdot r_p}| \sim |(\vec{k}-\vec{k'})\cdot r_p|$), we obtain
\begin{eqnarray}
&&\sigma \simeq\frac{\mu}{p}\int\int2\pi\frac{d^{3}\vec{k}}{\left(2\pi\right)^{3}}\frac{d^{3}\vec{k^{\prime}}}{\left(2\pi\right)^{3}}\left|\left(\vec{k}-\vec{k'}\right)\cdot\left\langle \psi_{\alpha,\vec{p}}^{+}\left|\vec{r_{p}}\right|\psi_{\beta}\right\rangle \right|^{2}\times\nonumber \\
 && \frac{\left(4\pi\right)^{2}\rho\left(\vec{k}\right)}{\left|\vec{k}^{\prime}-\vec{k}\right|^{4}}\delta\left(E_{\alpha}+\frac{p^{2}}{2\mu}+c\sqrt{k^{2}+c^{2}}-E_{\beta}-c\sqrt{k^{\prime2}+c^{2}}\right).\nonumber \\
\end{eqnarray}
Integrating over $dk^{\prime}$, we obtain
\begin{equation}
\bar{k}^{2} = \left( \frac{E_{\alpha} - E_{\beta} + \frac{p^{2}}{2\mu}}{c} + \sqrt{k^{2} + c^{2}} \right)^{2} - c^{2},
\end{equation}
which leads to
\begin{eqnarray}\label{cs_prot2}
&&\sigma\simeq\frac{\mu}{p}\int\int\frac{2\pi\left(4\pi\right)^{2}}{\left(2\pi\right)^{6}}\times \nonumber \\ &&\frac{\left|\left(k\vec{\hat{k}}-\bar{k}\vec{\hat{k}^{\prime}}\right)\cdot\left\langle \psi_{\alpha,\vec{p}}^{+}\left|\vec{r_{p}}\right|\psi_{\beta}\right\rangle \right|^{2}}{\left|\vec{k}^{\prime}-\vec{k}\right|^{4}}\rho\left(\vec{k}\right)k^{2}\bar{k}^{2}dkd\Omega_{\vec{\hat{k}}}d\Omega_{\vec{\hat{k}^{\prime}}}.
\end{eqnarray}
We further assume that $\rho\left(\vec{k}\right) = \rho\left(k\right)$ and integrate over the differential solid angle \(d\Omega_{\hat{\vec{k'}}} = \sin \theta' \, d\theta' \, d\phi'\), where the unit vector $\vec{\hat{k}}^{\prime}$ spans the unit sphere and points in the direction specified by the angular coordinates ($\theta'$, $\phi'$). The $z$-axis is aligned with the direction of $\vec{\hat{k}}$, and the $xz$-plane contains both $\vec{\hat{k}}$ and $\left\langle \psi_{\alpha,\vec{p}}^{+} \left| \vec{r}_{p} \right| \psi_{\beta} \right\rangle$.
If $\theta$ is the angle between $\vec{\hat{k}}$ and $\left\langle \psi_{\alpha,\vec{p}}^{+}\left|\vec{r}_{p}\right|\psi_{\beta}\right\rangle$, and $\theta^{\prime}$ is the angle between $\vec{\hat{k}}$ and $\vec{\hat{k}}^{\prime}$, then $\vec{\hat{k}}\cdot\left\langle \psi_{\alpha,\vec{p}}^{+}\left|\vec{r}_{p}\right|\psi_{\beta}\right\rangle =\left\Vert \left\langle \psi_{\alpha,\vec{p}}^{+}\left|\vec{r}_{p}\right|\psi_{\beta}\right\rangle \right\Vert \cos\theta$ and $\vec{\hat{k}}\cdot\vec{\hat{k}^{\prime}}=\cos\theta^{\prime}$ while $\vec{\hat{k}^{\prime}}\cdot\left\langle \psi_{\alpha,\vec{p}}^{+}\left|\vec{r_{p}}\right|\psi_{\beta}\right\rangle =\left\Vert \left\langle \psi_{\alpha,\vec{p}}^{+}\left|\vec{r_{p}}\right|\psi_{\beta}\right\rangle \right\Vert \left(\sin\theta\sin\theta^{\prime}\cos\phi^{\prime}+\cos\theta\cos\theta^{\prime}\right)$.
Then Eq. (\ref{cs_prot2}) reads:
\begin{eqnarray}
&\sigma \simeq\frac{\mu}{p}\int_{0}^{\infty}dk\int_{0}^{\pi}\sin\theta d\theta\int_{0}^{2\pi}d\phi\int_{0}^{\pi}\sin\theta^{\prime}d\theta^{\prime}\int_{0}^{2\pi}d\phi^{\prime}\frac{2\pi\left(4\pi\right)^{2}}{\left(2\pi\right)^{6}}\times\nonumber \\
&\frac{\left|k\cos\theta-\bar{k}\left(\sin\theta\sin\theta^{\prime}\cos\phi^{\prime}+\cos\theta\cos\theta^{\prime}\right)\right|^{2}}{\left|k^{2}+\bar{k}^{2}-2k\bar{k}\cos\theta^{\prime}\right|^{2}}\times \nonumber \\
& \left\Vert \left\langle \psi_{\alpha,\vec{p}}^{+}\left|\vec{r_{p}}\right|\psi_{\beta}\right\rangle \right\Vert ^{2}\rho\left(k\right)k^{2}\bar{k}^{2}dkd\Omega_{\vec{\hat{k}}}
,\nonumber\\
\end{eqnarray}
which, when integrated over $d\phi^{\prime}$, leads to
\begin{eqnarray}
&\sigma \simeq\frac{\mu}{p}\int_{0}^{\infty}dk\int_{0}^{\pi}\sin\theta d\theta\int_{0}^{2\pi}d\phi\int_{0}^{\pi}\sin\theta^{\prime}d\theta^{\prime}\frac{64\pi^{4}}{\left(2\pi\right)^{6}}\times\nonumber \\
&\frac{\left|k^{2}\cos^{2}\theta-2k\bar{k}\cos^{2}\theta\cos\theta^{\prime}+\frac{1}{2}\bar{k}^{2}\sin^{2}\theta\sin^{2}\theta^{\prime}+\bar{k}^{2}\cos^{2}\theta\cos^{2}\theta^{\prime}\right|}{\left|k^{2}+\bar{k}^{2}-2k\bar{k}\cos\theta^{\prime}\right|^{2}}\times\nonumber \\
& \left\Vert \left\langle \psi_{\alpha,\vec{p}}^{+}\left|\vec{r_{p}}\right|\psi_{\beta}\right\rangle \right\Vert ^{2}\rho\left(k\right)k^{2}\bar{k}^{2}dkd\Omega_{\vec{\hat{k}}}.
\end{eqnarray}
Let us integrate over $\vec{\hat{k}}$ (or over the variables $\phi$ and $\theta$). Observing that the integrand does not depend on $\phi$, we obtain
\begin{eqnarray}
&\sigma \simeq\frac{\mu}{p}\int_{0}^{\infty}dk\int_{0}^{\pi}\frac{128\pi^{5}}{\left(2\pi\right)^{6}}\sin\theta^{\prime}d\theta^{\prime}\times \nonumber\\ &\int_{0}^{\pi}\frac{\left|k^{2}\cos^{2}\theta-2k\bar{k}\cos^{2}\theta\cos\theta^{\prime}+\frac{1}{2}\bar{k}^{2}\sin^{2}\theta\sin^{2}\theta^{\prime}+\bar{k}^{2}\cos^{2}\theta\cos^{2}\theta^{\prime}\right|}{\left|k^{2}+\bar{k}^{2}-2k\bar{k}\cos\theta^{\prime}\right|^{2}}\sin\theta d\theta\times\nonumber\\
&\left\Vert \left\langle \psi_{\alpha,\vec{p}}^{+}\left|\vec{r_{p}}\right|\psi_{\beta}\right\rangle \right\Vert ^{2}\rho\left(k\right)k^{2}\bar{k}^{2}\nonumber\\
 & =\frac{\mu}{p}\int_{0}^{\infty}dk\int_{0}^{\pi}\frac{128\pi^{5}}{\left(2\pi\right)^{6}}\sin\theta^{\prime}d\theta^{\prime}\left\Vert \left\langle \psi_{\alpha,\vec{p}}^{+}\left|\vec{r_{p}}\right|\psi_{\beta}\right\rangle \right\Vert ^{2}\times\nonumber \\ &\int_{0}^{\pi}\frac{\left|\frac{2}{3}k^{2}-\frac{4}{3}k\bar{k}\cos\theta^{\prime}+\frac{2}{3}\bar{k}^{2}\sin^{2}\theta^{\prime}+\frac{2}{3}\bar{k}^{2}\cos^{2}\theta^{\prime}\right|}{\left|k^{2}+\bar{k}^{2}-2k\bar{k}\cos\theta^{\prime}\right|^{2}}\rho\left(k\right)k^{2}\bar{k}^{2}\nonumber \\
 & =\frac{\mu}{p}\int_{0}^{\infty}dk\int_{0}^{\pi}\frac{4}{3\pi}\frac{\left\Vert \left\langle \psi_{\alpha,\vec{p}}^{+}\left|\vec{r_{p}}\right|\psi_{\beta}\right\rangle \right\Vert ^{2}}{\left|k^{2}+\bar{k}^{2}-2k\bar{k}\cos\theta^{\prime}\right|}\sin\theta^{\prime}d\theta^{\prime}\rho\left(k\right)k^{2}\bar{k}^{2}\nonumber \\ & =\frac{\mu}{p}\int_{0}^{\infty}\frac{4\left\Vert \left\langle \psi_{\alpha,\vec{p}}^{+}\left|\vec{r_{p}}\right|\psi_{\beta}\right\rangle \right\Vert ^{2}}{3\pi}\log\left(\frac{\left|k^{2}+\bar{k}^{2}+2k\bar{k}\right|}{\left|k^{2}+\bar{k}^{2}-2k\bar{k}\right|}\right)\rho\left(k\right)\frac{k^{2}\bar{k}^{2}}{2k\bar{k}}.
\nonumber \\
\end{eqnarray}
Finally, using Eq. (\ref{assum}) we obtain
\begin{eqnarray}\label{kappa_exit}
&&\sigma\simeq\frac{1}{\mu p}\frac{2K}{3\pi\left(E_{\alpha}+\frac{p^{2}}{2m}-E_{\beta}\right)^{4}}\frac{2\pi\eta}{e^{2\pi\eta}-1}\times\nonumber\\ &&\int_{0}^{\infty}dk\log\left(\frac{\left|k^{2}+\bar{k}^{2}+2k\bar{k}\right|}{\left|k^{2}+\bar{k}^{2}-2k\bar{k}\right|}\right)\rho\left(k\right)k\bar{k},
\end{eqnarray}
where the electron density is $\rho\left(k\right)=\frac{2}{e^{\frac{c\sqrt{k^{2}+c^{2}}-\mu}{k_{\rm{B}}T}}+1}$, and one can use $K$ from Eq. (\ref{kappa}).
We emphasise that in Eq. (\ref{kappa_exit}), a different initial proton kinetic energy ($p^{2}/2m$) results in a different virtual photon energy and, consequently, a different momentum of the scattered electron in the continuum, $\overline{k}$, which leads to a different available phase space.
\begin{table*}[hbt!]
\caption{Proton-capture cross sections on $^{7}$Be.}
\label{tab:s_auger}
\centering
\setlength{\tabcolsep}{4pt}
\begin{tabular}{c c c c c c c c c}
\hline\hline
\multicolumn{3}{c|}{$k_{\rm{B}}T=$ 10 (keV)} & \multicolumn{2}{|c|}{30 (keV)} & \multicolumn{2}{|c|}{60 (keV)} & \multicolumn{2}{|c}{100 (keV)} \\
\hline
$E_p$ (keV) &
$\sigma_{0}$ &
$\sigma_{0}^{\mathrm{AU}}$ &
$\sigma_{0}$ &
$\sigma_{0}^{\mathrm{AU}}$ &
$\sigma_{0}$ &
$\sigma_{0}^{\mathrm{AU}}$ &
$\sigma_{0}$ &
$\sigma_{0}^{\mathrm{AU}}$ \\
\hline
2   & $1.915\times10^{-8}$ & $8.485\times10^{-18}$ &
      $1.915\times10^{-8}$ & $9.073\times10^{-18}$ &
      $1.915\times10^{-8}$ & $2.069\times10^{-14}$ &
      $1.915\times10^{-8}$ & $1.621\times10^{-12}$ \\
20  & $1.644\times10^{-5}$ & $4.480\times10^{-15}$ &
      $1.644\times10^{-5}$ & $4.760\times10^{-15}$ &
      $1.644\times10^{-5}$ & $1.081\times10^{-11}$ &
      $1.644\times10^{-5}$ & $8.429\times10^{-10}$ \\
40  & $2.912\times10^{-5}$ & $4.900\times10^{-15}$ &
      $2.912\times10^{-5}$ & $5.181\times10^{-15}$ &
      $2.912\times10^{-5}$ & $1.171\times10^{-11}$ &
      $2.912\times10^{-5}$ & $9.101\times10^{-10}$ \\
80  & $3.724\times10^{-5}$ & $2.781\times10^{-15}$ &
      $3.724\times10^{-5}$ & $2.940\times10^{-15}$ &
      $3.724\times10^{-5}$ & $6.553\times10^{-12}$ &
      $3.724\times10^{-5}$ & $5.069\times10^{-10}$ \\
100 & $3.864\times10^{-5}$ & $2.022\times10^{-15}$ &
      $3.864\times10^{-5}$ & $2.123\times10^{-15}$ &
      $3.864\times10^{-5}$ & $4.732\times10^{-12}$ &
      $3.864\times10^{-5}$ & $3.640\times10^{-10}$ \\
140 & $3.920\times10^{-5}$ & $1.101\times10^{-15}$ &
      $3.920\times10^{-5}$ & $1.151\times10^{-15}$ &
      $3.920\times10^{-5}$ & $2.557\times10^{-12}$ &
      $3.920\times10^{-5}$ & $1.957\times10^{-10}$ \\
160 & $3.920\times10^{-5}$ & $8.289\times10^{-16}$ &
      $3.920\times10^{-5}$ & $8.653\times10^{-16}$ &
      $3.920\times10^{-5}$ & $1.921\times10^{-12}$ &
      $3.920\times10^{-5}$ & $1.467\times10^{-10}$ \\
180 & $3.892\times10^{-5}$ & $6.357\times10^{-16}$ &
      $3.892\times10^{-5}$ & $6.609\times10^{-16}$ &
      $3.892\times10^{-5}$ & $1.465\times10^{-12}$ &
      $3.892\times10^{-5}$ & $1.117\times10^{-10}$ \\
200 & $3.864\times10^{-5}$ & $4.929\times10^{-16}$ &
      $3.864\times10^{-5}$ & $5.125\times10^{-16}$ &
      $3.864\times10^{-5}$ & $1.131\times10^{-12}$ &
      $3.864\times10^{-5}$ & $8.625\times10^{-11}$ \\
\hline
\end{tabular}
\tablefoot{Proton-capture cross sections on $^{7}$Be as a function of the incident proton kinetic energy $E_p$. The quantities $\sigma_0$ and $\sigma_0^{\rm AU}$ denote the cross sections computed without and with the Auger-like mechanism, respectively. All cross sections are expressed in barns (b).}
\end{table*}
Using Eq.~(\ref{kappa_exit}) for proton kinetic energies of 10, 30, 60, and 100 keV, and a plasma temperature in the range $1 < k_{\rm{B}}T < 100$ keV to determine the electron density at the nucleus, the resulting cross sections as a function of incident proton energy are reported in Tab.~(\ref{tab:s_auger}).
\begin{figure}[hbt!]
\centering
\includegraphics[width=0.9\columnwidth]{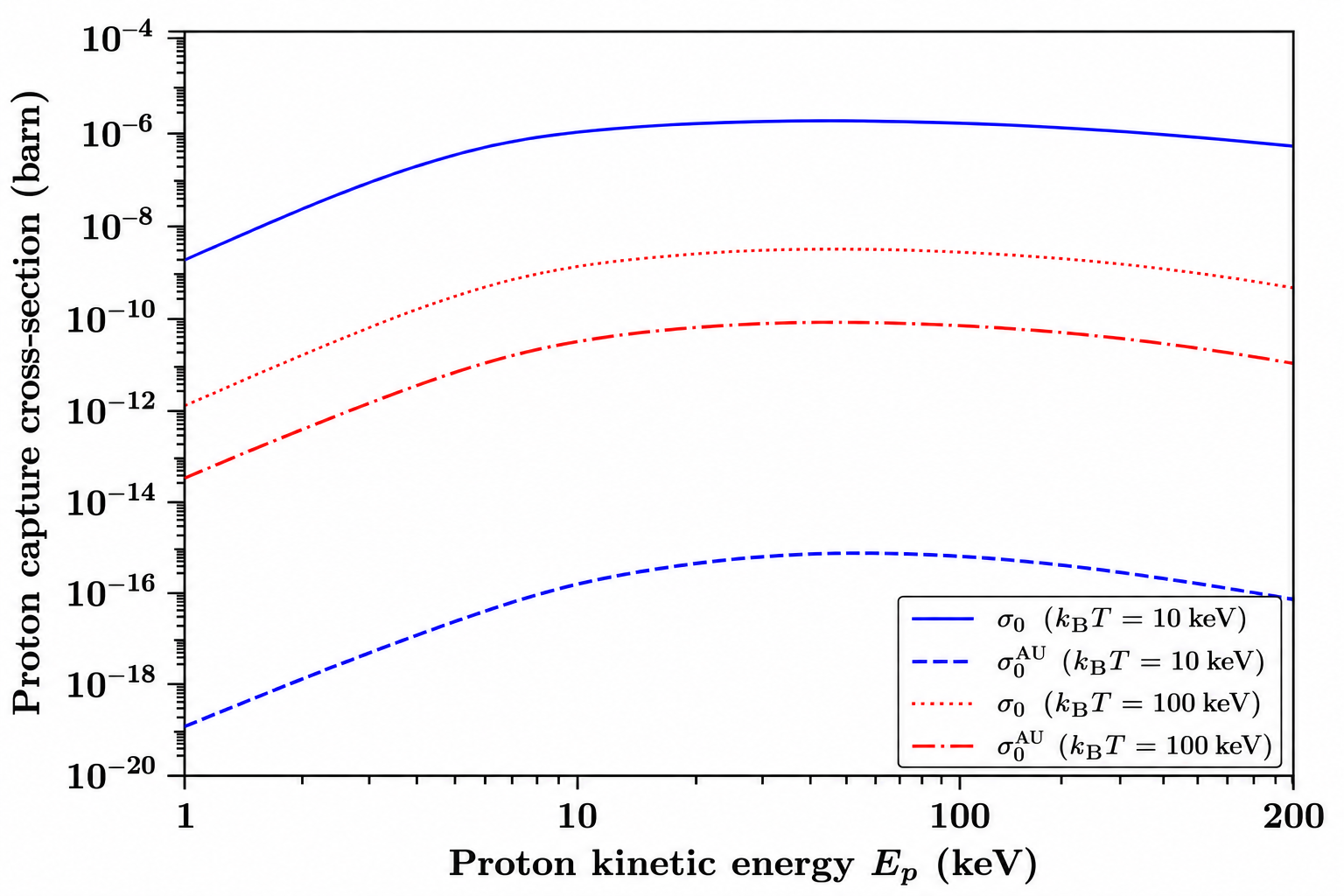}
\caption{
Proton capture cross sections on $^7$Be as a function of proton kinetic energy, $E_p$. The solid and dashed curves show the cross sections without ($\sigma_0$) and with ($\sigma_0^{\mathrm{AU}}$) the Auger mechanism at $k_{\rm{B}}T = 10\,\mathrm{keV}$, respectively. The dotted and dash-dotted curves show the corresponding cross sections at $k_{\rm{B}}T = 100\,\mathrm{keV}$. Cross sections are given in barn.
}
\label{fig:sigma_auger}
\end{figure}
These results are also shown in Fig.~(\ref{fig:sigma_auger}).\\
\indent
In particular, we note that at a plasma temperature of $k_{\rm{B}}T = 100~$keV ($n_e = 240\,\mathrm{a_0^{-3}}$), for a 2 keV kinetic energy of the protonic projectile, the contribution of the three-body Auger-type transition is approximately $10^{-4}$\% of the radiative PC cross section. In contrast, at $k_{\rm{B}}T = 10~$keV ($n_e = 2.64 \times 10^{-9}\,\mathrm{a_0^{-3}}$), for a proton with 80 keV kinetic energy, the Auger contribution is only about $\approx 10^{-10}$\% of the radiative PC cross section. The Auger-like mechanism does not significantly contribute to the total PC cross section at temperatures much below NSE, as the electron density near the nucleus is typically low, while its relative importance increases at higher temperatures and proton kinetic energy.

\section{Numerical solution of the radial Dirac equation}
\label{app:dirac_numerics}
The time-independent Dirac equation (\ref{matrix_solve_radial_2}) in radial coordinates, due to the spherically symmetric internuclear and interelectronic interaction potentials, can be written as
\begin{equation}\label{param_Dirac}
\frac{d\psi}{dr}=F_E(\psi,r),
\end{equation}
where $\psi=(u,d)^{T}$ is the four-component Dirac spinor and $F_E$ is a linear operator in both $\psi$ and the eigenvalue $E$. Equation (\ref{param_Dirac}) is a first order differential equation, which we solve on a real-space grid of points.
Although the DHF equations include a non-local Fock term, they remain linear in $\psi$.
Furthermore, in our method, we use an LDA approximation for the many-electron interaction (see Eq. (\ref{hartree+fock_pot})).\\
\indent
To solve the differential Eq. (\ref{param_Dirac}) we employ a collocation method, a variant of the
Runge--Kutta approach \citep{morresi2018relativistic}.
The radial interval $[0,R_{\max}]$ (where $R_{\max}$ depends on whether the orbital is bound or continuum) is divided into grid points
$0=r_0<\cdots<r_N=R_{\max}$, chosen by Gauss--Legendre sampling so that the same
weights can be used for nuclear-volume integrals.\\
\indent
Bound states are computed with $R_{\max}\simeq 25$~a.u., whereas continuum states
require $R_{\max}\simeq 100$~a.u.\\
\indent Each subinterval $[r_{j-1},r_j]$ contains $s_j$ collocation points
$x_{j,l}$, where the derivative appearing in Eq. (\ref{param_Dirac}) is approximated as
\begin{equation}
\frac{d\psi}{dr}\approx \sum_{l=1}^{s_j} q_{j,l}\,\phi'_{j,l}(r),
\end{equation}
with $\phi_{j,l}$ chosen such that $\phi'_{j,l}(x_{j,m})=\delta_{lm}$.
Defining
\begin{equation}
a_{j,lm}=\!\int_{r_{j-1}}^{x_{j,l}}\phi'_{j,m}(x)\,dx,
\qquad
b_{j,l}= \!\int_{r_{j-1}}^{r_j}\phi'_{j,l}(x)\,dx,
\end{equation}
the Dirac equation reduces to the block-triangular eigenvalue problem \citep{morresi2018relativistic}
\begin{equation}
\label{eq:triangular}
\big[c\,\gamma + (V+\gamma_0 mc^{2})A - E A \big] q = 0,
\end{equation}
where $q$ collects all coefficients $q_{j,l}$, $A$ is the sparse matrix
containing the $a_{j,lm}$ and $b_{j,l}$, and $\gamma$ is 2$\times$2 matrix.
A unitary transformation \(A \rightarrow AQ\) further reduces the bandwidth, enabling efficient eigenvalue extraction using an \(LDU\) decomposition.
The wave function follows from $\psi(r)=Aq$, and for continuum solutions the
normalisation is fixed at the outer boundary using the asymptotic Coulomb
functions,
\begin{equation}
u \sim \frac{\sqrt{\epsilon+c^{2}}}{\sqrt{\pi p}}\cos(pr+\delta),\qquad
v \sim -\frac{\sqrt{\epsilon-c^{2}}}{\sqrt{\pi p}}\sin(pr+\delta),
\end{equation}
with $\epsilon = \sqrt{p^{2}c^{2} + c^{4}}$ and $\delta$ as the phase shift due to the atomic Coulomb potential \citep{morresi2018relativistic}.
Because the normalisation oscillates with $R_{\max}$, we average it over one
full oscillation period of the continuum wave function.
More details on computational methods are given in Ref. \citep{morresi2018relativistic}.
\section{Particle densities and their chemical potentials}
\label{lep_bar_neu}
To calculate EC and PC rates, including SE, we need to determine the densities of the particles involved in the reactions. Below, we derive the analytical relation for calculating the densities of photons, electrons, positrons, neutrinos, neutrons, protons, and their respective antiparticles.
\subsection{Photon density}\label{photden}
Since the photon mass is \(0\), its energy \(E\) and momentum \(p\) are related by the dispersion relation \(E = pc\). The number of photons in an infinitesimal volume of phase space \(d^{3}q\,d^{3}p\), where \(q\) is the position, is
\begin{equation}
\label{infinitesimal_phase_space_be}
dn_{\gamma}=\frac{d^{3}qd^{3}p}{h^{3}}\frac{g}{e^{\beta E}-1},
\end{equation}
where \(\beta = \frac{1}{k_{\rm{B}}T_{\gamma}}\), \(h\) is the Planck constant, and \(g\) is the degeneracy number for photons. Photons are spin-1 particles and are characterised by two independent polarisations; therefore, for photons, \(g = 2\). The factor \(\frac{1}{e^{\beta E} - 1}\) ensures that photons are distributed according to Bose-Einstein statistics. If we divide expression (\ref{infinitesimal_phase_space_be}) by \(d^{3}q\), we obtain
\begin{equation}
\label{density_be}
dn_{\gamma}(p)=\frac{d^{3}p}{h^{3}}\frac{g}{e^{\beta E}-1}=\frac{4\pi p^{2}dp}{8\pi^{3}\hbar^{3}}\frac{g}{e^{\beta E}-1}=\frac{g}{2\pi^{2}\hbar^{3}}\frac{p^{2}dp}{e^{\beta E}-1},
\end{equation}
which is the infinitesimal volumetric variation of the photon density for momentum ranging from \(p\) to \(p+dp\) and $\hbar=h/2\pi$ is the reduced Planck constant. From \(E=pc\), it follows that \(dE=c\,dp\), and
\begin{equation}
\label{infinitesimal_photon_density}
dn_{\gamma}(E)=\frac{g}{2\pi^{2}\hbar^{3}c^{3}}k_{\rm{B}}^{3}T_{\gamma}^{3}\frac{\beta^{3}E^{2}dE}{e^{\beta E}-1}.
\end{equation}
The infinitesimal variation of the photon energy per unit volume is
\begin{equation}
\label{infinitesimal_photon_energy_density}
du_{\gamma}=Edn_{\gamma}(E)=\frac{g}{2\pi^{2}\hbar^{3}c^{3}}k_{\rm{B}}^{4}T_{\gamma}^{4}\frac{\beta^{4}E^{3}dE}{e^{\beta E}-1}.
\end{equation}
The total photon density and energy density are obtained by integrating the Eqs. (\ref{infinitesimal_photon_density}) and (\ref{infinitesimal_photon_energy_density}):
\begin{equation}
\label{total_photon_density}
n_{\gamma}(T)=\frac{g}{2\pi^{2}\hbar^{3}c^{3}}k_{\rm{B}}^{3}T_{\gamma}^{3}\int\frac{\beta^{3}E^{2}dE}{e^{\beta E}-1}=\frac{g}{2\pi^{2}\hbar^{3}c^{3}}k_{\rm{B}}^{3}T_{\gamma}^{3}2.404,
\end{equation}
\begin{equation}
\label{total_photon_energy_density}
u_{\gamma}(T_{\gamma})=\frac{g}{2\pi^{2}\hbar^{3}c^{3}}k_{\rm{B}}^{4}T_{\gamma}^{4}\int\frac{\beta^{4}E^{3}dE}{e^{\beta E}-1}=\frac{g}{2\pi^{2}\hbar^{3}}k_{\rm{B}}^{4}T_{\gamma}^{4}\frac{\pi^{4}}{15}.
\end{equation}
We note from Eq. (\ref{infinitesimal_photon_energy_density}) that we can define the energy density function for photons
\begin{equation}
\label{function_density_energy_photons}
\mathcal{U}(E)=\frac{du_{\gamma}}{dE}=\frac{g}{2\pi^{2}\hbar^{3}c^{3}}\frac{E^{3}}{e^{\beta E}-1},
\end{equation}
which states that the radiation of the early Universe has a black-body spectrum. As the Universe expanded and cooled, this radiation can still be detected and is now observed as the CMB radiation. The current spectrum corresponds to a temperature of \(2.725\) K \citep{Plank_coll}. Integration of Eq. (\ref{infinitesimal_photon_density}) for \(T_{\gamma} = 2.725\) K gives \(n_{\gamma} = 4.09649 \times 10^8\;\mathrm{photons}\cdot\mathrm{m}^{-3}\).
\subsection{Baryon densities}\label{baden}
\paragraph{Neutrons and protons} To determine the total rate of $^7$Be-involving electroweak and strong nuclear reactions, it is necessary to compute the population factors for electrons, baryons, and neutrinos after NSE freeze-out in the 10–100 keV temperature range. \\
\indent
In this context, at $t \sim 10^{-4}~\mathrm{s}$ after the Big Bang (or $k_{\rm{B}}T \sim 100~\mathrm{MeV}$, where $k_{\rm{B}}$ is the Boltzmann constant), as the temperature fell below the rest masses of the heavier baryons, their abundances became exponentially suppressed and they efficiently annihilated with their antiparticles. Thus, at NSE freeze-out, only protons, neutrons, and their respective antiparticles essentially remained \citep{Weinberg2008, Mukhanov2005}.
Taking into account the large matter–antimatter asymmetry, the net baryon density is
\begin{equation}
n_{b} = n_{\hat{b}} - n_{\overline{b}} = n_{p} + n_{n}\,,
\end{equation}
with \(n_{\hat{b}} = n_{\hat{p}} + n_{\hat{n}}\), \(n_{\overline{b}} = n_{\overline{p}} + n_{\overline{n}}\) and we define the net proton and neutron densities as follows:
\begin{equation}
n_{p} = n_{\hat{p}} - n_{\overline{p}}\,, \qquad
n_{n} = n_{\hat{n}} - n_{\overline{n}}\,,
\end{equation}
where \(n_{\hat{p}}\) and \(n_{\hat{n}}\) are the proton and neutron number densities, while \(n_{\overline{p}}\) and \(n_{\overline{n}}\) are those of their respective antiparticles.
After decoupling from radiation down to weak freeze-out, weak processes determine the equilibrium abundance of different isospin states of nucleons through EC and AC on protons to form neutrons, free neutron decay to protons, and symmetric reactions involving their antiparticles.
Denoting the neutron and proton masses by \(m_{n}\) and \(m_{p}\), the $Q$-value of the reaction
\(
Q = (m_{n}-m_{p})c^{2} = 1.293~\mathrm{MeV}
\)
is comparable to the temperature at the NSE epoch ($k_{\rm{B}}T \approx 700\,\mathrm{keV}$).
Boltzmann statistics then give
\begin{equation}
\label{weak_freeze_out_nucleons}
\frac{n_{\hat{n}}}{n_{\hat{p}}}
=\frac{\mathrm{e}^{-m_{n}c^{2}/(k_{\rm{B}}T)}}{\mathrm{e}^{-m_{p}c^{2}/(k_{\rm{B}}T)}}
=\mathrm{e}^{-Q/(k_{\rm{B}}T)} \, .
\end{equation}
At \(k_{\rm{B}}T \approx 700\) keV, after the isospin flip ceases and only the excess baryons remain, protons and neutrons are classical, non-relativistic particles that obey Boltzmann statistics. Very few neutrons had undergone \(\beta^{-}\) decay into protons (half-life \(\tau \approx 880.5\) s), as only \( t \lessapprox 3.5 \) seconds had elapsed since the Big Bang. The neutron ($n_{\hat{n}}$) to proton ($n_{\hat{p}}$) number density ratio at $t\lessapprox 3.5$ s can be determined as follows:
\begin{equation}
\label{ratio_init_prot_neut}
\frac{n_{\hat{n}}(t\approx 0)}{n_{\hat{p}}(t\approx 0)}=\frac{e^{-\frac{m_{n}c^{2}}{k_{\rm{B}}T}}}{e^{-\frac{m_{p}c^{2}}{k_{\rm{B}}T}}}=e^{-\frac{Q}{k_{\rm{B}}T}}=0.20.
\end{equation}
Between \(k_{\rm{B}}T = 700\) keV and NSE freeze-out (\(k_{\rm{B}}T = 100\) keV, approximately 300 s after the Big Bang), free neutrons can undergo \(\beta^{-}\) decay into protons, following the exponential law \(n_{\hat{n}}(t) = n_{\hat{n}}(0)e^{-\frac{t}{\tau}}\), where \(n_{\hat{n}}(0)\) is the initial number density of neutrons immediately after the Big Bang. On the other hand, the total number of nucleons \(n_{\rm{tot}}\) is conserved:
\begin{equation}
\label{decay_intermed_1}
n_{\rm{tot}} = n_{\hat{p}}(0)+n_{\hat{n}}(0)=\frac{n_{\hat{n}}(0)}{0.20}+n_{\hat{n}}(0)
=\frac{1.20}{0.20}n_{\hat{n}}(0)=6n_{\hat{n}}(0),
\end{equation}
\begin{equation}
n_{\hat{p}}(t)=n_{\rm{tot}}-n_{\hat{n}}(t)=6n_{\hat{n}}(0)-n_{\hat{n}}(0)e^{-\frac{t}{\tau}}.
\end{equation}
Therefore after \(\approx 300\) s the ratio is \citep{perkins}
\begin{equation}
\label{ratio_neutron_proton}
\frac{n_{\hat{n}}(300 ~\mathrm{s})}{n_{\hat{p}}(300 ~\mathrm{s})}=\frac{n_{\hat{n}}(0)e^{-\frac{300}{\tau}}}{6n_{\hat{n}}(0)-n_{\hat{n}}(0)e^{-\frac{300}{\tau}}}\approx 0.135.
\end{equation}
At NSE, \(\beta^{-}\) decay almost completely stops because nearly all neutrons are bound in \(\alpha\) particles.
After weak freeze-out, the subsequent decrease in \(n_{\hat{n}}/n_{\hat{p}}\) is governed by free neutron \(\beta^{-}\) decay (half-life $\tau \simeq 880.5$ s), following the exponential law derived in Eq.~(\ref{ratio_neutron_proton}):
\begin{equation}
\label{beta_exponential}
\frac{n_{\hat{n}}(t)}{n_{\hat{p}}(t)}
=\frac{\mathrm{e}^{-t/\tau}}{6-\mathrm{e}^{-t/\tau}}\, ,
\end{equation}
until NSE, after which almost all neutrons are bound in \(\alpha\) particles and can no longer decay, so the ratio \(n_{\hat{n}}/n_{\hat{p}}\) becomes fixed.\\
\indent
Using Eq.~(\ref{beta_exponential}) from weak freeze-out down to NSE at \(T \lesssim 100\) keV, the neutron-to-proton abundance evolves over time due to isospin flip reactions and neutron $\beta^-$ decay to
\begin{equation}\label{NSEepoch}
\frac{n_{\hat{n}}}{n_{\hat{p}}} = \frac{n_{n}}{n_{p}} \approx 0.135\,.
\end{equation}
Below the NSE epoch (\(100\, \mathrm{keV} \gtrsim k_{\rm{B}} T \gtrsim 10\,\mathrm{keV}\)), we can safely assume that all antibaryons have annihilated, while isospin flip reactions between \litiosette\ and \berilliosette\ become significant.
To determine the EC and PC rates, we must assess the fraction of free electrons and protons; that is, we need to subtract from their total density the number of protons and electrons bound in the nuclear species present at the NSE epoch. In this context, the Saha equation provides the equilibrium ratio between the product of the densities of free electrons and free protons (which are equal under the assumption of an electrically neutral Universe) and the total number of baryons (neutrons plus protons), including those in bound nuclear states, in a plasma. Therefore, it determines the density of protons relative to the total baryon density, and indirectly indicates how many protons are bound in nuclei and how many remain free.
However, the abundance of light isotopes such as protium, deuterium, tritium, and $^3$He is negligible in NSE, with only $^4$He present \citep{fuller_smith}. In fact, at NSE ($k_{\rm{B}}T \approx 0.1$ MeV), almost all protons are either free or bound in $\alpha$ particles, while nearly all neutrons are bound in $^4$He.
\begin{table*}
\caption{Particle densities and electron chemical potential during Big Bang nucleosynthesis.}\label{denden}
\centering
\begin{tabular}{c c c c c c}
\hline\hline
$t$ (s) & $k_{\rm B}T$ (keV) & $n_p \times 10^{7}$ & $n_{\hat{e}^{-}}$ & $n_{\hat{e}^{+}}$ & $\mu_e-m_ec^2$ \\
        &                    & ($10^{-7}\,a_0^{-3}$) & ($a_0^{-3}$)     & ($a_0^{-3}$)     & (a.u.) \\
\hline
3.156      & 87.233 & 16.8120 & 88.3470137     & 88.3470121     & -18778.865 \\
7.964      & 54.913 & 4.1938  & 1.26692624     & 1.26692582     & -18778.865 \\
20.098     & 34.568 & 1.0461  & 2.47121238e-03 & 2.47110776e-03 & -18778.838 \\
50.719     & 21.760 & 0.2609  & 2.10035855e-07 & 1.83939709e-07 & -18725.819 \\
127.990    & 13.698 & 0.0651  & 6.50968418e-09 & 9.70e-19       & -13083.666 \\
322.989    & 8.623  & 0.0162  & 1.62384586e-09 & 1.03e-37       & -8491.340 \\
815.074    & 5.428  & 0.0041  & 4.05069692e-10 & 6.44e-68       & -5504.808 \\
2,056.872  & 3.417  & 0.0010  & 1.01044983e-10 & 4.34e-116      & -3546.005 \\
5,190.598  & 2.151  & 0.0003  & 2.52057531e-11 & 1.35e-192      & -2281.927 \\
13,098.677 & 1.354  & 0.0001  & 6.28759452e-12 & 5.14e-314      & -1472.696 \\
31,560.000 & 0.872  & 0.00002  & 1.68120328e-12 & $\approx 0.0$   & -970.181  \\
\hline
\end{tabular}
\tablefoot{Proton ($n_p$), electron ($n_{\hat{e}^{-}}$), and positron ($n_{\hat{e}^{+}}$) number densities (in atomic units $a_0^{-3}$, where $a_0$ is the Bohr radius), as well as the electron chemical potential relative to the electron mass ($\mu_e - m_e c^2$, with $m_e c^2 = 18,775.13$ Hartrees), are given as functions of indicative cosmic time since the Big Bang (first column, time in seconds) and the corresponding temperature in keV (second column). Note that $n_p$ has been rescaled by $10^7$.}
\end{table*}
\begin{figure}
\centering
\includegraphics[width=0.4\textwidth]{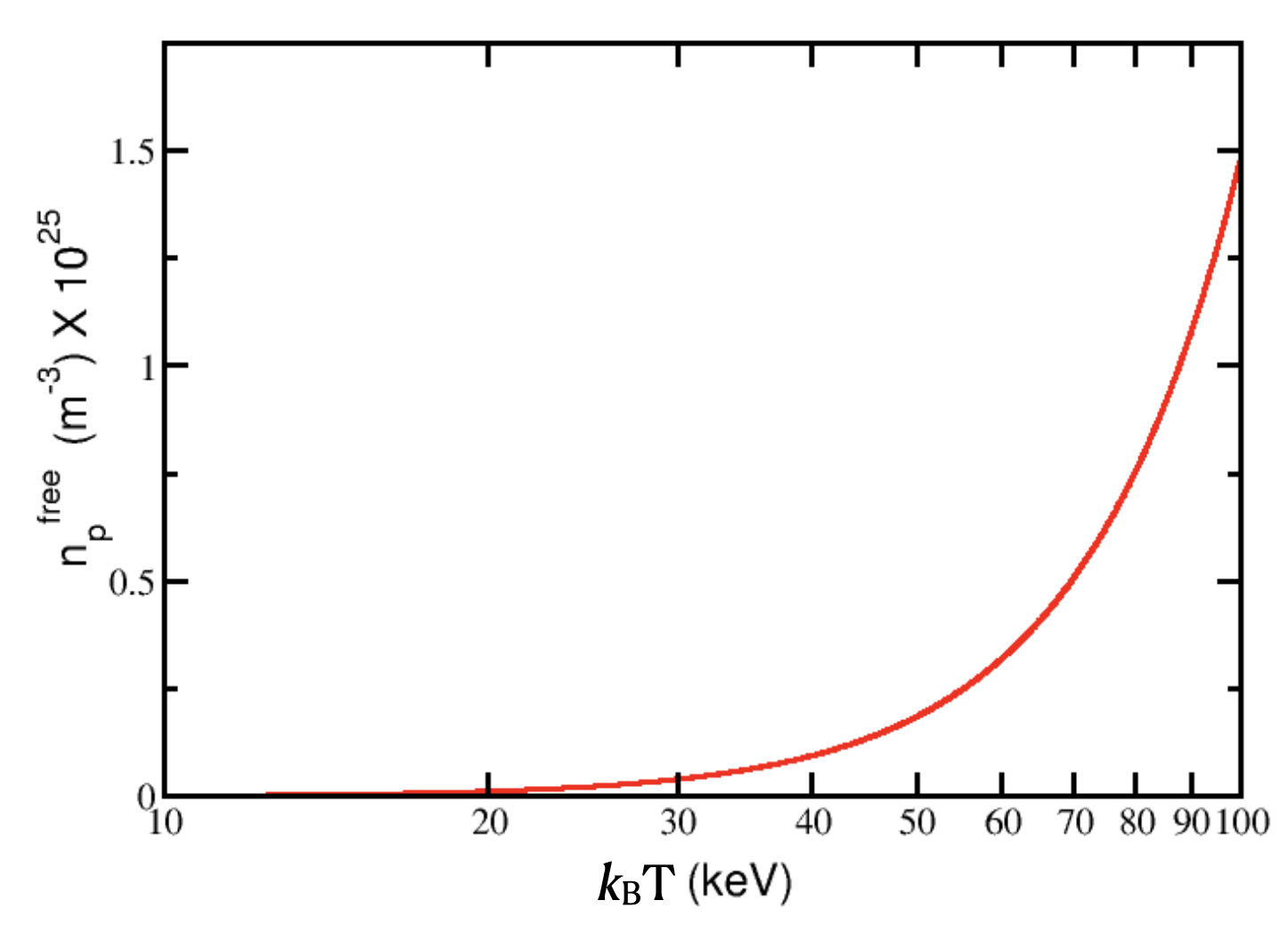}
\caption{Free proton density (m$^{-3}\times 10^{25}$) in total baryonic matter as a function of temperature $k_{\rm{B}}T$ (keV) for $n_{b} = \eta \times n_{\gamma}$ ($10~\mathrm{keV} < k_{\rm{B}}T < 100~\mathrm{keV}$) baryons/m$^{3}$, according to the Saha equation.}
\label{saha}
\end{figure}
In particular, applying the Saha equilibrium equation under NSE conditions shows that:
\begin{equation}
\label{proton_density_temperature}
n_{p}(T) \approx \frac{15}{17} n_b(T) \approx \eta\,\frac{15}{17}\,n_{\gamma}(T)\,,
\end{equation}
where the last relation follows from the WMAP-observed baryon-to-photon number density ratio \(\eta = n_{b}/n_{\gamma} = 6.11 \times 10^{-10}\) \citep{bar_den}, assuming that \(\eta\) remains essentially constant after BBN.
Using Eq. (\ref{NSEepoch}), the neutron density is
\begin{equation}\label{neutron_density_temperature}
n_{n}(T)\approx 0.135 \times n_p(T) \approx \eta\,\frac{2}{17}\,n_{\gamma}(T)\,.
\end{equation}
We note that the temperature dependence of the total photon density, $n_{\gamma}(T) \propto T^3$ (see Eq. (\ref{total_photon_density}) in the Appendix \ref{photden}), can be derived by assuming a Bose-Einstein distribution for massless particles, which implies that the radiation in the early Universe had a black-body spectrum.
Hence, from Eqs. (\ref{proton_density_temperature}) and (\ref{neutron_density_temperature}), at the NSE epoch, out of 17 baryons, about 2 are neutrons and 15 are protons, either free or bound in nuclei. Moreover, assuming that light nuclei (D, $^3$He, $^3$H, etc.) are negligible and only \(\alpha\) particles exist at the NSE epoch, among 17 baryons, 2$p$ + 2$n$ are in $^4$He, while 13 are free protons. Finally, the free proton density is
\begin{equation}
\label{free_proton_density}
n_{p}^{\mathrm{free}}(T)  \approx \frac{13}{17} n_b = \eta\,\frac{13}{17}\,n_{\gamma}(T).
\end{equation}
For $10\,\mathrm{keV} < k_{\rm{B}}T < 100\,\mathrm{keV}$, we have $1.4813\times10^{22}\ \mathrm{m^{-3}} < n_p^{\mathrm{free}} < 1.4813\times10^{25}\ \mathrm{m^{-3}}$, as shown in Fig. (\ref{saha}), where the fraction of free protons is plotted as a function of increasing temperature. In Tab. (\ref{denden}), the net number densities of protons at selected thermal energies relevant to the post-NSE epoch are provided.
\subsection{Lepton densities}
\paragraph{Electrons} Similarly, we define the net electron number density as
\begin{equation}
n_{e}=n_{\hat{e}^{-}}-n_{\hat{e}^{+}}\, ,
\end{equation}
where \(n_{\hat{e}^{-}}\) and \(n_{\hat{e}^{+}}\) are the electron and positron number densities, respectively. Assuming the large-scale charge neutrality of the Universe \citep{CapriniFerreira2005,Cameron2022} implies that
\begin{equation}
\label{electron-protons_equality}
n_{e} = n_{p}\, .
\end{equation}
Thus, once the net proton density \(n_{p}\) is known at a given epoch (see Eq. (\ref{proton_density_temperature})), the net electron density is also determined and can be used to compute the EC rate at that time.
\paragraph{Neutrinos}
Another process that may contribute to the destruction of \berilliosette~is AC, Eq. (\ref{eqn2}). The relevant question is whether a significant population of antineutrinos is present during the NSE epoch. After weak decoupling, neutrino–antineutrino pairs cease to interact with the $(e^{-}e^{+})$ plasma and subsequently free-stream, analogous to CMB photons after photon decoupling. Similar to the CMB, relic neutrinos retain a thermal distribution, but unlike photons, they obey Fermi–Dirac statistics:
\begin{equation}
\label{infinitesimal_neutrino_density}
\mathrm{d}n_{\nu}(E)=\frac{g}{2\pi^{2}\hbar^{3}c^{3}}k_{\rm{B}}^{3}T_{\nu}^{3}\frac{\beta^{3}E^{2}\,\mathrm{d}E}{e^{\beta E}+1},
\end{equation}
where $\beta=kT_{\nu}$.
The energy dispersion is given by \(E = pc\) due to the small mass of the neutrino. Note that for neutrinos, \(g = 1\) since they interact via the weak interaction, which only couples to left-handed neutrinos. The total density of neutrinos is
\begin{equation}
\label{total_density_neutrino}
n_{\nu}(E)=\frac{g}{2\pi^{2}\hbar^{3}c^{3}}k_{B}^{3}T_{\nu}^{3}\int\frac{\beta^{3}E^{2}\,\mathrm{d}E}{e^{\beta E}+1}=\frac{g}{2\pi^{2}\hbar^{3}c^{3}}k_{\rm{B}}^{3}T_{\nu}^{3}\frac{3}{4}2.404.
\end{equation}
In Eq. (\ref{total_density_neutrino}), $T_{\nu}$ differs from the photon temperature $T_{\gamma}$, as neutrinos decouple from photons once they decouple from electrons and are therefore no longer in thermal equilibrium.\\
\indent The relic neutrino number density is obtained from Eq.~(\ref{total_density_neutrino}) and may be compared directly with the photon number density in Eq.~(\ref{total_photon_density}) to give:
\begin{equation}
\label{ratio_den_neutrino_photon}
n_{\nu} = \frac{3}{4}\frac{g_{\nu}}{g_{\gamma}}\left(\frac{T_{\nu}}{T_{\gamma}}\right)^3 n_{\gamma},
\end{equation}
where $g_{\gamma}=2$ and $g_{\nu}=1$ for a single helicity state (electron neutrinos and antineutrinos each carry one helicity state, and thus we used $g_{\nu}=2$ for the combined $\nu_{e}-\bar{\nu}_{e}$ population). Moreover, after $e^{-}e^{+}$ annihilation, photon and neutrino temperatures are related by \citep{Dolgov2002, Weinberg2008Cosmology}
\begin{equation}
\label{neutrino_photon_temp}
T_{\nu} = \left(\frac{4}{11}\right)^{1/3} T_{\gamma},
\end{equation}
which, using Eq.~(\ref{ratio_den_neutrino_photon}), yields
\begin{equation}
\label{ratio_den_neutrino_photon2}
n_{\nu} = \frac{g_{\nu}}{g_{\gamma}} \frac{3}{11} n_{\gamma},
\end{equation}
for each of the neutrino flavours, including muon ($\nu_{\mu}$) and tau ($\nu_{\tau}$) neutrinos.
Electron neutrino–antineutrino densities are reported in Tab.~(\ref{tab_neutrino_dens_BBN}). These values may be multiplied by three to include the $\nu_{\mu}-\bar{\nu}_{\mu}$ and $\nu_{\tau}-\bar{\nu}_{\tau}$ contributions because, before their annihilation, $\mu^{\pm}$ and $\tau^{\pm}$ pairs were also in thermal equilibrium with the radiation field.
\begin{table}[hbt!]
    \caption{Neutrino and antineutrino number densities at the epoch of NSE.}
    \centering
    \begin{tabular}{c c c}
        \hline
        \hline
        $k_{\rm{B}}T (\mathrm{keV})$
        & \(n_{\gamma}\;(\mathrm{m^{-3}})\)
        & \(n_{\nu+\bar{\nu}}\;(\mathrm{m^{-3}})\) \\
        \hline
        10  & \(3.17027\times10^{31}\) & \(8.64618\times10^{30}\) \\
        20  & \(2.53621\times10^{32}\) & \(6.91695\times10^{31}\) \\
        30  & \(8.55972\times10^{32}\) & \(2.33447\times10^{32}\) \\
        40  & \(2.02897\times10^{33}\) & \(5.53356\times10^{32}\) \\
        50  & \(3.96283\times10^{33}\) & \(1.08077\times10^{33}\) \\
        60  & \(6.84778\times10^{33}\) & \(1.86758\times10^{33}\) \\
        70  & \(1.08740\times10^{34}\) & \(2.96564\times10^{33}\) \\
        80  & \(1.62318\times10^{34}\) & \(4.42685\times10^{33}\) \\
        90  & \(2.31112\times10^{34}\) & \(6.30307\times10^{33}\) \\
        100 & \(3.17027\times10^{34}\) & \(8.64618\times10^{33}\) \\
        \hline
    \end{tabular}
\tablefoot{Density of neutrino+antineutrino pairs $n_{\nu}+n_{\bar{\nu}}$ at the epoch of NSE calculated from the photon number density $n_{\gamma}$ (see Eq. (\ref{ratio_den_neutrino_photon2}) in the main text). To obtain the number density of neutrinos (or antineutrinos) only, divide the third column by 2, as we set \(\mu_{\nu} = 0\).}
    \label{tab_neutrino_dens_BBN}
\end{table}
Both $\nu_{\mu}$ and $\nu_{\tau}$ decouple from the primordial plasma while their associated charged leptons are still present in significant abundance. Consequently, they undergo thermal freeze-out with a number density comparable to that of electron neutrinos. Subsequently, the muons and tauons decay into their respective associated neutrinos.
\subsection{Proton and electron chemical potentials}
Once \( n_p(T) = n_e(T) \) is known, we can determine the chemical potentials of these species at NSE, assuming that a system of e$^-$ and e$^+$ pairs can be modelled as a Fermi gas in thermal equilibrium with the $\gamma$ radiation, thus characterised by their temperature and a single chemical potential via a relativistic energy-momentum dispersion relation.
While in the upper part of our temperature range ($k_{\rm{B}}T \geq 50$ keV), electrons and positrons remain in pair equilibrium ($\mu_{\gamma} = 0$, $\mu_{e^+} = -\mu_{e^-} \equiv -\mu_e$), at lower temperatures positrons progressively freeze out and annihilate. In particular, the assumption of pair equilibrium breaks down below $k_{\rm{B}}T \approx 20$ keV ($\mu_{\gamma} \neq 0$). In this temperature regime, in principle $\mu_{e^+} \neq -\mu_{e^-}$, and one should determine the chemical potentials of particles and their respective antiparticles using kinetic equations. This task is well beyond the scope of this contribution; therefore, we assumed that the Fermi gas is in equilibrium and that a single chemical potential describes both species.\\
\indent
In this context, matter–antimatter asymmetry can be described as a non-zero chemical potential for particle extraction from, or insertion into, the Fermi sea. Consider a gas of positron–electron pairs. Decoupling of electrons occurs at $k_{\rm{B}}T \approx 1\,\mathrm{MeV}$ ($t \approx 1.71\,\mathrm{s}$), after the weak decoupling of neutrinos. At this epoch, electrons are relativistic, and their number density is given by
\begin{equation}
\label{example_density_calc}
n_{\hat{e}}(T) = \frac{g_{i}}{2\pi^{2}\hbar^{3}c^{3}}
\int \frac{\sqrt{E^{2}-m_{e}^{2}c^{4}}\,E\,\mathrm{d}E}
{\exp\left[\beta\left(E-\mu_{e}\right)\right]+1}\, ,
\end{equation}
where $E$ is the relativistic kinetic energy, $\mu_e$ is the electron (or positron) chemical potential, and $m_e$ is the electron mass.\\
\indent
Similarly, the antiparticle number density is
\begin{equation}
\label{example_density_calc_anti}
n_{\overline{e}}(T) = \frac{g_{i}}{2\pi^{2}\hbar^{3}c^{3}}
\int \frac{\sqrt{E^{2}-m_{e}^{2}c^{4}}\,E\,\mathrm{d}E}
{\exp\left[\beta\left(E+\mu_{e}\right)\right]+1}\, .
\end{equation}
Owing to the particle-antiparticle asymmetry, the electron net number density is the difference between Eqs. (\ref{example_density_calc}) and (\ref{example_density_calc_anti})
\begin{eqnarray}
\label{electron_density_integral_chemic_pot}
n_{e}(T)
&=& \frac{g_{e}}{2\pi^{2}\hbar^{3}c^{3}}
\int \frac{\sqrt{E^{2}-m_{e}^{2}c^{4}}\,E\,\mathrm{d}E}
{e^{\frac{E-\mu_{e}}{k_{\rm{B}}T}}+1} \nonumber \\
&-& \frac{g_{e}}{2\pi^{2}\hbar^{3}c^{3}}
\int \frac{\sqrt{E^{2}-m_{e}^{2}c^{4}}\,E\,\mathrm{d}E}
{e^{\frac{E+\mu_{e}}{k_{\rm{B}}T}}+1} = n_{p}(T),
\end{eqnarray}
where the last equality implies charge neutrality, as in Eq.~(\ref{electron-protons_equality}).
In Eq. (\ref{electron_density_integral_chemic_pot}), while electrons become mildly relativistic in the range 10 keV < $k_{\rm{B}}T$ < 100 keV and therefore the relativistic particle dispersion law is used, protons remain deeply non-relativistic over the entire temperature range considered here.
In Tab. (\ref{denden}), the number densities of electrons and positrons at selected thermal energies relevant to the post-NSE epoch are reported.
Given \(n_{p}(T)\) from Eq. (\ref{proton_density_temperature}), we can, in principle, obtain \(\mu_{e}(T)\) by inverting Eq. (\ref{electron_density_integral_chemic_pot}). Once \(\mu_{e}\) is known, the densities of electrons and positrons, as well as their population factors entering the EC rate, can be obtained directly from the respective Fermi–Dirac distributions in Eq. (\ref{electron_density_integral_chemic_pot}) for particles and antiparticles. Although the electron number density reaches values well above solid density (approximately three orders of magnitude higher; see the fourth column in Tab. (\ref{denden})), the high temperatures considered here keep the leptonic component in the non-degenerate regime. However, degeneracy increases as temperature decreases (\(0.006 < n_e\lambda_T^3 < 0.2\), where \(\lambda_T \propto T^{-1/2}\) is the thermal de Broglie wavelength).\\
\indent
In the last column of Tab. (\ref{denden}), we present the electron chemical potential over the temperature range in which isospin flip drives the \litiosette--\berilliosette\ equilibrium via Eq. (\ref{eqn1}), using a bisection method to invert Eq. (\ref{electron_density_integral_chemic_pot}). As the temperature decreases, the electron chemical potential \(\mu_e\) increases (becomes less negative; see the last column of Tab. (\ref{denden}); please note that \(\mu_e\) is defined relative to the electron rest mass \(m_e c^2= 18,775.13\) Hartrees), reflecting the gradual increase in the degeneracy parameter throughout the evolution. Moreover, the positron and electron number densities (see Tab. (\ref{denden})) are nearly identical at high temperatures ($k_{\rm{B}}T \geq 50$ keV), implying that the net electron number density \(n_{e}\) is almost zero. In contrast, in the low-temperature regime (below 20 keV), a non-zero \(\mu_{e}\) (see Tab. (\ref{denden})) produces the required particle–antiparticle imbalance imposed by global charge neutrality. Therefore, at the higher end of our temperature range, electrons and positrons remain almost in pair equilibrium, while at lower temperatures, positrons progressively freeze out and annihilate, disrupting pair equilibrium.\\
\indent
Similarly, one can obtain the chemical potential of protons at NSE from Eq. (\ref{electron_density_integral_chemic_pot}), even though we do not treat them as relativistic particles in our calculations.
\subsection{Neutrino chemical potential}\label{neu_chem}
In contrast to electrons and positrons, we assume a vanishing neutrino chemical potential, $\mu_{\nu}=0$.
Before weak freeze-out, neutrinos remain in thermal equilibrium with photons and $e^{-}e^{+}$ pairs through
\begin{equation}
\label{before_weak_decoupling}
\gamma+\gamma \longleftrightarrow e^{-}+e^{+} \longleftrightarrow \nu_{e}+\bar{\nu}_{e}.
\end{equation}
This assumption is justified by the very small lepton–antilepton asymmetry at weak freeze-out $(k_{\rm{B}}T \simeq 3~\mathrm{MeV},\ t \simeq 0.2~\mathrm{s})$. In particular, the difference between the electron and positron number densities is negligible:
\begin{equation}
\label{delta_n_3_mev}
\left( n_{\hat{e}^{-}} - n_{\hat{e}^{+}} \right) |_{kT = 3~\mathrm{MeV}} \approx 0.
\end{equation}
Since electrons and positrons have equal abundances at that stage, neutrinos and antineutrinos also freeze out with equal number densities, which implies $\mu_{\nu_e}\approx 0$ for all neutrino flavours.
\newpage
\section{$^7$Be to $^7$Li decay rates}
\label{app:rates_final}
\begin{table}[hbt!]
\caption{Excited-to-excited $^7$Be$\rightarrow{}^7$Li decay rates.}
\label{tab:rate_Be7_ee}
\centering
\begin{tabular}{r r r r r}
\hline\hline
$t$ & $k_{\rm{B}}T$ & $R_{e^-}$  & $R_{e^+}$ & $R_{\bar{\nu}}$ \\
(s) & (keV) & (s$^{-1}$)  & (s$^{-1}$) & (s$^{-1}$) \\
\hline
3.156 & 87.23 & 3.693e-08 & 3.089e-06 & 3.053e-06 \\
7.964 & 54.91 & 3.445e-08 & 1.831e-06 & 1.797e-06 \\
20.10 & 34.57 & 1.046e-07 & 2.471e-03 & 2.471e-03 \\
50.72 & 21.76 & 2.610e-08 & 2.100e-07 & 1.839e-07 \\
127.99 & 13.70 & 6.510e-09 & 6.510e-09 & 9.698e-19 \\
322.99 & 8.623 & 1.624e-09 & 1.624e-09 & 1.029e-37 \\
815.07 & 5.428 & 4.051e-10 & 4.051e-10 & 6.435e-68 \\
5190.6 & 2.151 & 2.521e-11 & 2.521e-11 & 1.349e-192 \\
13099 & 1.354 & 6.288e-12 & 6.288e-12 & 5.143e-314 \\
31560 & 0.8723 & 1.681e-12 & 1.681e-12 & 0.0 \\
\hline
\end{tabular}
\tablefoot{Excited to excited decay rates as a function of temperature (keV) or indicative time since the Big Bang (s). $R_{e^-}$: electron capture; $R_{e^+}$: positron decay; $R_{\bar{\nu}}$: antineutrino capture.}
\end{table}
\begin{table}[hbt!]
\caption{Excited-to-ground $^7$Be$\rightarrow{}^7$Li decay rates.}
\label{tab:rate_Be7_eg}
\centering
\begin{tabular}{r r r r r}
\hline\hline
$t$ & $k_{\rm{B}}T$ & $R_{e^-}$  & $R_{e^+}$ & $R_{\bar{\nu}}$ \\
(s) & (keV) & (s$^{-1}$)  & (s$^{-1}$) & (s$^{-1}$) \\
\hline
3.156 & 87.23 & 3.862e-07 & 2.449e-06 & 1.016e-06 \\
7.964 & 54.91 & 5.259e-09 & 2.579e-06 & 2.078e-07 \\
20.10 & 34.57 & 1.027e-11 & 2.630e-06 & 4.529e-08 \\
50.72 & 21.76 & 9.104e-16 & 2.630e-06 & 1.027e-08 \\
127.99 & 13.70 & 3.622e-17 & 2.630e-06 & 2.422e-09 \\
322.99 & 8.623 & 8.789e-18 & 2.630e-06 & 5.807e-10 \\
815.07 & 5.428 & 2.239e-18 & 2.630e-06 & 1.407e-10 \\
5190.6 & 2.151 & 2.315e-19 & 2.630e-06 & 8.572e-12 \\
13099 & 1.354 & 7.445e-20 & 2.630e-06 & 2.118e-12 \\
31560 & 0.8723 & 2.815e-20 & 2.630e-06 & 5.665e-13 \\
\hline
\end{tabular}
\tablefoot{Excited to ground decay rates as a function of temperature (keV) or indicative time since the Big Bang (sec). $R_{e^-}$: electron capture; $R_{e^+}$: positron decay; $R_{\bar{\nu}}$: antineutrino capture.}
\end{table}
\begin{table}[hbt!]
\caption{Ground-to-ground $^7$Be$\rightarrow{}^7$Li decay rates.}
\label{tab:rate_Be7_gg}
\centering
\begin{tabular}{r r r r r}
\hline\hline
$t$ & $k_{\rm{B}}T$ & $R_{e^-}$  & $R_{e^+}$ & $R_{\bar{\nu}}$ \\
(s) & (keV) & (s$^{-1}$)  & (s$^{-1}$) & (s$^{-1}$) \\
\hline
3.156 & 87.23 & 5.233e-07 & 0.0000 & 3.628e-07 \\
7.964 & 54.91 & 7.332e-09 & 0.0000 & 2.814e-09 \\
20.10 & 34.57 & 4.533e-11 & 0.0000 & 1.027e-11 \\
50.72 & 21.76 & 1.030e-14 & 0.0000 & 9.104e-16 \\
127.99 & 13.70 & 2.422e-15 & 0.0000 & 3.622e-17 \\
322.99 & 8.623 & 5.807e-16 & 0.0000 & 8.789e-18 \\
815.07 & 5.428 & 1.407e-16 & 0.0000 & 2.239e-18 \\
5190.6 & 2.151 & 8.572e-18 & 0.0000 & 2.315e-19 \\
13099 & 1.354 & 2.118e-18 & 0.0000 & 7.445e-20 \\
31560 & 0.8723 & 5.665e-19 & 0.0000 & 2.815e-20 \\
\hline
\end{tabular}
\tablefoot{Ground to ground decay rates as a function of temperature (keV) or indicative time since the Big Bang (sec). $R_{e^-}$: electron capture; $R_{e^+}$: positron decay; $R_{\bar{\nu}}$: antineutrino capture.}
\end{table}
\begin{table}[hbt!]
\caption{Ground-to-excited $^7$Be$\rightarrow{}^7$Li decay rates.}
\label{tab:rate_Be7_ge}
\centering
\begin{tabular}{r r r r r}
\hline\hline
$t$ & $k_{\rm{B}}T$ & $R_{e^-}$  & $R_{e^+}$ & $R_{\bar{\nu}}$ \\
(s) & (keV) & (s$^{-1}$)  & (s$^{-1}$) & (s$^{-1}$) \\
\hline
3.156 & 87.23 & 6.444e-08 & 0.0000 & 4.523e-10 \\
7.964 & 54.91 & 1.277e-09 & 0.0000 & 2.035e-12 \\
20.10 & 34.57 & 2.471e-12 & 0.0000 & 1.027e-15 \\
50.72 & 21.76 & 2.100e-16 & 0.0000 & 9.104e-20 \\
127.99 & 13.70 & 6.510e-18 & 0.0000 & 3.622e-21 \\
322.99 & 8.623 & 1.624e-18 & 0.0000 & 8.789e-22 \\
815.07 & 5.428 & 4.051e-19 & 0.0000 & 2.239e-22 \\
5190.6 & 2.151 & 2.521e-20 & 0.0000 & 2.315e-23 \\
13099 & 1.354 & 6.288e-21 & 0.0000 & 7.445e-24 \\
31560 & 0.8723 & 1.681e-21 & 0.0000 & 2.815e-24 \\
\hline
\end{tabular}
\tablefoot{Ground to excited decay rates as a function of temperature (keV) or indicative time since the Big Bang (sec). $R_{e^-}$: electron capture; $R_{e^+}$: positron decay; $R_{\bar{\nu}}$: antineutrino capture.}
\end{table}
\end{appendix}
\end{document}